\def\nh{{n_{\rm H}}}
\def\nh2{{n(\rm H_2)}}
\def\h2{${\rm H_2}$}

\def\3cm{\rm {cm^{-3}}}
\def\2cm{\rm {cm^{-2}}}
\def\s-1{\rm {s^{-1}}}

\def\kms {\hbox{${\rm km\,s}^{-1}$}}
\def\Kkms {\hbox{${\rm K}\,{\rm km}\,{\rm s}^{-1}$}}

\def\twco{\hbox{$^{12}$CO}}
\def\thco{\hbox{$^{13}$CO}}
\def\oi{\hbox{\rm {[O {\scriptsize I}]}}}

\def\ci{\hbox{\rm {[C {\scriptsize I}]}}}
\def\cii{\hbox{\rm {[C {\scriptsize II}]}}}
\def\thcii{\hbox{\rm {[$^{13}$C {\scriptsize II}]}}}

\def\hii{\hbox{\rm {H {\scriptsize II}}}}
\def\c18o{\hbox{C$^{18}$O}}
\def\cilo{\hbox{\rm {[C {\scriptsize I}]}~$^3P_1\to{^3P_0}$}}
\def\ciup{\hbox{\rm {[C {\scriptsize I}]}~$^3P_2\to{^3P_1}$}}
\def\ciitrans{\hbox{\rm {[C {\scriptsize II}]}~$^2P_{3/2}\to{^2P_{1/2}}$}}

\def\oilo{\hbox{\rm {[O {\scriptsize I}]}~$^3P_2\to{^3P_1}$}}
\def\oiup{\hbox{\rm {[O {\scriptsize I}]}~$^3P_0\to{^3P_1}$}}
\def\Xlev{\hbox{\rm {X$^1\Sigma_g^+$}}}
\def\Lylev{\hbox{\rm {B$^1\Sigma_u$}}}
\def\Werlev{\hbox{\rm {C$^1\Pi_u$}}}

\documentclass[traditabstract]{aa}
\usepackage[varg]{txfonts}
\usepackage{natbib}

\usepackage{amsmath}
\bibpunct{(}{)}{;}{a}{}{,} 
\usepackage{multirow,tabularx}
\usepackage{caption}
\usepackage{subcaption}
\usepackage{algorithm}
\usepackage{algpseudocode}
\makeatletter
\renewcommand{\ALG@name}{adaptive}
\makeatletter\newcommand\l@adaptive{\@dottedtocline{1}{1.5em}{2.3em}}\makeatother

\usepackage{graphicx}

\usepackage{ulem}
\usepackage{chemformula}
\let\ce\ch
\usepackage{epsfig}
\usepackage{amsmath,amssymb}

\usepackage{psfrag}
\usepackage{units}
\usepackage{textgreek}
\newcommand{\Kt}{KOSMA-{\texttau}}
\usepackage{hyperref}
\usepackage{relsize}  

\usepackage{makecell} 
\usepackage{longtable} 
\usepackage[online]{threeparttablex}
\usepackage{supertabular} 
\usepackage{adjustbox}
\usepackage{xcolor}
\usepackage{xfrac}
\usepackage{alphalph}
\usepackage{float}
\usepackage{placeins}

\newcommand*\ruleline[1]{\par\noindent\raisebox{.8ex}{\makebox[\linewidth]{\hrulefill\hspace{1ex}\raisebox{-.8ex}{#1}\hspace{1ex}\hrulefill}}}
\algblockdefx[Foreach]{Foreach}{EndForeach}[1]{\textbf{foreach} #1 \textbf{do}}{\textbf{end foreach}}

\begin{document}
\title{The KOSMA-\texttau\, PDR Model}
\subtitle{I. Recent updates to the numerical model of photo-dissociated regions}
\author{M. R\"ollig\inst{1}
\and
V. Ossenkopf-Okada\inst{1}
}
\offprints{M. R\"ollig}
\institute{
 I. Physikalisches Institut der Universit\"at zu K\"oln, Z\"ulpicher Stra\ss e 77, 50937 K\"oln, Germany\\
 \email{roellig@ph1.uni-koeln.de} 
}
\date{Received  / Accepted  }
%
\titlerunning{The KOSMA-\texttau\, PDR Model}


%
\abstract{Numerical models of Photodissociation Regions (PDRs) are an essential tool to quantitatively understand observations of massive star forming regions through simulations. Few mature PDR models are available and the Cologne \Kt{}  PDR model is the only sophisticated model that uses a spherical cloud geometry thereby allowing us to simulate clumpy PDRs. We present the current status of the code as reference for modelers and for observers that plan to apply {\Kt} to interpret their data.
For the numerical solution of the chemical problem we present a  superior Newton-Raphson stepping algorithm and discuss strategies to numerically stabilize the problem and speed up the iterations. The chemistry in \Kt{} is upgraded to include the full surface chemistry in an up-to-date formulation and we discuss  a novel computation of branching ratios in chemical desorption reactions. The high dust temperature in PDRs leads to a selective freeze-out of oxygen-bearing ice species due to their higher condensation temperatures and we  study changes in the
ice mantle structures depending on the PDR parameters, in particular the impinging UV field. Selective freeze-out can produce enhanced \ce{C} abundances and higher gas temperatures resulting in a fine-structure line emission of atomic carbon {\ci} enhanced by up to 50\% if surface reactions are considered. We show how recent ALMA observations of \ce{HCO+} emission in the Orion Bar with high spatial resolution on the scale of individual clumps can be interpreted in the context of non-stationary, clumpy PDR ensembles.
Additionally, we introduce \texttt{WL-PDR}, a simple plane-parallel PDR model written in Mathematica to act as numerical testing environment of PDR modeling aspects.}    
 
\keywords{(ISM:) photon-dominated region (PDR)
--- Astrochemistry
--- ISM: molecules
--- ISM: kinematics and dynamics
--- ISM: abundances
--- Methods: numerical}
\maketitle
\section{Introduction}
Numerical models of Photodissociation Regions (PDRs) are a valuable tool to model their chemical and thermodynamic structure and to simulate their line and continuum emission \citep{TH85,htt91,leBurlot93,sternberg1995,kaufman99,lepetit2006,comparison07,roellig2013dust}. This is particularly relevant, since ``all neutral atomic hydrogen gas and a large fraction of the molecular gas in the Milky Way Galaxy and external galaxies lie in PDRs'' \citep{ht97}. 

The transition from the atomic to the molecular phase in the interstellar medium (ISM) is controlled by the balance between photo-dissociation by far ultra-violet radiation (FUV) and the recombination of atomic hydrogen on grain surfaces. Due to its nature as a line absorption process \citep{federman79}, photo-dissociation of H$_2$ is subject to efficient shielding that leads to a sharp transition from H to H$_2$ . This is particularly true for dense gas ($n>10^5$~cm$^{-3}$). This transitions has also been studied analytically and compared with corresponding numerical results \citep{sternberg1988,krumholz2008,krumholz2009,mckee2010,sternberg2014,bialy2016,bialy2017,sternberg2021}.

\begin{figure*}[htb]
\resizebox{\hsize}{!}{\includegraphics{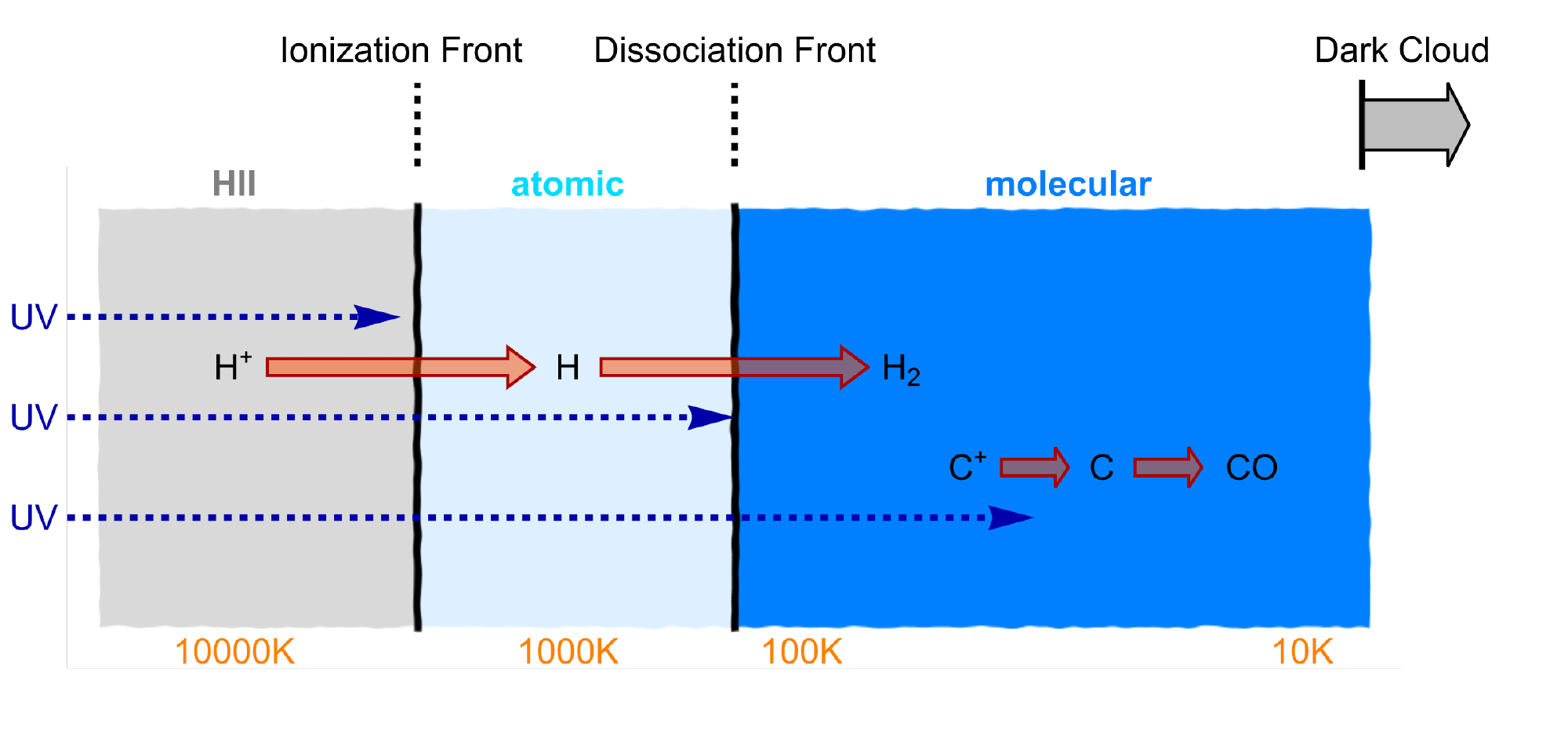}}
\caption{Simplified structure of a PDR. UV radiation (coming from the left) is progressively attenuated while penetrating a gas cloud. This leads to a chemical stratification and a strong temperature gradient.}
\label{fig:structure}
\end{figure*}

Numerical models of the physical and chemical conditions of the ISM compute the balance between formation and destruction of the considered chemical species. The respective reaction rates depend on the local temperature which is the result of balancing local heating and cooling processes. Cooling is predominantly done by spectral line emission of energetically excited states of atomic and molecular species. 
In PDRs the heating is dominated by photo-electric heating, vibrational de-excitation of electronically excited H$_2$ and cosmic ray heating. Both, heating and cooling, depend on the chemical abundances of the relevant species, which numerically asks for an iterative solution. 

Astrochemical models of the ISM chemistry date back to the early 1950s \citep{bates1951} and have been used to explain the abundance of early observed molecules. They have been constantly extended since then and most modern astrochemical ISM models also simulate chemical kinetic processes on dust-grain surfaces with various degrees of complexity \citep{tielens1982,hasegawa1993,garrod2013}. The first species that has been recognized as almost exclusively formed in space on grain surfaces is molecular hydrogen \citep{gould1963,hollenbach1971} and all PDR models have to include this reaction to yield physically reasonable results. For an extended review of grain surface models we refer the reader to \citet{cuppen2017}.

The first PDR models were presented almost 50 years ago to explain fine-structure emission observed with sub-millimeter instruments \citep{melnick1979, russel1980,russel1981,stacey1983,crawford85}.  The earliest models \citep{glassgold1974,glassgold1975,glassgold1976,langer1976,black1977,clavel1978,dejong1980} have been applied to low gas density $n<10^3$~cm$^{-3}$ illuminated by the ambient UV radiation. Higher gas density models with more intense radiation have been performed by \citet{tielens1985, sternberg1989, leBurlot93,vandishoeck1986, black1987,vandishoeck1988,sternberg1995,draine1996}. Some of these models have been further developed since then reaching a high level of sophistication. Here, we can only name a few prominent ones: The PDR~Toolbox\footnote{\url{http://dustem.astro.umd.edu/}} \citep{kaufman99,kaufman06,pound2008,hollenbach2009} provides an extensive set of computed model grids and user friendly tools to apply the model results to data. The Meudon PDR code\footnote{\url{https://ism.obspm.fr/}}  \citep{lepetit2006,goicoechea2007,lebourlot2012,bron2014} is  a  plane-parallel code with a particular focus on a detailed treatment of physical and chemical processes, in particular radiative transfer and shielding processes. It computes the line intensities of tens of atomic and molecular species. The chemistry includes hundreds of species with reactions in gas phase and on surfaces. The code is publicly available together with an extensive online database of model grids and analysis tools. The \Kt{} group collaborates with both teams to integrate the model results into their online databases and tools. 

The \texttt{CLOUDY}\footnote{\url{https://gitlab.nublado.org/cloudy}} code  is a spectral synthesis code designed to simulate conditions in interstellar matter under a broad range of conditions \citep{ferland2013,ferland2017}. With an initial focus on the ionized gas it has been extended to other regimes including PDRs \citep{vanhoof2004,shaw05,abel05, abel06a,abel06b}. 
PDR model physics has also been included in a number of other numerical models. \texttt{ProDiMo}\footnote{\url{https://www.astro.rug.nl/~prodimo/}}, a model of protoplanetary disks including gas phase, X-ray and UV-photo-chemistry, gas heating and cooling balance, disk structure and (dust and line) radiative transfer, is actively developed and extended \citep{woitke2009,kamp2010,woitke2016,kamp2017,thi2019,thi2020}. Other disk model are for instance \texttt{DALI}, a thermo-chemical models of protoplanetary disk atmospheres  \citep{bruderer2009a,bruderer2009b,bruderer2010,bruderer2012,bruderer2013}. A general model for the thermochemical structure of disks presented by \citet{gorti2004,gorti2008} was recently updated to include  the effects of gas-grain chemistry in protoplanetary disks using a three-phase approximation \citep{ruaud2019}. \citet{agundez2018} developed a generic model of protoplanetary disk, with a focus to the photochemistry, and applied it to a T Tauri and a Herbig Ae/Be disk.

 \texttt{UCL\_PDR}\footnote{\url{https://uclchem.github.io/ucl_pdr/}} is a publicly available, one-dimensional and time-dependent PDR model \citep{bell2005,bell2006,priestley2017} that has been extended into the first 3-dimensional PDR code \texttt{3D-PDR}\footnote{\url{https://uclchem.github.io/3dpdr}} \citep{bisbas2012,gaches2019}. The \Kt{} team used their clumpy PDR model as building blocks for KOSMA-\texttau-3D, to model arbitrary 3-dimensional geometries \citep{cubick08,andree2017,yanitski2022}.
 
 Most numerical models assume stationary conditions and do not solve the time-dependence, however the importance of dynamical effects in the ISM lead to a number of new non-stationary models. A first approach was to couple (magneto) hydro-dynamical (MHD) models with PDR models to post-process their output \citep[e.g.][]{levrier2012,bisbas2015} and as sub-grid models \citep{li2018}. With the recent growth of computing power MHD models started to numerically solve PDR physics and small chemical networks in parallel \citep{nelson1997,glover2010,grassi2014,walch2015,seifried2016,girichidis2016,valdivia2016,seifried2017,franeck2018,inoue2020,bisbas2021}. Some PDR codes solve the PDR physics and chemistry self-consistently with the dynamics but in a reduced dimensionality \citep{kirsanova2009,akimkin2015,akimkin2017,bron2018,kirsanova2020}.  

Because of the high model complexity, the large variety in model assumptions and numerical approximations it remains difficult to compare results of different PDR model codes. The PDR benchmark was a first attempt to cross-calibrate different PDR model codes against each other \citep{comparison07}. The {\Kt} model is one of the models that participated in the benchmark study. It is the only PDR model employing a spherical model geometry and has been applied to a large variety of studies in the recent 20 years.

PDR models are constantly improved and calibrated against observations of Galactic and Extragalactic PDRs. Major constraints were obtained in the last years by a growing number of observations of the main PDR cooling lines, in particular the fine-structure lines of \cii{} and \oi{}.
Modern instruments such as (up)GREAT \citep[German REceiver for Astronomy at Terahertz Frequencies ][]{heyminck2012,risacher2018} on board the Stratospheric Observatory for Infrared Astronomy \citep[SOFIA,][]{young2012} allow to observe Galactic PDRs with high spatial and very high spectral resolution showing complex line profiles due to source kinematics, optical thickness effects, and foreground absorption \citep[e.g.][]{pabst2019, kirsanova2019,schneider2021,kabanovic2022}. 
Similar studies for the Large Magellanic Clouds (LMC) \citep{okada2019b} together with the first detection of velocity resolved \thcii\ emission in the LMC \citep{okada2019a} show the potential of these observations for Extragalactic studies \citep[other examples include][]{roellig2016,cormier2019,madden2020,bigiel2020,tarantino2021}.
Future PDR models may also be applied to ALMA observations of protostars and hot cores and corinos which reveal a rich complex organic chemistry (COM) that current PDR models have yet to explain. Data from future instruments such as the James Webb Space Telescope \citep[JWST, ][]{gardner2006} and FYST \citep{parshley2018}, will challenge present PDR model predictions. The near- and mid-infrared (IR) instruments on board of JWST allow for very detailed studies of the hot gas and dust on PDR surfaces and the detailed knowledge of the \ce{H2} excitation and the PAH energetics \citep{okada2013} will help to further calibrate these models \citep{berne2022}. Consequently, an active development of sophisticated PDR models is essential to deal with the wealth of data from current and near-future observations. The \Kt{} model will try to contribute with its particular strengths.

In this paper we report on recent major developments of the model and give an updated overview over the model properties. In particular, we present an upgrade of the model chemistry in \Kt{} to account for the full grain-surface chemistry in a quasi three-phase model  allowing us to study the interplay between ice formation on grains and PDR physics.

\section{Models of PDRs - General properties}
\begin{figure*}[htb]
	\resizebox{\hsize}{!}{\includegraphics{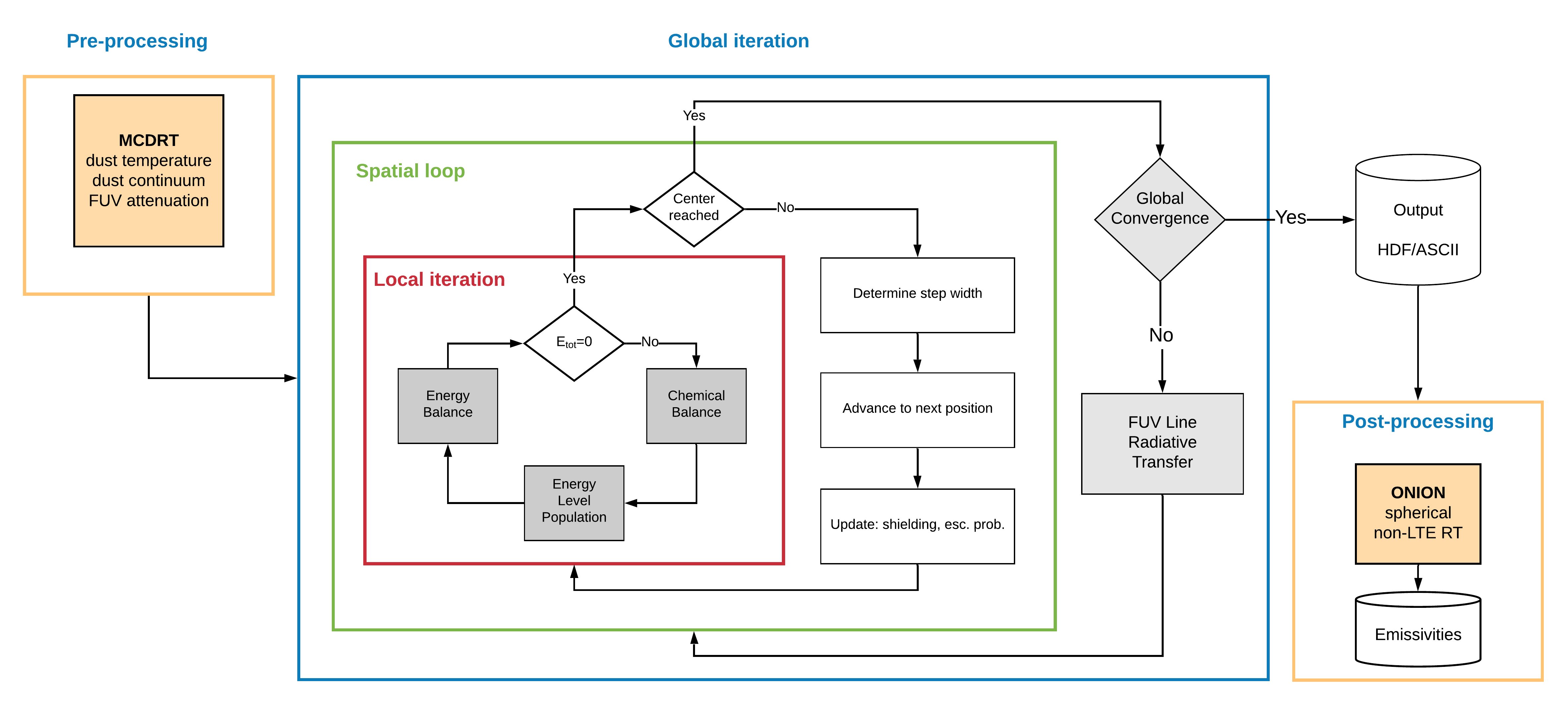}}
	\caption{General numerical scheme that needs to be solved in \Kt. Local iterations (red) are performed on every spatial grid point, spatial iterations (green) are performed over all spatial grid points in the model, and global iterations (blue) repeat the spatial loop if necessary until numerical convergence is reached. }
	\label{fig:numscheme}
\end{figure*}

Figure~\ref{fig:structure} shows the general structure of a PDR. Attenuation of FUV photons produces a chemical stratification with the more stable chemical species more abundant closer to the surface of the PDR, while species that are easily dissociated can only survive if sufficiently shielded from the FUV field. The chemical stratification is accompanied by a strong temperature gradient from the surface to the shielded parts of the PDR because the main heating takes place via ejection of photo-electrons from dust grains and subsequent collisional energy transfer from the ejected hot electrons to the gas particles \citep{bt94,WD01PEH}. A decreased FUV flux reduces the photo-electric heating (PEH) significantly and results in lower temperatures. FUV attenuation similarly affects heating by vibrational de-excitation of \ce{H2} which dominates over PEH at higher densities \citep{roellig06}. In this paper we denote total FUV field strength with $\chi$ in units of the Draine field $\chi_D=2.6\times 10^{-3}$~erg~s$^{-1}$~cm$^{-2}$ (integrated from 91.2 to 200 nm) \citep{draine78}.\footnote{\citet{habing1968} provided an alternative description which is taken as the reference with scaling factor $G_0$ in some models. To convert between the two descriptions use $\chi_D=1.7 G_0$.   }

Fig.~\ref{fig:numscheme} shows the numerical scheme any PDR model has to solve. It can be divided into four distinct problems (gray boxes). Three of them are local but the fourth one couples them non-locally, asking for a multi-layer iteration scheme. The local problems are: \textbf{1a)} the local chemical problem, i.e. solving the chemical balance equations at a given position in the PDR. \textbf{1b)} the local excitation, i.e. solving the energy level population of all species relevant for cooling and for comparison with observations at a given position in the PDR. \textbf{1c)} the local energy balance, i.e. computing the gas and dust temperature at a given position in the PDR. Those problems are mutually coupled asking for an iterative local solution. As they have to be solved at every spatial point of the model, they are embedded in a spatial loop. The fourth problem is \textbf{2)}, the solution of the radiative transfer (RT) equation. This step provides quantities such as the photo-dissociation rates of important species, the FUV intensity across the model cloud, and others. Step \textbf{2)} non-locally couples all positions within a model PDR. Each individual point requires the solution of all other points. Hence a global iterative solution is needed. Problems \textbf{1a)}-\textbf{1c)}  are solved iteratively at every spatial location in the cloud. Using this set of solutions the radiative transfer problem \textbf{2)} is solved over the whole cloud geometry.  As a consequence,  the 
local physical conditions computed in the previous iteration will be updated due to revised absorption and emission of IR and UV photons, which results in modified local cooling and heating capabilities leading to a new temperature solution and therefore a new chemical solution. Global iteration of the scheme ( \textbf{1a)}-\textbf{1c)} - \textbf{2)} ) is then performed until a predefined convergence criterion is met. An exception to this scheme are semi-infinite plane-parallel  models. In these models, radiation can only enter and escape from one side of the cloud. As a consequence, the UV transfer is computed as the code advances along the spatial grid without the need for separate loops.

Any other model geometry with the possibility of radiation entering and escaping from different directions requires repeated iterations on the spatial grid. The global as well as the local numerical iterations are stopped after predefined convergence criteria are met. 

Typical input quantities of PDR models are
\begin{itemize}
	\item model geometry, e.g. maximum visual extinction $A_\mathrm{V,max}$, total radius $R_\mathrm{tot}$, disk size, directed or isotropic illumination, inclination angle, etc.
	\item  thermodynamic properties of the gas, e.g. isobaric or isochoric conditions with fixed or variable density profile 
	\item macro-physical parameters, e.g. total gas density $n$, pressure $p$, illuminating FUV spectrum and intensity $\chi$, cosmic ray (CR) and X-ray (XR) ionization rate $\zeta_\mathrm{CR,XR}$
	\item heating and cooling  processes to be used \citep[e.g.][]{goldsmith12}
	\item dust properties, e.g. composition, distribution, optical properties. Common choices are described by \citet{mrn} and \citet{wd01}. 
	\item chemical composition and elemental abundances \citep[e.g.][]{simon-diaz2011}
	\item set of chemical reaction rates. Common databases include KIDA\footnote{\url{http://kida.astrophy.u-bordeaux.fr/}} \citep{kida}, UdfA12\footnote{\url{http://udfa.ajmarkwick.net/}} \citep{udfa06} and OSU\footnote{\url{http://faculty.virginia.edu/ericherb/research.html}} \citep{garrod2008}.
	\item micro-physical parameters, e.g. H$_2$ formation efficiency, grain surface binding energies, turbulent Doppler line width and velocity distribution, magnetic pressure
	\item atomic and molecular data, e.g. collision rates, level energies, transition frequencies, Einstein A values. Good starting points are the LAMDA Database\footnote{\url{https://home.strw.leidenuniv.nl/~moldata/}} and the Cologne Database for Molecular Spectroscopy CDMS\footnote{\url{https://cdms.astro.uni-koeln.de/}} \citep{cdms,lamda}. \ce{H2} can be treated either with full rotational-vibrational level structure, simplified vibrational level structure only or using parameterized approximations \citep[][and forthcoming paper]{roellig06}
\end{itemize}

Typical output quantities of PDR models are
\begin{itemize}
	\item spatial structure of the model cloud per cloud position
	\begin{itemize}
		\item physical conditions: gas and dust temperature, local FUV radiation field strength, local photo-dissociation and photo-ionization rates, heating and cooling rates, energy level population of species of interest
		\item chemical conditions: gas phase abundances of all included chemical species, possibly grain surface abundances and ice mantle composition, local chemical formation and reaction rates
	\end{itemize}
	\item integrated quantities of the model cloud
	\begin{itemize}
		\item spectral line emission: IR and FIR line emission (either spatially resolved or averaged over an assumed telescope beam), possibly UV and FUV emission
		\item continuum emission: dust continuum emission (spectrally resolved or total), possibly UV continuum emission 
		\item spectral line optical depths
	\end{itemize}
	\item possibly time evolution of the model cloud
\end{itemize}	
A comparison with experimental data is usually limited to integrated model quantities, such as intensities or emission line ratios. The limited spatial resolution of IR and FIR telescopes and the generally long distance to massive star forming regions allows only in very few cases to directly compare the modeled PDR structure to real observations. In terms of model calibration this is a substantial complication.

In the following we will present the \Kt{} code. Section 3 lays out the details of the code structure and the implemented physics. Section 4 discusses the details of the numerical solution of the chemical problem. Section 5 discusses the energy solution and the occurrence of multiple solutions. Section 6 presents the new model predictions of the surface chemistry and compares with predictions by other models and with observations. Section 7 finally demonstrates the practical value of the clumpy PDR modeling approach.

\section{The \Kt{} Code}

The \Kt{}\ PDR model is based on the plane-parallel PDR code by \citet{sternberg1988}. The spherical \Kt{}\ model has been developed at the University of Cologne in collaboration with Tel Aviv University and the first results have been published by \citet{gierens92} and \citet{stoerzer1996,stoerzer2000b}. Historically, the model geometry was driven by the finding that observations showed molecular clouds to be clumpy, porous or fractal \citep[e.g.][]{stutzki1988}. The fractal properties of many clouds can be described by a fractional Brownian motion structure and such a structure can be reproduced by an ensemble of clumps with a well defined power-law mass spectrum and mass-size relation \citep{stutzki1998,zielinsky2000}. Theory and numerical simulations of the ISM also show that turbulent flows naturally lead to a highly fragmented structure \citep[e.g.][]{gammie2003}. \citet{gong2011} and \citet{gong2015} showed that dense cores form from supersonic turbulent converging flows and \citet{guszejnow2018} argue that the observed power-law scaling is the generic result of scale-free structure formation where the different scales are uncorrelated. For a review on clump formation see \citet{ballesterosparedes2020}. Consequently it is necessary to include the effect of clumpiness and fragmented structures in any PDR modeling as far as possible. 

The fundamental difference between the spherical model geometry and plane-parallel models is the higher ratio of surface to mass (or volume). The infinite extent of plane-parallel models parallel to the surface is not real but a necessity of the 1-D setup.
Two-sided plane-parallel models slightly improve on that aspect but have to introduce an additional model parameter, the cut-off $A_V$. The simplest possible model with a finite configuration is the spherical clump, but as for the two-sided models, it comes at the cost of an additional model parameter, the clump mass. The spherical model is also a strong simplification since molecular clouds are never spherical but it allows for a better description of the clumpy structure of the ISM with a large fraction of internal surfaces. By matching the parameters to the observed clump-mass spectra we can approximate the fractal structure in PDRs, something that is impossible with plane-parallel configurations. 

The model geometry naturally explains some observations that are hard to explain otherwise. \citet{bolatto1999} and \citet{roellig06} showed that the observed metallicity influence on the emission ratio of \cii{}/\twco\  and \ci{}/\twco\  can be explained by a spherical (i.e. finite mass) cloud model. 
\citet{izumi2021} observed unusually high ratios of \cilo\ to \twco{}(1-0) emission at high extinction  that can not be explained by standard plane-parallel PDR models and suggests highly clumpy gas. 
\citet{schneider2021} performed a detailed PDR multi-line model analysis of a globule in Cygnus~X  and showed that the observed line emission can only be explained with a two-component PDR: a clumpy internal and a non-clumpy external PDR. Other studies that used \Kt{}\ model predictions to successfully analyze and interpret PDR observations include \citet{mookerjea2006,schulz07, kramer08,pineda08,cubick08,sun08,roellig2011,roellig2012,schneider2016,roellig2016,schneider2018,garcia2021, nayak2021,mookerjea2021}. 
\citet{andree2017} investigated the spatial variations of PDR emission lines across the Orion Bar
and showed that the observed spatial profiles can be explained by dense PDR clumps embedded in a thinner inter-clump medium.

The spherical geometry requires more computational power due to the need of angular averaging and the additional mass parameter. The clump ensemble approach requires the computation of large parameter grids with the consequence of a high computational effort. To compensate for this, physical complexity was reduced in certain calculations, for instance the full ro-vib structure of \ce{H2} was approximated by 15 virtual vibrational levels only. As computing power becomes more readily available, we are gradually increasing the model complexity again.

The general model iteration is preceded by the model setup and the pre-computation of the FUV continuum radiative transfer. Using the multi-component dust radiative transfer code MCDRT \citep{szczerba97} we compute the spectrally resolved FUV radiation field in the model clump together with the dust temperature (fully resolved for all dust components and dust sizes) \citep{roellig2013dust}. In a post-processing step following the global model convergence, we then compute the detailed line \& continuum emission, spatially resolved as well as model clump averaged \citep{gierens92}. 

\subsection{Geometry}
\Kt{} is a 1-dimensional model with spherical geometry using the cloud depth $z=R_\mathrm{tot}-r$ as spatial coordinate, with the total radius $R_\mathrm{tot}$ and the radius $r$ from 0 to $R_\mathrm{tot}$. We usually represent the spatial coordinate by the optical extinction $A_V$ along this coordinate.\footnote{If not stated otherwise we always use the perpendicular $A_V$ along the central line of sight (in contrast to e.g. the effective $A_V$ \citep{comparison07}).} The total gas density $n$ is described by
\begin{equation}\label{eq:density_profile}
n=n_0\begin{cases} 
(r/R_\mathrm{tot})^{-\alpha} & R_\mathrm{core}\le r/ R_\mathrm{tot} \le 1 \\
R_{core}^{-\alpha}=const. & 0\le r/ R_\mathrm{tot} \le R_\mathrm{core}
\end{cases} 
\end{equation}
$n_0$ is the total gas density at the surface ($r=R_\mathrm{tot}$) in cm$^{-3}$. The standard parameters are: $\alpha=1.5$, $R_\mathrm{core}=0.2$, approximating the structure of Bonnor-Ebert spheres with a density contrast from edge to center of 11.2. Note, that similar density contrasts are found in isobaric PDR models \citep{joblin2018,Wu2018,bron2018}.
In general we use the clump mass instead of the clump radius as the model parameter characterizing the size of the clumps.

The spatial loop starts at the surface $z=0$ and proceeds into the model cloud until the cloud center is reached.  A 1-dimensional setup in spherical coordinates automatically implies an isotropic FUV radiation field providing radiation from all directions. This assumption is approximately true for the average ambient FUV field in the Galaxy and for nearby FUV sources if the PDR is embedded in a diffuse medium because of the strong scattering of the interstellar dust in the FUV range. With albedos between 0.3 and 0.4 \citep{wd01} and the FUV extinction being about three times stronger than at visible wavelengths, a dust column providing an $A_V\la 1$  is sufficient to convert a directed FUV field into an isotropic field. Observations and corresponding models seem to support this assumption. \citet{choi2015} showed that the FUV emission of the \hii{} region around \textzeta~ Oph is dominated by dust grain scattering. Similar findings have been reported for the Spica nebula \citep{choi2013}, the Orion–Eridanus Superbubble \citep{jo2011,jo2012}, and the Taurus-Perseus-Auriga Complex \citep{lim2013}. A similar scattering effect has been proposed for diffuse H\textalpha\ emission outside of bright \hii\ regions \citep{seon2012}.
Therefore the assumption of an isotropic FUV illumination is a good approximation for low to intermediate mass (size) clumps. It becomes questionable for very large model clumps where the radiation would have to be redistributed over parsec scales. However, the very massive clumps are actually a good approximation to the common plane-parallel setup. In \citet{comparison07} we could show
that for these cases our spherical model agrees well with predictions from plane-parallel PDR codes with a perpendicular FUV irradiation producing comparable PDR structures. This is due to the fact that the emission of very massive structures typically arises from a thin surface layer due to optical thickness effects. This layer itself is dominated by the FUV radiation falling perpendicular onto it. Radiation from other directions and the backside is quickly absorbed in the large columns of material. Even if the approximation of the isotropic illumination breaks down in those cases it has no measurable effect because only the perpendicular contribution affects the large-clump model.

To assure a sufficient numerical resolution of the dissociation front and the transition from atomic carbon to \ce{CO} we do not assume a fixed spatial model grid. Instead we rely on an adaptive spatial gridding using the Bulirsch-Stoer method \citep[][Sect. 16.4]{numericalrecipes} to determine the next step width during each spatial iteration.

\subsubsection{Radiative transfer}
Solving the radiative transfer (RT) problem is one of the key aspects in any PDR model. The local FUV photon density determines the heating efficiency as well as the strength of the local photo-chemistry. The problem is complicated by the fact that some of these processes are affected primarily by the FUV continuum dust shielding, while others depend on the combination of dust and spectral line shielding \cite[e.g.][]{wd01pe,heays2017}. Solving the RT fully self-consistently is very time-consuming and many PDR codes compromise on certain aspects of the RT to reduce the computational efforts.
The RT in \Kt{} is currently divided into 3 parts.
\begin{enumerate}
\item The dust RT is done in a pre-processing step using the \texttt{MCDRT}-code \citep[details described in][]{roellig2013dust}. \texttt{MCDRT} computes the continuum RT within the spherical clump for a given dust composition based on the respective optical properties of the dust components and an assumed dust size distribution. The result of this computation is the internal FUV field, the emitted continuum spectrum, and the dust temperature distribution in the clump. These are used as input quantities for the \Kt{} calculation. \texttt{MCDRT} doesn't know the gas structure of the clump yet and can not account for FUV absorption by \ce{H2} or \ce{CO}. 
\item During the \Kt{} iterations the line RT to compute the local photo-processes (dissociation and ionization)  are computed during each iteration via a ray tracing scheme that is described below. The line cooling is computed per cooling line based on the local energy level excitation using a spherical escape probability formalism.
\item The final clump line emission is computed in a post-processing step based on the final chemical and physical structure of the clump using the spherical RT code \texttt{ONION} fully accounting for non-LTE (local thermal equilibrium) effects \citep{gierens92}. These computations are performed per line ignoring any line overlap or line-pumping between different molecules. The error due to the inconsistent treatment of line cooling and final line emission is small and can be ignored (example \cii{} 158{\textmu}m: median $\Delta T_\mathrm{ex}/T_\mathrm{ex}=3.1\%$ over full parameter grid).
\end{enumerate}

\texttt{MCDRT} uses the Mie theory \citep{bohren83} to compute the extinction efficiency $Q_\mathrm{ext}$ and albedo $\omega$ for each dust component, assuming the scattering properties of spherical grains. For silicates and graphite we use the dielectric constants from \citet{draine2003a}, while for very small grains we followed the approach given by \citet{li2001}. Scattering by dust particles is currently assumed to be isotropic. More details are given in \citet{roellig2013dust}.

Computing the continuum radiative transfer only once in a pre-processing step is a reasonable approximating due to the weak thermodynamic coupling between gas and dust.
The same approach is not possible for the FUV-line radiative transfer because the chemical structure changes during the iterations. Therefore, \Kt{} calculates the line radiative transfer for all relevant line transitions. This includes all absorption processes where spectral line absorption is important, i.e. the photo-dissociation of $\mathrm{H}_2$  \citep{vandishoeck1987,sternberg1989}, atomic carbon  $\mathrm{C}$, and all isotopologues of $\mathrm{CO}$ \citep{vandishoeck1988,visser2009}. Line shielding might be important for other species as well, e.g. for $\mathrm{N}_2$ \citep{heays2017}, but is not yet implemented into \Kt. 

For all photo-processes, the relevant physical quantity is the local photon flux above critical energy available for the respective process. The flux is determined by the initial unattenuated FUV field and the column of absorbing material which the photons penetrated. Given the spherical geometry and the isotropic radiation field the column density of absorbing material is a function of local position and direction of the incoming photons. The final local flux is the result of the angular average over the full solid angle. To compute the attenuation along all directions we use the method described in \citet{gierens92}.

\citet{sternberg2014} showed that the absorption of UV photons in Werner and Lyman \ce{H2} lines can dominate the continuum absorption under certain circumstances. Treating these cases fully self-consistently requires a re-computation of the dust continuum radiative transfer and all dust properties after each iteration. During each iteration we compute the pumping of \ce{H2} levels and its dissociation as well as the dissociation of carbon monoxide. We use the local dust column densities as function of angle and the effective UV continuum absorption cross section $\sigma_D$ \citep{sternberg1989} to compute the continuum absorption in all directions and integrate over all angles (see Sect.\ref{sect:ray_tracing}).  $\sigma_D$ depends on the current dust size distribution and is computed by \texttt{MCDRT}.

\Kt{} computes the self-shielding of all $\mathrm{CO}$ isotopologues using the shielding function provided by \citet{visser2009} accounting for self-shielding, mutual shielding (line-overlap) and shielding by atomic and molecular hydrogen. The shielding factors $f_{sh}(N(X))$ are parameterized as function of \ce{H2} and \ce{CO} column densities. The CO shielding varies with the assumed Doppler line width and \citet{visser2009} report a 26\% increase of the \ce{CO} photodissociation rate for an increase in the Doppler broadening $b_{\ce{CO}}$ from 0.3 to 3~km~s$^{-1}$. We use CO shielding functions computed for $b_{\ce{CO}}=3$~km~s$^{-1}$, $T_\mathrm{ex}(\ce{CO})=20$~K, and a $^{12}\mathrm{C}/^{13}\mathrm{C}$ ratio of 69 and neglect variations with $b_{\ce{CO}}$.
The \ce{H2} self-shielding for the radiative pumping by FUV photons is computed line by line based on the shielding prescription by \citet{federman79} accounting for the Voigt profile absorption lines. \citet{draine1996} provided \ce{H2} shielding functions that account for line overlap but do not allow for a line-by-line treatment. \Kt{} offers the option to use their shielding prescription for the computation of the \ce{H2} photodissociation rate instead.

In this study we do not resolve the rotational energy levels of \ce{H2} but only consider transitions between vibrational levels in the ground state \Xlev\ (15 levels) and the electronically excited Lyman \Lylev\ (24 levels) and Werner \Werlev\ (10 levels) bands. The vib-level population in the ground state is computed from the balance between collisions, spontaneous decay and UV pumping \citep{allison1969,dalgarno1970,stephens1972,stephens1973,abgrall1992}. The photodissociation rate is computed from decay of Lyman and Werner states, excited by UV pumping, to an unbound  state \citep{sternberg1988}.

Continuum processes that depend on the local FUV intensity, such as the photoelectric heating, are computed based on the intensity results of \texttt{MCDRT} and neglect the effect of line absorption on the local mean FUV intensity. We do not account for line-overlap and treat the radiative transfer line by line.

\subsubsection{Ray tracing scheme\label{sect:ray_tracing}}
A ray tracing approach is applied to determine the local photon flux inside the model cloud in order to compute photo-processes such as photo-dissociation and photo-ionization. It is also applied to compute the final emission characteristic of the model cloud. \Kt{} is a 1-D code therefore the angular gridding has to be derived from the existing spatial grid.   

\begin{figure}
	\resizebox{\hsize}{!}{\includegraphics{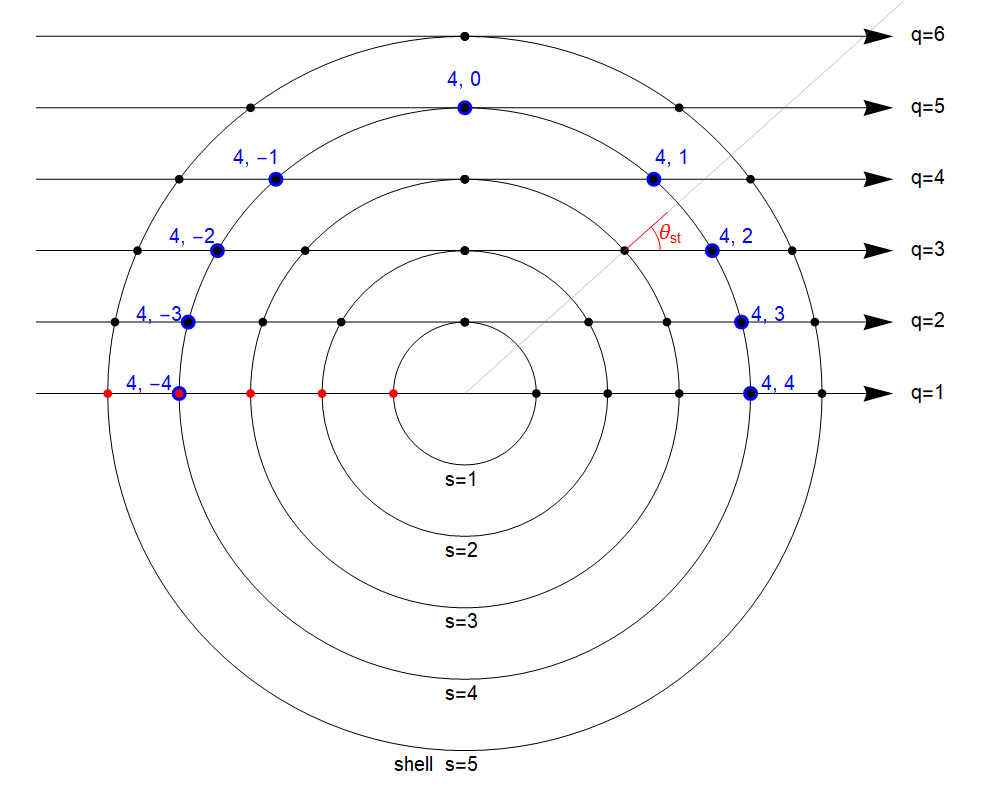}}
	\caption{Structure of the shell ray tracing setup. The shells are arranged equidistantly for clarity. Real spatial grids are not equidistant. The angle index $t$ is the second index given at the blue labeled points. }
	\label{fig:ray_tracing_1}
\end{figure}

The model consists of a series of concentric, non-equidistant shells with radius $r_i$. The local conditions are constant across a particular shell, e.g. all positions indicated by the blue points in Fig.~\ref{fig:ray_tracing_1} share the same physical and chemical conditions. The spatial computation iterates over all shells, i.e. computes all red points in the Fig.. The cloud is intersected by a series of parallel rays $q_i$, one across the center and the others tangential to the shells. The crossing points between the shell boundaries and the intersecting rays (black in Fig.~\ref{fig:ray_tracing_1}) and the angles of incidents at these points constitute the spatial and angular grid that is used to perform the radiative transfer computations. An angle index $t=0$ corresponds to the tangential point between ray and shell. The number of spatial grid points $\mathcal{N}$ and therefore shells and rays in real calculations is in the order of a few hundred depending on the cloud parameters. 

The grid points are labeled with their shell index $s$ and their angular index $-s\le t \le +s$ as shown in Fig.~\ref{fig:ray_tracing_1}. Intensities and relevant quantities such as column densities are computed point by point along a ray and for all rays $q$. We use the radiative transfer equation:
\begin{equation}
\label{eq:transfer_equation}
I_{s,t}=I_{s',t'}\exp(-\kappa l)+S[1-\exp(-\kappa l)] 
\end{equation} 
where $s,t$ and $s',t'$ are adjacent points on the same ray and $\kappa$ and $S$ are the opacity and source function in the shell between the points. $l$ is the distance between the two points.  It is convenient to reorder the grid points $s,t \rightarrow q,t$ where $q=s - |t| + 1$ is the corresponding ray index. Along each ray $q$ the angle index runs from $t_\mathrm{min}=-\mathcal{N}+q-1$ to $t_\mathrm{max}=\mathcal{N}-q+1$.  Using the transfer equation we can recursively compute the intensities along a given ray $q$
\begin{equation}
	\label{eq:transfer_equation_ray}
	I_{q,t}= 
	\begin{cases}
	I_\mathrm{bg},    &t=t_\mathrm{min} \\
	I_{q,t-1}e^{-\kappa l_{s,s-1}}+S[1-e^{-\kappa l_{s,s-1}}],   &t_\mathrm{min}< t\le t_\mathrm{max} 
	\end{cases}
\end{equation} 
with $ t=-\mathcal{N}+q-1,...,+\mathcal{N}-q+1$ and $s,s-1$ the corresponding shell indices. $I_\mathrm{bg}$ is the background intensity. Accordingly, the intensity at left surface points in Fig.~\ref{fig:ray_tracing_1} is $I_{q,t}=I_\mathrm{bg}=B_\nu(T_\mathrm{bg})$ with Planck's radiation law $B_\nu$ at the background temperature $T_\mathrm{bg}=2.73$~K, while for the rest of the points we have a step-wise integration of the radiative transfer equation. 
The detailed energy level population of a species is obtained from the corresponding set of non-LTE rate equations which are solved with a generalized Newton-Raphson technique. We derive the line source function $S_l$ from the corresponding excitation temperatures. $S=S_l+S_c$ where $S_c$ is the continuum contribution from the local dust temperature  also contributing to the pumping of the quantum levels. Similarly, we compute the total  opacity from their line and continuum contributions $\kappa=\kappa_l+\kappa_c$. Contributions from external dust continuum are not considered. 
We use 16 frequency points to resolve the (Gaussian) line profile and apply a Gauss-Hermite quadrature for the frequency integral. Collision rates, A-values and line frequencies are taken from the Leiden Atomic and Molecular Database \citep{lamda,lamda2020}. 

The final result is the emergent intensity for each ray and the intensity inside the cloud for all shell-ray intersections. For any given cloud depth, i.e. shell, this allows us to compute the angular average over intensities coming from all directions:
\begin{align}
\label{eq:angular_average}
J_s=&\frac{1}{2}\int_0^\pi I(r_s,\theta)\sin(\theta)\,d\theta \;\; or\\
J_s=&\frac{1}{2}\sum_{t=-s,s}I_{s,t}\sin(\theta_{s,t})
\end{align}
where $I(r_s,\theta)$ are the intensities across the shell $s$ (e.g. at all blue points in Fig.~\ref{fig:ray_tracing_1}) and $\theta_{s,t}$ are the angles at the points $s,t$, $\theta_{s,t}=\arccos(r_{s-|t|}/r_s])$. 

A similar ray-tracing scheme is also used to compute e.g. the \ce{H2} and \ce{CO} self-shielding of the UV absorption lines and the corresponding \ce{H2} pumping along each ray \citep{stoerzer1996}.
In our model geometry column densities are functions of the angle  $\theta_{s,t}$. We compute $N(X)_{s,t}$ the column density of species \ce{X} at point $s,t$ using a simple trapezoidal rule.
Equivalent to Eq.~\ref{eq:transfer_equation}, we compute the attenuation of all photo-reactions in the chemistry by using the angle-dependent columns in the attenuation factor $\exp(-\gamma A_V)$ from Eq.~(\ref{eq:photo_reaction}) where we expressed column density by $A_V$ in order to compute the reaction-specific local photo-rate. To compute the clump averaged emission we compute:
\begin{equation}
\label{eq:clump_average}
I_{clump}=\frac{2}{R_{tot}^2}\int_0^{R_{tot}}I(p)r\,dp
\end{equation} 
For more details on the radiative transfer scheme see \citet{gierens92}. Note, that Eq.~(\ref{eq:transfer_equation}) and (\ref{eq:angular_average}) have to be computed for all relevant radiative transitions. For species with many transitions this introduces a significant numerical complexity. In particular the solution to the full rotational-vibrational structure of the \ce{H2} molecule, including the UV transitions, involves almost 30000 radiative transitions: Thus, the described scheme has to be solved for 30000 transitions along  several hundred rays at each spatial step. In the past this was a prohibitive effort, but we recently extended the code to include the full \ce{H2} problem. This will be described in detail in a subsequent  paper \citep{roellig2022}.

\subsection{Gas Phase Chemistry}
The chemistry in \Kt{} is modular. Chemical species can easily by added or removed from the chemistry and the code selects all participating reactions from the chemical database in use. 
The standard database is based on the 2012 edition of the UMIST Database for Astrochemistry (UDfA12) \citep{udfa12} \footnote{\url{http://udfa.ajmarkwick.net/}} including a number of modification and updates:
\begin{itemize}
	\item isotopologues including $^{13}\mathrm{C}$ and $^{18}\mathrm{O}$ have been added \citep{roellig2013fract}
	\item new branching ratios from \citet{chabot2013} are used
	\item addition of l-type isomers (UMIST only contains c-type)
	\item updated fractionation reaction rates \citep{mladenovic2014} 
	\item refitted low-temperature rates \citep{roellig2011refit}
\end{itemize}
From the list of reactions \Kt{} constructs a set of ordinary differential equations (ODE):
\begin{multline}\label{eq:chemical_rate_equation}
\mathcal{F}_i=\frac{dn(i)}{dt}=\\ \sum_{j,j'}k_{jj'} n(j) n(j') + \sum_l k_l n(l)  - n(i)\left[ \sum_m k_{im} n(m) +\sum_{m'} k_{m'}  \right] 
\end{multline}
The first two terms sum over all two-body processes forming species $i$ and all photo-processes and/or cosmic-ray processes leading to the formation of species $i$, respectively. The final two terms are the equivalent sums over all processes leading to the destruction of species $i$ by two-body reactions (index $m$) and photo-processes (index $m'$). The pre-factor $k$ is called the rate coefficient and tabulated in parameterized form in chemical databases.  We use lower-case $k$ and upper-case $K$ to distinguish gas-phase and surface reaction rate coefficients, respectively. We list the implemented gas-phase reaction types in the Appendix~\ref{sect:chem_appendix}.

The set of $N$ rate equations (Eq.~\ref{eq:chemical_rate_equation}), one for each of the $N$ chemical species is complemented by a set of  conservation equations per chemical element $M$:
\begin{equation}\label{eq:conservation_equation}
n(M)=\sum_i n(i) c_i^M
\end{equation}
where $c_i^M$ is the number of atoms of element $M$ in species $i$. Analogously we also find a charge conservation equation:
\begin{equation}\label{eq:charge_conservation_equation}
n(e^-)=\sum_i n(i) c_i^e
\end{equation} 
where $c_i^e$ is the charge of species $i$ in units of the elemental charge where  $c_i^e$ can be negative in contrast to $c_i^M$. Charge exchange between grains and gas is currently not considered.

\subsection{Surface chemistry}
A large number of observations of star formation regions show that observed gas-phase abundances of species other than \ce{H2}such as \ce{NH3}, \ce{N2H+} and \ce{CH3OH} can not be explained with pure gas phase chemistry \citep[e.g.][]{geppert2005,garrod2006,herbst2009,bottinelli2010,oeberg2011,boogert2015}. The freeze-out of gas-phase species in the dark region of molecular clouds also removes important tracer species from the observable content of the ISM and modifies the IR/FIR line emission of the clouds. 

Historically, the only grain surface chemistry included in PDR codes was the formation of \ce{H2} followed by immediate desorption to the gas-phase. Based on measurements in the diffuse medium a mean \ce{H2} formation rate of $3 \times 10^{-17}~\mathrm{cm^3\,s^{-1}}$ was found \citep{jura1974}. This formation rate or comparable temperature dependent parametrizations \citep[e.g.][]{tielens1985,sternberg1995} was implemented in astrochemical models as gas phase reaction simulating the surface process. Later, other prescriptions have been suggested to simulate the formation of \ce{H2} more precisely, including considering Langmuir-Hinshelwood and Eley-Rideal processes \citep[e.g.][]{lebourlot2012}, physisorption and chemisorption of \ce{H2} \citep{cazaux2002,cazaux2004}, as well as stochastic treatment of dust temperatures and grain populations \citep[e.g.][]{barzel2007,bron2014}. 
\Kt{}  implements the \ce{H2} formation formalism described by \citet{cazaux2002,cazaux2004}. Details are described in \citet{roellig2013dust}.

It is yet unclear whether surface chemistry takes place on very small particles such as polycyclic aromatic hydrocarbons (PAHs) \citep{boschman2015, andrews2016, foley2018}. 
To estimate the chemically active grain surface we use the \ce{H2} formation rate observed in the diffuse medium.  \citet{cazaux2016} derive the grain surface per H-atom that is available to \ce{H2} formation as $4\times 10^{-21}$~cm$^2$/H-atom. We use this value to compute the minimum grain radius for surface chemistry, integrating over the dust size distributions from \citet{wd01} that apply to diffuse gas conditions. Assuming that the limiting factor is the thermal stability of the dust grains, carbonaceous dust can be approximately 20\% larger than corresponding silicate grains \citep{li01b}. From these two conditions we find minimum radii of $24 \text{\r{A}} $ and $29 \text{\r{A}}$ for silicate and carbon dust by integrating over dust size distributions suitable for diffuse gas. This is comparable to a threshold of $20 \text{\r{A}}$  given by \citet{hollenbach2009}. Using this as lower integration limit we find total surface areas $A_\mathrm{sil}=2.6\times 10^{-21}$~cm$^2$/H-atom and $A_\mathrm{carb}=1.4\times 10^{-21}$~cm$^2$/H-atom for silicate and carbonaceous grains, respectively.

Other surface reactions have been introduced to numerical PDR models with a primary focus on formation routes of chemical species observed in PDRs, such as \ce{H2O} and \ce{H2CO} \citep[for example][]{hollenbach2009, guzman2011,lebourlot2012,esplugues2016,putaud2019}.
The updated chemistry in \Kt{} now includes all relevant surface processes in a quasi-three-phase  model (gas + surface + inert ice bulk).
For the surface chemistry we follow the rate equation approach as described in \citet{hasegawa1992, hasegawa1993} including the competition between different processes as described by \citet{chang2007} and \citet{garrod2011}. In our quasi-three-phase model we limit the mobility of surface species to the top surface layer following \citet{cuppen2017} and  desorption to the top two ice mantle monolayers (MLs) (see Eq.~\ref{eq:desorbable_fraction}). For surface reactions with an activation energy barrier we account for competing processes such as diffusion and desorption following \citet{garrod2011}.  Including surface chemistry increases the number of chemical species in the network because gas-phase species and surface species must be considered as two different chemical species for all mathematical aspects involved. In the following, the symbol $n_s$ denotes chemical species on grain surfaces. The rate equations of the gas-phase chemistry Eq.~\ref{eq:chemical_rate_equation} is extended and modified. We get an additional set of equations governing the surface processes:
\begin{multline}\label{eq:surfac_rate_equation}
\frac{dn_s(i)}{dt}= \sum_{j,k}K_{jk} n_s(j) n_s(k) + \sum_l K_l n_s(l)  \\ - n_s(i)\left[ \sum_m K_{im} n_s(m) +\sum_n K_n  \right] \\+ k_\mathrm{acc,i} n(i) - K_\mathrm{des,i} n_s(i) 
\end{multline}
where, $K_l$ and $K_{jk}$ denote the surface reaction rate coefficients for one-body and two-body reactions, respectively. We also have to extend the gas-phase chemistry Eq.~\ref{eq:chemical_rate_equation} by the additional accretion and desorption terms:
\begin{equation}\label{eq:chemical_rate_equation2}
\frac{dn(i)}{dt}=(...)- k_\mathrm{acc,i} n(i) +  K_\mathrm{des,i} n_s(i) 
\end{equation}
where $(...)$ corresponds to the right hand side of Eq.~\ref{eq:chemical_rate_equation}.
Here $K_\mathrm{des}$ describes all thermal and non-thermal desorption processes, i.e. a conversion of a surface species to a gas-phase species while $k_\mathrm{acc}$ is the rate coefficient for accretion (freeze-out) which converts gas-phase species into their surface equivalent.

\begin{table}
	\caption{\label{tab:binding_energies} Modified desorption energies used in the surface chemistry}
	\centering
	\begin{tabular}{ll}
		\hline\hline
		species&$E_D$ [K]\\
		\hline
		\ce{C}, \ce{^{13}C}&14000\\
		\ce{O}, \ce{^{18}O}&1440\\
		\ce{N} &400\\
		\hline
	\end{tabular}
	\tablebib{
		\citet{shimonishi2018}
	}
\end{table}

\begin{table}
	\caption{\label{tab:binding_energies_bare} Desorption energies for binding on bare grain surfaces.}
	\centering
	\begin{tabular}{ll}
		\hline\hline
		species&$E_D$ [K]\\
		\hline
		\ce{H}&500\\
		\ce{H2}&300\\
		\ce{O2}&1250\\
		\ce{OH}&1360\\
		\ce{HCO}&830\\
		\ce{CO}, \ce{H2CO}, \ce{CH3O}, \ce{CH3OH}&1100\\
		\ce{CO2}&2300\\
		\ce{N}&720\\
		\ce{N2}&790\\
		\hline
	\end{tabular}
	\tablebib{
		\citet{esplugues2016}
	}
\end{table}

We assume a surface density of binding sites on the grains $n_\mathrm{site}=1.5\times10^{15} \mathrm{cm^{-2}}$ \citep{tielens1987}. For tunneling through activation energy barriers we assume an energy barrier width of $a=2 \text{\AA}$ \citep{garrod2011}. For the desorption (or binding) energies $E_D$, we use values provided by KIDA \citep{wakelam2017binding} with the modifications given in Table~\ref{tab:binding_energies}. Typically, $E_D$ is given for species bound to a \ce{H2O} ice surface as the most abundant ice component. Desorption energies for other surfaces may differ significantly \citep{esplugues2016} and we include different desorption energies for species bound directly to a carbonaceous surface if they are known (see Table~\ref{tab:binding_energies_bare}).   
For a more thorough discussion on desorption energy choice and the resulting model sensitivity we refer the reader to e.g. \citet{penteado2017} and \citet{kamp2017}. We use the notation \ce{J(X)} to distinguish surface species from their gas-phase counterpart \ce{X}. 

In Appendix~\ref{sect:surface_appendix} we summarize all surface reactions that are currently implemented in \Kt. Here, we only report the new chemical desorption framework that we included in the surface chemistry. 

\paragraph{\ce{H2} ice}
A significant population of \ce{H2} ice in the ISM has been proposed already almost 30 years ago based on early astrochemical models  \citep{sandford1993}. However, no clear observational evidence could be detected to date. In the context of numerical models, freeze-out of gas-phase \ce{H2} is potentially problematic because it can lead to a catastrophic freeze-out of all molecular gas onto dust grains. 
Several physical processes have been discussed to prevent nonphysical \ce{H2} ice populations and we include encounter desorption for \ce{J(H)} and \ce{J(H2)} as suggested by \citet{hincelin2015}. The underlying idea is based on the significantly lower desorption energy for hydrogen on an \ce{H2} substrate compared to other substrates \citep[e.g.][]{vidali1991,cuppen2007,das2021}.
For \ce{J(H)} or \ce{J(H2)} located next to a \ce{J(H2)} molecule we decrease $E_D$ from its canonical value to $E_D=23$~K \citep{cuppen2007} greatly enhancing the possibility for a desorption of \ce{J(H)} and \ce{J(H2)}. As a consequence, the further accretion of \ce{H2} ice is strongly suppressed after the build-up of the first ML.

\paragraph{Dust model}
The dust properties are computed by MCDRT assuming a  dust composition and dust size distribution, for example as given by \citet{mrn} or \citet{wd01}. The dust radiative transfer and dust properties are computed per dust size bin together with corresponding mean values, e.g. for dust temperature, dust surface area and dust density. Details are described in \citet{roellig2013dust}.  Presently, \Kt{} uses a single surface weighted  average grain temperature $T_d$ for all types of grains independent of size and grain composition.\footnote{In a future update we will split the surface chemistry computation into dust size bins with individual values of $T_d$ and surface area for each bin.} For the \Kt{} computations presented in this paper we assume the dust model \#7 from \citet{wd01} consisting of 4 dust types: carbonaceous grains, silicates, PAHs and ionized PAHs. 
In this model the average radii for silicates and carbon grains (including VSGs and PAHs) are $\langle\langle a\rangle\rangle_\mathrm{sil}=4.5$~{\textmu}m and $\langle\langle a\rangle\rangle_\mathrm{carb}=4.0$~{\textmu}m, when $\langle\langle \,\rangle\rangle$ indicates a surface weighted average. For the photo-electric heating and grain recombination cooling we use the prescription from \citet{WD01PEH} corresponding to the assumed dust size distribution. This implicitly includes the computation of the grain charge distribution. We neglect any feedback of line UV absorption on the dust temperature.

\paragraph{Cosmic-ray induced desorption}

Cosmic rays (CR) hitting dust grains deposit an energy of $E_\mathrm{CR}\approx 0.4$~MeV into dust particles heating an average grain of 0.1~{\textmu}m radius \citep{leger1985} to approximately $T_\mathrm{CR,max}=70$~K \citep{hasegawa1993}, sufficiently hot to induce significant thermal desorption. Cooling of the grain takes place via radiative cooling $\dot{E}_\mathrm{rad}$ and via evaporation cooling $\dot{E}_\mathrm{subl}$ due to the sublimation of bound ice particles. The cooling time scale of a grain in s is given by 
\begin{equation}
\label{eq:tau_col}
\tau_\mathrm{cool}^\mathrm{CR}=\frac{E_\mathrm{CR}}{\dot{E}_\mathrm{rad}+\dot{E}_\mathrm{subl}}
\end{equation}
where the cooling rate due to sublimation is given by:
\begin{equation}
\label{eq:sublimation}
\dot{E}_\mathrm{subl}=\frac{f_\mathrm{des}}{\langle n_\mathrm{dust}\rangle} \sum_i E_b(i) n_s(i) \nu(i) \exp{\left(-\frac{E_b(i)}{T_\mathrm{CR,max}}\right)}
\end{equation}
where the sum includes all ice species, $\langle n_\mathrm{dust}\rangle$ is the average number density of the dust particles, and $f_\mathrm{des}$ gives the fraction of surface species that are candidates for desorption so that $f_\mathrm{des}n_s(i)/\langle n_\mathrm{dust}\rangle$ gives the number of desorbable molecules of species $i$ per dust grain. $\nu(i)~\approx 3\times 10^{12}$~s$^{-1}$ is the desorption attempt frequency given by the surface vibrations \citep{hasegawa1993}. Radiative cooling is computed by $\dot{E}_\mathrm{rad}=4\pi q_\mathrm{abs} \langle a_D\rangle^3 \sigma T_\mathrm{CR,max}^6$ with the Stefan-Boltzmann constant $\sigma$ and $q_\mathrm{abs}=0.13$~K$^{-2}$cm$^{-1}$ for silicate grains. Cooling is always fast compared to the frequency of CR hits for any dust grain.

Assuming that most desorption occurs at about $T_\mathrm{CR,max}=70$~K  we can approximate the CR-induced desorption rate coefficient as:
\begin{equation}
	K_{\mathrm{CR-des,}i}=f_{CR} K_{\mathrm{evap,}i}(T_\mathrm{CR,max})
\end{equation}
where $f_\mathrm{CR}=\tau_\mathrm{cool}^\mathrm{CR}/\tau_\mathrm{heat}^\mathrm{CR}$ is the fraction of time spent by grains at the temperature $T_\mathrm{CR,max}$, given by the cooling time relative to the frequency of CR hits. $K_{\mathrm{evap,}i}(T_\mathrm{CR,max})$ is the thermal desorption rate coefficient of species $i$ at $T_\mathrm{CR,max}$.
Assuming $\tau_\mathrm{heat}^\mathrm{CR}=3.16\times 10^{13}$~s  \citet{hasegawa1993} find $f_\mathrm{CR}=3.16\times 10^{-19}$ assuming $\zeta_\mathrm{CR}=5\times 10^{-17}$~s$^{-1}$. 
To use this formalism we can distribute the total available dust surface in our model across virtual average grains of radius $\langle a_D\rangle=0.1$~{\textmu}m and find an average grain density of $\langle n_\mathrm{dust}\rangle=3.17\times 10^{-12}$ per H-atom. Using the same description of $0.1$~{\textmu}m grains \citet{sipilae2021} published an updated estimate based on time-dependent computations of the heating and cooling due to CRs. Their heating depends on $\zeta_\mathrm{CR}$ and the local $A_\mathrm{V}$ and can be approximated in the range $\zeta_\mathrm{CR}=[1.3\times 10^{-17},10^{-16}]$~s, and $A_\mathrm{V}=[0,100]$~mag by:

\begin{equation}
    \label{eq:tau_heat}
    \log_{10}\tau_\mathrm{heat}^\mathrm{CR}=0.19A_\mathrm{V}-1.6\times 10^{-4}A_\mathrm{V}^2+0.01 z A_\mathrm{V}  + 0.045z^2
\end{equation}
where $z=log_{10} \zeta_\mathrm{CR}$. Energy deposition by CRs directly into the ice mantle \citep{Wakelam2021} is not considered but will be implemented in a future code update.

\paragraph{Chemical desorption}\label{sect:chem_desorption}
\citet{minissale2016} and \citet{cazaux2016} presented an analytical expression for an additional desorption mechanism using the released binding energy from exothermic reactions in the ice-phase to desorb  species $i$ to the gas-phase first presented by \citet{dulieu2013}. They fitted a set of experimentally measured desorption rates through a probability that the reaction product has an energy higher than the desorption energy $E_{D,i}$:
\begin{equation}
\label{eq:chemisorption_probability}
\mathcal{P}_{CD,i}=\exp\left(- \frac{E_{D,i}}{\epsilon_i \Delta H_R/d_f}\right)
\end{equation}
where $\Delta H_R$ is the exothermicity of the formation of the product(s) and $d_f$ specifies the degrees of freedom that the energy is distributed among. A good fit was obtained when using a description that is equivalent to an elastic collision with a grain mass of $M=120$ amu where the energy fraction obtained by the product is $\epsilon_i=(M-m_i)^2/(M+m_i)^2$. This approach, however, is not applicable if multiple reaction products are generated that can desorb. In this case the released energy $\Delta H_R$ must be distributed over all reaction products. A modification of Eq.~(\ref{eq:chemisorption_probability}) is needed which accounts for the energy distribution between multiple reaction products:
\begin{equation}
\label{eq:new_chemisorption_probability}
\mathcal{P}_{CD,i}=\exp\left(- \frac{E_{D,i}}{\epsilon_i \eta_i \Delta H_R/d_f}\right)
\end{equation}
where $\eta_i$ specifies the fraction of $\Delta H_R$ that is available to species $i$. 
Unfortunately, no laboratory data seem to be available for the desorption rate in case of more than one reaction product so that $\eta_i$ cannot be determined from a fit to the experimental data. \citet{Wakelam2021} used an equal split between all atoms within the reaction products, assigning equal energy fractions to all atoms within the molecules. We propose a different approach based on the nature of the chemical reaction that produces the exothermicity. We assume that the fraction $\eta_i$ of the binding energy that is transferred to the individual products is proportional to the number of open shell electrons $e_i$ that contribute to the formation of each product (ignoring closed shells). Taking for example the reaction \ce{J(O) + J(HCO) -> CO2 + H} this gives the following $e_i$: $e_\mathrm{H}=1$, $e_\mathrm{C}=2$, $e_\mathrm{O}=4$.
\begin{equation}
\label{eq:energy_distribution}
\eta_i=\frac{E_i}{\Delta H_R}\approx\frac{e_i}{\sum_j e_j}
\end{equation}
where the sum is over all atoms involved in the reaction. For each product this quantifies the probability for desorption $\mathcal{P}_{CD,i}$ and for remaining on the surface $(1-\mathcal{P}_{CD,i})$. In the general case of two reaction products we can compute the branching ratios for all 4 possible reaction branches:
\begin{equation}\label{eq:branching_rates}
\ce{R_{s,1} + R_{s,2} -> } \begin{cases}
 P_{s,1} + P_{s,2} & \mathrm{BR}=(1-\mathcal{P}_{CD,1})\times(1-\mathcal{P}_{CD,2})\\
P_{s,1} + P_2& \mathrm{BR}=(1-\mathcal{P}_{CD,1})\times \mathcal{P}_{CD,2}\\
P_1 + P_{s,2}& \mathrm{BR}=\mathcal{P}_{CD,1}\times(1-\mathcal{P}_{CD,2})\\
P_1 + P_2& \mathrm{BR}=\mathcal{P}_{CD,1}\times\mathcal{P}_{CD,2}
\end{cases}
\end{equation}
where \ce{R_{1,2}} and \ce{P_{1,2}} stands for two reactants and products respectively, and the subscript $s$ denotes species on the surface or in the ice mantle. Take for example the reaction $\ce{\text{J}(O) + \text{J}(HCO) -> CO2 + H}$: we find $\eta_\mathrm{H}=1/11$, $\eta_{\ce{CO2}}=10/11$ and $\mathcal{P}_{CD,\ce{H}}=0.695$, $\mathcal{P}_{CD,\ce{CO2}}=0.1188$. Following Eq.~\ref{eq:branching_rates} gives the branching ratios (Values in parenthesis would result from the application of Eq.~\ref{eq:chemisorption_probability} to all products, this means a value of $\eta_i=1$ for all products.):
\begin{align*}
&\ce{\text{J}(O) + \text{J}(HCO) -> CO2 + H}  \hspace{1cm}& BR=0.08\, (0.14) \\
&\ce{\text{J}(O) + \text{J}(HCO) -> \text{J}(CO2) + H }\hspace{1cm}&BR=0.61\, (0.82)\\
&\ce{\text{J}(O) + \text{J}(HCO) -> CO2 + \text{J}(H)} \hspace{1cm}&BR=0.04\, (0.01)\\ 
&\ce{\text{J}(O) + \text{J}(HCO) -> \text{J}(CO2) + \text{J}(H) }\hspace{1cm}&BR=0.27\, (0.03) 
\end{align*}
The final reaction rate coefficient for the chemical desorption of $i$ resulting from the reaction between species $j$ and $k$ is:
\begin{equation}
K_\mathrm{chem-des,i}=K_{jk}\times \mathrm{BR}  
\end{equation}
with $K_{jk}$ from Eq.~(\ref{eq:surface_rate}). For the computation of the BR we always use the $E_{D,i}$ for a \ce{H2O} substrate. Tests have shown that the chemistry is not very sensitive to variations in the assumed desorption energies. The list of all chemical desorption reactions and their branching ratios is given in Appendix~\ref{sect:surface_appendix}, Table~\ref{tab:chemical_desorption1}.

\begin{table}[htb]
	\centering
	\caption{Heating and cooling processes implemented in {\Kt}. \label{tab:heating_cooling}}
	\renewcommand{\arraystretch}{1.1} 
	\begin{tabular}{ccl}
		\hline \hline
		\vrule width 0pt height 2.2ex
		ID & \sfrac{h}{c} & notes \\
		\hline
		\vrule width 0pt height 2.2ex
		1 & h/c & collisional de-excitation of vibrationally excited \ce{H2}\\
		2 & h & photo-dissociation heating of \ce{H2}\\
		3 & h & \makecell[lt]{\ce{H2}-formation heating using 1/3 of the binding energy \\ of 4.48 eV }\\
		4 & c & {\rm{[O {\scriptsize I}]}~$^3P_1\to{^3P_2}$} (63\textmu m) line cooling\\
		5 & c & {\rm{[O {\scriptsize I}]}~$^3P_0\to{^3P_2}$} (44\textmu m) line cooling (negligible)\\
		6 & c & {\rm{[O {\scriptsize I}]}~$^3P_0\to{^3P_1}$} (146\textmu m) line cooling\\
		7 & h & cosmic ray heating \citep{glassgold2012}\\
		8 & h/c & grain photoelectric (PE) heating (minus recombination)\\
		9 & c & \ce{^{12}CO} line cooling (J=0-49)\\
		10& c & \makecell[lt]{\ciitrans (158 \textmu m) line cooling \\(\ce{^{13}C+} implemented but not used)}\\
		11 & c & {\rm {[C {\scriptsize I}]}~$^3P_1\to{^3P_0}$} line cooling\\
		12 & c & {\rm {[C {\scriptsize I}]}~$^3P_2\to{^3P_0}$} line cooling(negligible)\\
		13 & c & {\rm {[C {\scriptsize I}]}~$^3P_2\to{^3P_1}$} line cooling\\
		14 & c & {\rm {[Si {\scriptsize II}]}~$^2P_{3/2}\to{^2P_{1/2}}$} (35\textmu m) line cooling\\
		15 & c & \ce{^{13}CO} line cooling (J=0-49)\\
		16 & c & H {\scriptsize I} Lyman-\textalpha\, cooling \citep{spitzer1978}\\
		17 & c & \ce{H2O} line cooling \citep{neufeld1987}\\
		18 & h/c & gas-grain collisions\\	
		19 & c & \ce{OH} line cooling (including lowest 16 energy levels)\\  
		20 & c & O {\scriptsize I} 6300 \AA\, cooling \citep{bt94}\\
		21 & c & \makecell[lt]{\ce{H2} photo-dissociation kinetic cooling \\ \citep{lepp83}}\\
		22 & h & \makecell[lt]{carbon photo-ionization heating \\\citep{tielens1985}}\\
		\hline	
	\end{tabular}
	\tablefoot{Notes: h and c denote heating and cooling processes respectively. Some process may either cool or heat. The ID corresponds to the internal storage order in the code and to the subscripts of the heating and cooling rates $\Gamma_\mathrm{ID}$, $\Lambda_\mathrm{ID}$.}
\end{table}

\subsection{Thermal Balance}
Thermal balance is the local balance between all cooling $\Lambda_k$ and heating $\Gamma_h$ processes:
\begin{equation}\label{eq:energybalance}
E_\mathrm{tot}(T)^{\,i}_x=\sum_h\Gamma_h(\vec{c}^{\,i}_x)-\sum_k\Lambda_k(\vec{c}^{\,i}_x)=0
\end{equation}
where the efficiency of any individual process depends on the local conditions $\vec{c}^{\,i}_x=(\vec{n},\vec{N}(\vec{\theta}),T_{g/d},\chi_\mathrm{FUV},\zeta_\mathrm{CR},...)$, such as the chemical vector $\vec{n}=(n_1,...,n_N)$, the column density vector $\vec{N}(\vec{\theta })$ which is a function of direction $\vec{\theta}$, gas and dust temperature $T_{g/d}$, FUV intensity $\chi_\mathrm{FUV}$ (in units of the Draine field \citet{draine78}), the cosmic ray ionization rate $\zeta_\mathrm{CR}$, and possibly others. The superscript $i$ is the global iteration counter, the subscript $x$ enumerates the positions.

Eq.~\ref{eq:energybalance} is coupled to the problems of chemical balance, local excitation and radiative transfer (see Fig.~\ref{fig:numscheme}). The importance of various heating and cooling mechanisms as a function of position in the PDR has been widely discussed in the past \citep[e.g.][]{TH85, htt91, bt94, sternberg1995, roellig06, woitke2009}. At low $A_V$ photo-electric heating and \ce{H2} vibrational de-excitation are the dominant heating, at high $A_V$, cosmic ray heating (CR) and gas-grain collisions are the most important heating processes. Cooling at the surface of the PDR is provided by {\cii} and {\oi} fine-structure line cooling, by CO rotational line cooling at high $A_V$, and by dust \citep{ossenkopf2015}. 
The photo-electric heating depends on the grain properties. For an assumed dust distribution  given by \citet{wd01} we implemented the formalism given by \citet[][their Eqs.~(44),(45) and Table~2 and 3]{wd01pe}. Details of our implementation are described in \citet{roellig2013dust}. If the user selects a MRN dust distribution \citep{mrn} we use the photoelectric heating given by \citet[][their Eqs.~(41),(43),(44)]{bt94}. 
Table~\ref{tab:heating_cooling} lists all heating and cooling processes implemented in \Kt{} and Details are discussed in Appendix~\ref{sect:heatingcooling_appendix}. 

\subsection{Local convergence}
Solving the local problem is equivalent to finding a local gas temperature such that the local chemistry and all the local thermodynamics are in equilibrium. To solve the chemistry as well as the heating and cooling we need to compute the local radiation field, the dust temperature and the local photon escape probability (see Local iteration loop in Fig.~\ref{fig:numscheme}). In plane-parallel PDR models this can usually be achieved by assuming exponential attenuation along the line of sight. In spherical PDRs this is complicated by the fact that we assume an isotropic FUV irradiation and that absorbing columns of gas and dust depend on the direction.  To compute the local intensity in a spherical cloud we then need to average over all directions:
\begin{equation}\label{eqn:angle_average}
	I(z)=\frac{1}{4\pi}\int_\Omega I_{in}(\theta)\exp(-\gamma_\nu A_{V}(z,\theta))d\Omega 
\end{equation}  
 Along any given line-of-sight the attenuation of radiation at a given frequency can be described in terms of $\exp(-\gamma_\nu A_\mathrm{V})$.  $\gamma_\nu$ relates the attenuation at a frequency $\nu$ to the visual extinction.
Using the relation $A_\nu=1.086 \tau_\nu$  it is common practice to use the visual extinction $A_\mathrm{V}$ as 'optical' coordinate. A quantity such as the local optical depth $\tau_\nu=\sigma_\nu N$ \citep{sternberg1989}, with frequency $\nu$ dependent cross section $\sigma_\nu$ and the related column density of the absorber $N$, has to be considered for all directions $\theta$:
\begin{equation}\label{eqn:tau_theta}
\tau_\nu(\theta)=\sigma_\nu N(\theta) \,.
\end{equation} 

The  flow of the local iteration is summarized in \textbf{Algorithm: Local iteration}~\ref{alg:local}.

\setcounter{algorithm}{0}
\begin{algorithm}[thb]
	\floatname{algorithm}{Algorithm: Local iteration}
	\caption{} \label{alg:local}
	\begin{algorithmic}
		\State \textbf{Initial values:}
		\State $z \leftarrow z_\mathrm{new}$, $T_\mathrm{g} \leftarrow T_\mathrm{g,old}$\textbf{} 
\vspace{3pt}
		\State Compute $T_\mathrm{d}(z)$
		\ruleline{\textbf{angular average of radiative quantities}}\vspace{3pt}
		\State Compute $\chi_\mathrm{FUV}(z)$
		\State reaction specific dust attenuation\; $\frac{1}{4\pi}\int_\Omega e^{-\gamma_i A_{V}}d\Omega$
		\State self shielding \ce{CO} \; $\frac{1}{4\pi}\int_\Omega f_{SS}(N_{\ce{CO}},N_{\ce{H2}})d\Omega$
		\State \ce{H2} UV shielding and pumping\vspace{3pt}
		\ruleline{ \textbf{Compute $T_\mathrm{g}(z)$ }}\vspace{3pt}
		\While{$E_\mathrm{tot}\ne 0$}
		\State $T \leftarrow T_\mathrm{g}$
		\State solve level population: \ce{H2},\ce{CO},\ce{O},\ce{C}, ...
		\State solve chemistry
		\State compute local heating \& cooling $\Gamma(T)$, $\Lambda(T)$
		\State $E_\mathrm{tot}\leftarrow\Gamma(T)-\Lambda(T)$
		\State choose root finding algorithm
		\State $T\leftarrow  \mathop{\arg \min}\limits_{T \in [T_\mathrm{min},T_\mathrm{max}]} |E_\mathrm{tot}(T)|$ 
		\EndWhile
		\State $T_\mathrm{g}(z)\leftarrow T$
	\end{algorithmic}
\end{algorithm}

\subsection{Spatial loop}\label{sect:spatial-loop}
The physical and chemical conditions in a PDR are subject to non-linear effects such as the photo-dissociation of \ce{H2} and \ce{CO} and their respective shielding. The  photo-dissociation rates show a sudden drop once the absorption lines become optically thick. As a consequence, we find a steep density gradient of \ce{H2} and \ce{CO} in these transition regions covering many orders of magnitude in volume density. To avoid numerical problems it is important to use a sufficiently dense spatial grid across these transition regions and the method described below performs well across the full parameter range.

During each spatial iteration \Kt{} performs   an adaptive spatial gridding using the Bulirsch-Stoer method \citep[][Sect. 16.4]{numericalrecipes} to determine the next step width such that the numerical solution of the set of $\mathcal{N}$ coupled first-order differential equations for the functions $y_i; i = 1,2,\ldots,\mathcal{N}$, having the general form
\begin{equation}\label{coupled_ODEs}
\frac{d y_i(z)}{dz}=f(z,y_1,y_2,...,y_\mathcal{N})
\end{equation}
is sufficiently accurate. In the context of solving the PDR structure we identify $y_i(z)=N_i(z)$ as the perpendicular column density of species $i$ from the surface to depth $z$ in the cloud so that $dy_i(z)/dz=n_i(z)$. This extends to the column densities of energy level populations of species of interest, in particular \ce{CO}, \ce{C+}, \ce{C},\ce{O}, and \ce{H2}. From a given position and column density vector, the Bulirsch-Stoer stepper routine returns the next step-width $\Delta h$ together with the respective values for $n_i(z+\Delta h)$ and $N_i(z+\Delta h)$ as well as an estimate of which step-width to attempt next. This successively propagates the solution into the cloud until the cloud center is reached.

\subsubsection{Global convergence}\label{sect:global_convergence}
\Kt{} uses the final column density vector, i.e. the column density of all species measured from the edge of the cloud to the cloud center, $\vec{N}$ to test for convergence. Global convergence is met if
\begin{equation}\label{eq:global_convergence_crit}
\frac{N^\mathrm{old}_i-N^\mathrm{new}_i}{N^\mathrm{new}_i}<\epsilon_\mathrm{iter} \;\;\forall\, i\,\text{with}\; N^\mathrm{new}_i>10^{12}\,\mathrm{cm^{-2}}
\end{equation}
The standard value applied is $\epsilon_\mathrm{iter}=0.01$. The minimum number of iterations is 2. We also specify a maximum number of iterations (default of 60) before stopping further iterations (Most models converge within less than 30 iterations). If more than 40 iterations are needed we incrementally relax the 0.01 criterion in steps of 0.01 to 0.1. We try another 10 iterations with $\epsilon_\mathrm{iter}=0.1$. If no convergence is found in 60 iterations the result of the last iteration is stored because for certain applications the results might still be sufficient. It is for example in some cases numerically difficult to find a stable solution for massive dense clouds. There, the low $J$  transitions of \ce{CO} are the only cooling options for the gas in the central parts because of the low gas temperatures. The respective optical depths however are so high that Eq.~(\ref{eq:energybalance}) becomes strongly non-linear and  small changes in $T_g$ lead to large differences in the \ce{CO} population numbers that can produce oscillating solutions for the population column density vector. However, this happens for such high optical depths that the effect is not observable and has no impact on the total \ce{CO} emission of the model cloud. 

In general, we find that for a given number of chemical species  the code converges faster for low densities, low masses (i.e. low optical depths) and not too high FUV fields. The total computation time for one cloud model, i.e. one specific set of model parameters range between few minutes and 10-20 hrs on an Intel Core i7 CPU.

\subsubsection{Chemical time scales}\label{sect:time_scales}

To assess the limits of the model assumption of stationary chemistry we have to compare the involved chemical time scales with other relevant time scales, such as the dynamical time scale of the clump or its total lifetime. The model looses predictive power if those times are comparable or larger than the time the chemistry needs to reach steady-state. We derive chemical times from the total formation or destruction rate of a species divided by its total number density. Taking the inverse is the relevant formation/destruction time per particle and it will depend on the local conditions. Generally, the chemistry is very fast close to the surface with significant local photon densities and high temperatures and slows down with $A_V$. It typically peaks once $A_V>1-3$. The chemistry is slowest for low density and FUV illumination  (e.g. $n=10^3$~cm$^{-3}$ and $\chi=1$) with gas-phase time scales $\lesssim 10^6$~yr and fastest for high density and FUV illumination  (e.g. $n=10^6$~cm$^{-3}$ and $\chi=10^6$) with gas-phase time scales $\lesssim 10^4$~yr. The surface chemistry is considerably slower once $A_V>10$ with values of $10^8$~yr and longer. Our model showed the time scales for \ce{J(CO)} and \ce{J(CH4)} to be about at least 1-2 orders of magnitude faster compared to \ce{J(H2O)} and \ce{J(CO2)}.

Typical dynamic time scales are of the order of $10^5-10^6$~yr which is larger  than the gas-phase time scales under typical PDR conditions and of the same order for low density and FUV conditions. However, considering that PDRs typically evolve from cold molecular gas it is reasonable to assume a total life time of up to $\sim 10^7$~yr which is sufficient to establish chemical steady-state independent of $A_V$. For the surface chemistry even this time may be too short to reach a stationary state for e.g. water ice deep inside the clump ($A_V\gtrsim 10$). We note, that our models sometimes show a significant deviation between total formation and destruction rates of some ice species, due to our small network of surface reactions. We conclude that the assumption of steady-state is justified for the gas-phase chemistry under most PDR conditions and may become problematic in low $n_0$ and low $\chi$ environments. The ice chemistry can reasonably well be approximated with a stationary chemistry for $A_V \lesssim 5-10$ and even deeper in the cloud for some ice species, e.g. \ce{J(CO)} \citep[see also][]{hollenbach2009}. On the other hand, time-dependent computations by \citet{esplugues2019} show that \ce{J(H2O)} abundance at strong FUV illumination and $n=10^5$~cm$^{-3}$ reaches a plateau at $A_V>7$ after $10^6-10^7$~yr.
Note, that previous studies showed the existence of multiple chemical solutions \citep{lebourlot93bistability,lebourlot1995,lee1998,charnley2003,boger2006,dufour2019} as well as the existence of sustained chemical oscillations \citep{roueff2020} in gas-phase models of the ISM.

\section{Chemical Solution}

\subsection{Solving the chemical equations}

\Kt{} computes an equilibrium solution to Eq.~\ref{eq:chemical_rate_equation} and Eq.\ref{eq:chemical_rate_equation2}, i.e. the solution to the system
\begin{equation}\label{eq:linear_system}
	\mathcal{F}_i=\frac{dn(i)}{dt}=0
\end{equation}
using the Newton-Raphson method with the Jacobi matrix $\mathbb{Q}$. This is a locally convergent method that is guaranteed to converge for a starting point sufficiently close to the root. The elements of $\mathbb{Q}$ are:
\begin{equation}\label{eq:jacobian}
	Q_{i,j} = \frac{d\mathcal{F}_i}{dn(j)}
\end{equation}

For the unknown densities $n_i$ of $\mathcal{N}$ species built up from $\mathcal{M}$ elements we find an over-determined set of $\mathcal{N}+\mathcal{M}+1$ equations.  The default method in \Kt{} is to replace $\mathcal{M}+1$ equations from the set of chemical rate equations (\ref{eq:chemical_rate_equation} and \ref{eq:chemical_rate_equation2}) with the corresponding elemental and charge conservation equations (\ref{eq:conservation_equation} and \ref{eq:charge_conservation_equation}).
This can always be justified because the conservation equations should be strictly fulfilled while all rate equations are in any way only approximately known. Moreover, it improves the numerical stability of the iterative solution.

The Newton-Raphson method approaches the solution by using an (old) approximate solution (or initial guess) $n_i^\mathrm{old}$ to compute an (new) improved approximate solution $n_i^\mathrm{new}$ in the next step:
\begin{equation}\label{eq:newton-raphson}
	\sum_j -Q_{i,j}^\mathrm{old} \left( n(j)^\mathrm{new}-n(j)^\mathrm{old}\right) = \mathcal{F}_i^\mathrm{old}
\end{equation}
This is a linear(ized) system of the form $\mathbb{Q}\cdot \delta\vec{n}=\vec{\mathcal{F}}$. To solve Eq.~(\ref{eq:newton-raphson}), we need to invert $\mathbb{Q}$ in order to determine the new density vector $\vec{n}^{new}$. 
In standard Newton-Raphson algorithms, the new solution-step is computed as $n(i)^\mathrm{new}=n(i)^\mathrm{old}+{\delta}n_i$ and used as updated  $n(i)^\mathrm{old}$ value in the subsequent iteration step. The steps can be expressed as relative changes: $\eta={\delta}n_i /{n(i)^\mathrm{old}}$.  \Kt{} solves Eq.~\ref{eq:newton-raphson} by LU decomposition but has alternative solvers implemented.

\subsection{Newtonian stepping strategies}	

The Newton-Raphson algorithm is guaranteed to provide fast convergence if one is close to solution, but depending on the starting values and the overall topology of the Newtonian vector field $\delta \vec{n}$, i.e. the solution to the system $-\mathbb{Q}\cdot\delta\vec{n}=\mathcal{F}$, the Newton-Raphson algorithm may also diverge or become trapped in a local minimum leading to infinite oscillations. Adaptive stepping strategies may prevent this.

We can express the step in the Newton-Raphson algorithm, $n(i)^\mathrm{new}=n(i)^\mathrm{old}+{\delta}n_i$, as a relative change, so that $n(i)^\mathrm{new}=n(i)^\mathrm{old}\cdot f_\mathrm{step}$ with 
\begin{equation}\label{eq:newton_stepper}
		f_\mathrm{step,N}(\eta)=\left(1+\eta\right) \;\;\mathrm{with} \;\; \eta=\frac{{\delta}n_i }{n(i)^\mathrm{old}} 
	\end{equation}
For large steps, this approach can be problematic because it does not prohibit negative solutions, i.e. negative densities. In general, Newton stepping through the negative domain is not problematic as long as we can guarantee that the steps will finally provide positive densities. However, this is problematic because the algorithm may converge on a local minimum involving negative $n_i$. One way to avoid this is to prevent the Newton steps from producing sign switches in $\vec{n}$.
	
The $\tanh()$ allows for a convenient construction of such a general stepping function whose symmetry and limits can be controlled with few numerical parameters: 
\begin{equation}\label{eq:damping_symlogtanh}
	f_\mathrm{step,Tanh}(\eta)=
	\begin{cases}
		\left(1 + (\omega_- - 1) \lambda \tanh\left(-\eta \frac{1}{\omega_- - 1}\right)\right)^{-1}	& \text{if $\eta<0$} \\
		1+(\omega_+ - 1) \lambda \tanh\left(\frac{1}{\omega_+ - 1} \eta \right)  & \text{if $\eta\ge 0$} 
	\end{cases}
\end{equation}
$\omega_+>1$ and $\omega_->1$. For small $\eta$ Eq.~\ref{eq:damping_symlogtanh} approaches Eq.~\ref{eq:newton_stepper} and levels off for large arguments preventing too large steps. We implement an adaptive choice of $\omega_+, \omega_-$, and $\eta$, with details described in the Appendix~\ref{sect:efficient_stepping}.

\begin{table*}[htb]
	\centering
	\caption{Chemical species in the surface test models.}\label{tab:surface_chemistry_test}
	\begin{tabular}{llllllllll}
		\hline \hline
		\vrule width 0pt height 2.2ex
 \ce{e-} & \ce{H} & \ce{H2} & \ce{He+} & \ce{He} & \ce{H+} & \ce{H2+} & \ce{H3+} & \ce{C} & \ce{CH} \\
 \ce{C+} & \ce{CH+} & \ce{CH2+} & \ce{CH2} & \ce{CH3+} & \ce{CH3} & \ce{CH4+} & \ce{CH4} & \ce{CH5+} & \ce{CN} \\
 \ce{CN+} & \ce{HCN} & \ce{HCN+} & \ce{CO} & \ce{CO+} & \ce{HCO+} & \ce{HCO} & \ce{CO2} & \ce{CO2+} & \ce{H2CO} \\
 \ce{H2CO+} & \ce{^{13}C} & \ce{^{13}CH} & \ce{^{13}CO} & \ce{^{13}C+} & \ce{^{13}CH+} & \ce{^{13}CH2+} & \ce{^{13}CO+} & \ce{H^{13}CO+} & \ce{N} \\
 \ce{N2} & \ce{N2+} & \ce{N2H+} & \ce{NO} & \ce{NO+} & \ce{O} & \ce{O+} & \ce{OH} & \ce{OH+} & \ce{H2O} \\
 \ce{H2O+} & \ce{H3O+} & \ce{O2} & \ce{O2+} & \ce{SO} & \ce{SO+} & \ce{SO2} & \ce{SO2+} & \ce{HSO2+} & \ce{S} \\
 \ce{S+} & \ce{HS} & \ce{HS+} & \ce{H2S+} & \ce{CS} & \ce{CS+} & \ce{HCS+} & \ce{^{13}CS} & \ce{^{13}CS+} & \ce{H^{13}CS+} \\
 \ce{Si} & \ce{Si+} & \ce{SiH} & \ce{SiH+} & \ce{SiH2+} & \ce{SiO} & \ce{SiO+} & \ce{SiOH+} & \ce{^{13}CO2} & \ce{^{13}CO2+} \\
 \ce{H^{13}CO} & \textbf{\ce{CH3OH+}} & \textbf{\ce{CH3O}} & \textbf{\ce{CH3OH}} & \textbf{\ce{H2O2}} & \textbf{\ce{J(H)}} & \textbf{\ce{J(H2)}} & \textbf{\ce{J(C)}} & \textbf{\ce{J(CH)}} & \textbf{\ce{J(CH2)}} \\
 \textbf{\ce{J(CH3)}} & \textbf{\ce{J(CH4)}} & \textbf{\ce{J(CN)}} & \textbf{\ce{J(HCN)}} & \textbf{\ce{J(CO)}} & \textbf{\ce{J(HCO)}} & \textbf{\ce{J(CO2)}} & \textbf{\ce{J(H2CO)}} & \textbf{\ce{J(CH3O)}} & \textbf{\ce{J(CH3OH)}} \\
 \textbf{\ce{J(N)}} & \textbf{\ce{J(N2)}} & \textbf{\ce{J(NO)}} & \textbf{\ce{J(O)}} & \textbf{\ce{J(OH)}} & \textbf{\ce{J(H2O)}} & \textbf{\ce{J(H2O2)}} & \textbf{\ce{J(O2)}} & \textbf{\ce{J(SO)}} & \textbf{\ce{J(SO2)}} \\
 \textbf{\ce{J(S)}} & \textbf{\ce{J(^{13}C)}} & \textbf{\ce{J(^{13}CO)}} & \textbf{\ce{J(H^{13}CO)}} & \textbf{\ce{J(^{13}CO2)}} &   &   &   &   & \\ \hline
	\end{tabular}
  \tablefoot{\ce{J(X)} indicates surface species \ce{X}. Species written in bold font are not included in the gas-phase model computations.}
\end{table*}

\section{Energy Solution}
The numerical results presented in this section have been computed with WL-PDR, a simplified, plane-parallel PDR model written in Mathematica \citep{Mathematica}. WL-PDR is designed to act as numerical testing environment of PDR modeling aspects. A brief explanation of the code is given in Appendix~\ref{sect:wl-pdr}. 

\subsection{Numerical Iteration Scheme}

The key part in Fig.~\ref{fig:numscheme} is the local solution module. A local solution in this context is the local balance between all cooling $\Lambda_k$ and heating $\Gamma_h$ processes (Eq.~\ref{eq:energybalance}), 
where the efficiency of any individual process depends on the local conditions $\vec{c}_x^{\,i}$. 
\begin{figure}[htb]
	\resizebox{\hsize}{!}{\includegraphics{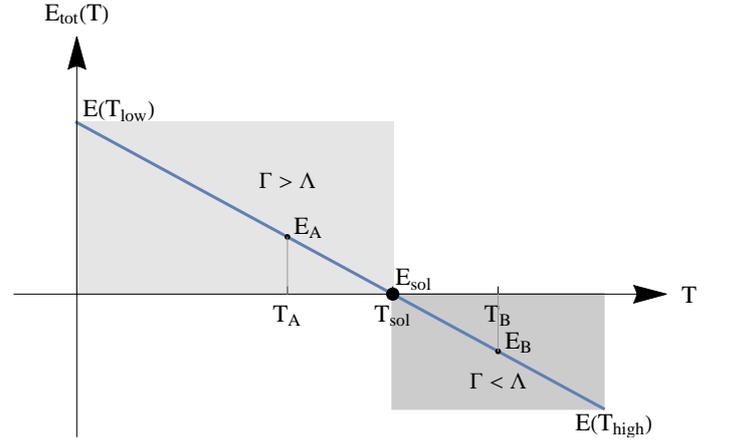}}
	\caption{Idealized form of Eq.~\ref{eq:energybalance} with a single root.  Numerically, a temperature solution $T_\mathrm{sol}$ is determined by finding the root of the energy balance equation $E_\mathrm{sol}=E_\mathrm{tot}(T_\mathrm{sol})=0$.}
	\label{fig:rootscheme}
\end{figure}  
Eq.~\ref{eq:energybalance} is coupled to the problems of chemical balance, local excitation and radiative transfer (see Fig.~\ref{fig:numscheme}). Figure~\ref{fig:rootscheme} shows an idealized form of Eq.~\ref{eq:energybalance}. At any given position in the PDR, every point on the curve $E_\mathrm{tot}(T)$ corresponds to a different chemical vector and a different temperature. At the solution $E_\mathrm{tot}(T_\mathrm{sol})=E_\mathrm{sol}=0$ heating and cooling are balanced. If $E_\mathrm{tot} > 0$ (gray rectangle left of $T_\mathrm{sol}$) we have  heating excess $ \Gamma > \Lambda$. In the opposite case $E_\mathrm{tot} < 0$ (gray rectangle right of $T_\mathrm{sol}$) we have a cooling excess $\Gamma < \Lambda$ \footnote{The assignment of positive values to heating is arbitrary and can be reversed.}. Physics ensures that $\lim_{T\rightarrow 0} E(T) > 0$ because all cooling process become inefficient at very low temperatures and heating is less dependent on $T$. Conversely, it is also ensured, that $E(T_\mathrm{high})=\lim_{T\rightarrow T_\mathrm{max}} E(T) < 0$. Here $T_\mathrm{max}$ corresponds to the upper temperature that is considered reasonable in a PDR and which depends on the parameter range that the model needs to cover. Typical values are $T_\mathrm{max}\approx 10000-20000$~K. Even though many coolants are inefficient at very high temperatures -- CO and H$_2$ for instance are chemically destroyed -- other cooling processes become strong. Typical candidates are Lyman \textalpha, O {\scriptsize I} 6300 \AA\ cooling \citep{bt94, spitzer1978} and  gas-grain collisional cooling. These processes ensure that $E(T_\mathrm{high})<0$.  If $E_\mathrm{tot}$ is a continuous function the intermediate value theorem states that at least one root must lie in the interval $[0,T_\mathrm{max}]$ and a physical solution of the problem exists.

$T_\mathrm{sol}$ is a stable solution if $\partial E_\mathrm{tot}/\partial T <0$. Imagine the gas at $T_\mathrm{sol}$ is slightly perturbed towards higher temperatures, e.g.  toward $E_B$. In this regime, the cooling exceeds the heating and  drives the temperature back to  $T_\mathrm{sol}$. The same happens for perturbation toward $E_A$. The stronger heating  drives the temperature back into the equilibrium. 

\begin{figure}[htb]
	\resizebox{\hsize}{!}{\includegraphics{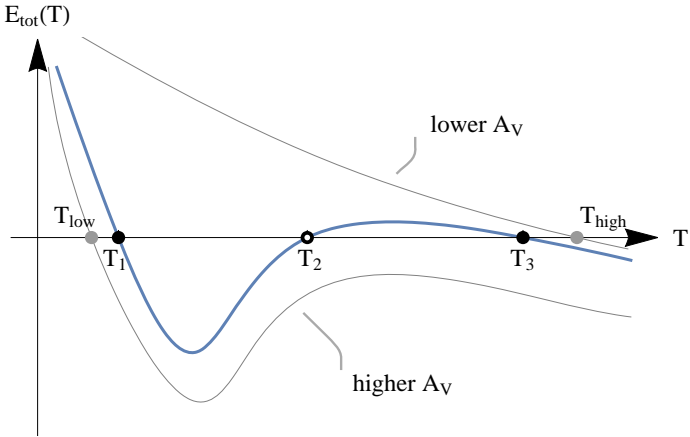}}
	\caption{Idealized form of Eq.~\ref{eq:energybalance} with multiple roots. Filled points indicate a stable solution, open points are unstable against perturbations. The energy balance for two different positions in the PDR are schematically drawn in gray together with their equilibrium temperatures $T_\mathrm{high}$ and $T_\mathrm{low}$.}
	\label{fig:rootscheme2}
\end{figure}

\subsection{Multiple temperature solutions in PDR models\label{sect:multipleroots}}

The energy brackets $E(T_\mathrm{low})>0$ and $E(T_\mathrm{high})<0$ guarantee that at least one root exits, but it does not prohibit the existence of multiple roots.  For numerical root finding algorithms this poses a serious problem, because they  usually find only one root. Which one depends on the choice of the  algorithm, the initial starting temperature and the temperature range where the root is searched. 

\begin{figure*}[hbt]
    \begin{subfigure}[t]{.48\textwidth}
    	\centering
     	\includegraphics[width=\linewidth]{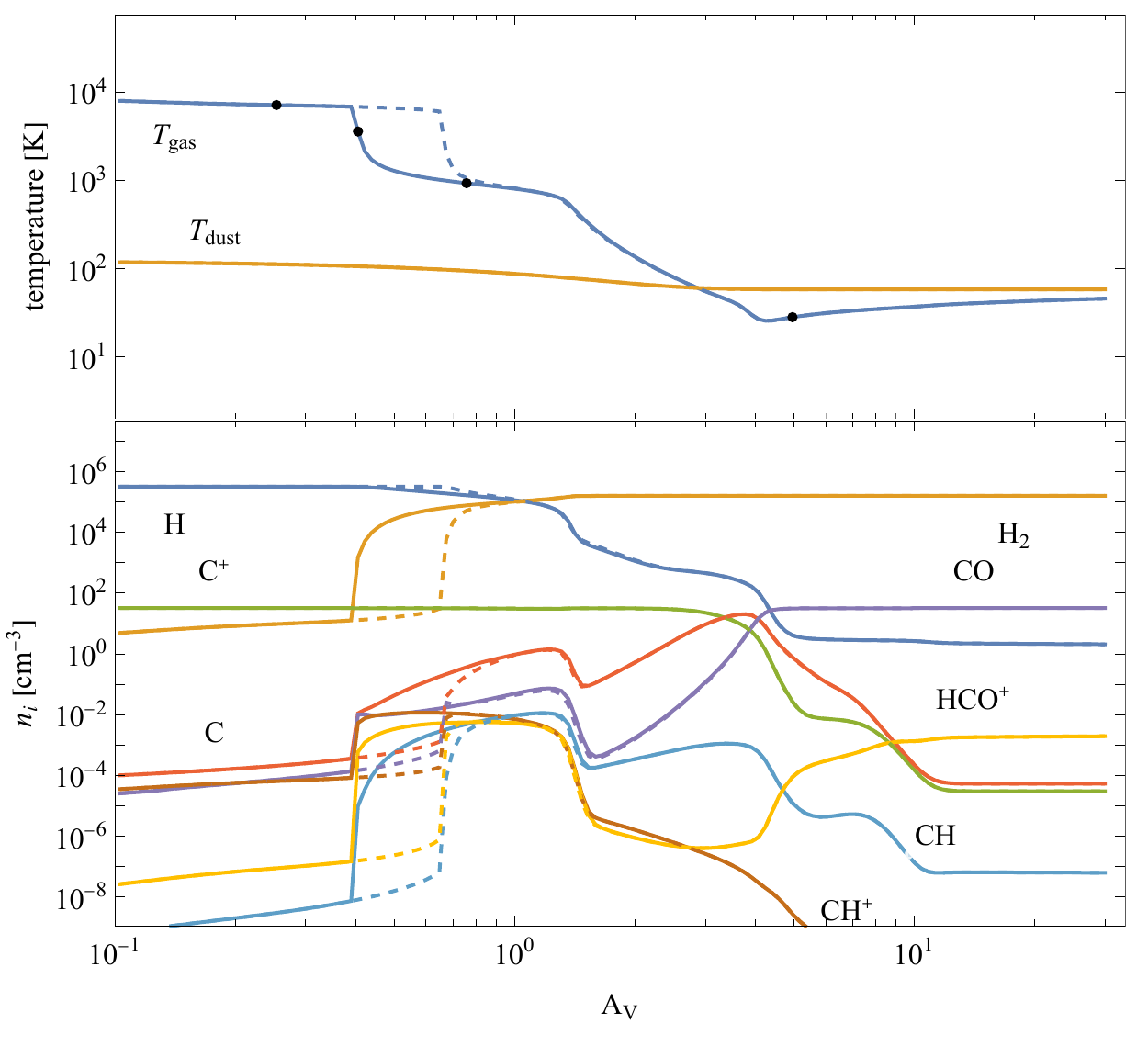}
    	\caption{PDR structure}\label{fig:chem-V4}
    \end{subfigure}
    \hfill
    \begin{subfigure}[t]{.48\textwidth}
 		\centering
 		\includegraphics[width=\linewidth]{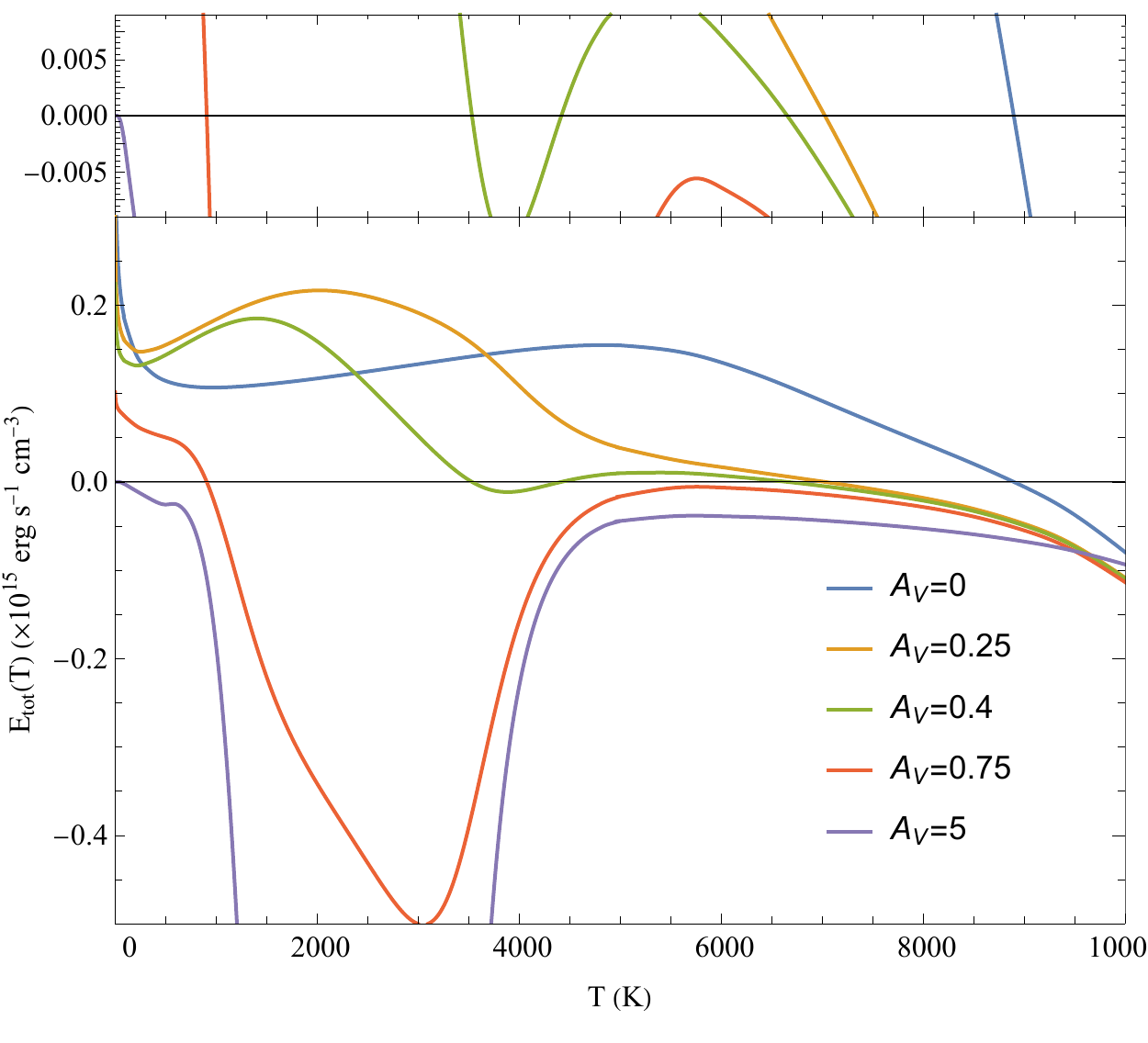}
		\caption{Energy balance}\label{fig:etot-V4}
    \end{subfigure}
\caption{Two-temperature solution for the  V4 benchmark model ($n=10^{5.5}\,\mathrm{cm}^{-3},\chi=10^5$). \textbf{Left:} $T_{g/d}$ (top) and $n_i$(bottom) profile. Solid and dashed lines correspond to the low and high temperature solutions (WMM and HAM. i.e. $T_1$ and $T_3$ from Fig.~\ref{fig:rootscheme2}). 
 \textbf{Right:} $E_\mathrm{tot}(T)$ at selected positions in the PDR (marked by black points in Fig.~\ref{fig:chem-V4}). The top panel magnifies the range around $E_\mathrm{tot}=0$.}\label{fig:V4-benchmark-model}
\end{figure*}

The form of $E_\mathrm{tot}(T)$ can change when moving through the PDR. In Fig.~\ref{fig:rootscheme2} we sketch such a scenario. At low values of $A_V$ we find a high temperature solution $T_\mathrm{high}$ that is the root of the upper, gray curve in Fig.~\ref{fig:rootscheme2}.  We denote the high and low temperature solutions as hot atomic medium (HAM) and warm molecular medium (WMM), respectively, as we will see in Fig.~\ref{fig:etot-V4} that the high temperature solution is usually also associated with significantly less H$_2$ than the low temperature solution.  At higher values of $A_V$  the local conditions allow for an additional cooling process to become efficient in an intermediate temperature range producing a local minimum of the thick line in the Fig.. We  compute the PDR structure from the outside to the inside. After determining $T_\mathrm{high}$ at the low $A_V$ value we move to a deeper cloud position and search the temperature root there. To speed up root search the new temperature root is usually searched in the neighborhood of the previous solution. Assuming that the spatial gridding is fine enough, we expect the temperature to change slowly between subsequent numerical steps. However, this strategy is doomed to always find the solution $T_3$ and will not reach $T_1$ (or $T_2$).

If we continue the computation steps at even higher $A_V$ we have to find a significantly lower root at $T_\mathrm{low}$. In this three-step picture we have a large temperature jump from $T_3$ to $T_\mathrm{low}$. In real PDR model computations this is a sudden transition from temperatures of a few thousand K to few hundred K across a small spatial range, especially in models with high gas density and high FUV illumination. Depending on the search range of the applied temperature finding algorithm this sudden temperature jump can lead to severe numerical instabilities and prohibit global convergence.

\subsection{Physics of multiple PDR temperatures} \label{sect:multiple_root_physics}
The described multiple temperature solutions were first reported by \citet{burton90} for $n=10^6$~cm$^{-3}$ and $G_0=10^3$. They described two temperature solutions with different heating and cooling balances. We find similar effects for even higher FUV fields. It is important to include this behavior in the PDR code to evaluate knock-on effects for the chemical structure and the observable line intensities.
 Fig.~\ref{fig:V4-benchmark-model} summarizes the model results (computed with \texttt{WL-PDR}) for the benchmark case V4 ($n=10^{5.5}$cm$^{-3}$, $\chi=10^5$). In panel a)  we show the temperature and chemical abundance profiles. The dashed  curves corresponds to the HAM solution, the solid curves to the WMM. In panel b)  we  plot the total energy balance vs. the gas temperature. At low $A_V$ values we find a  strong contribution by vibrational H$_2$ de-excitation heating (pump heating) and H$_2$ line cooling. They inherit the non-monotonous temperature dependence from the \ce{H2} abundance with a weak maximum around 5000~K. H$_2$ formation heating is also strong and even remains so at much higher temperatures while the pump heating and line cooling drops off beyond 7000-8000~K.

Moving even deeper into the PDR, H$_2$ cooling becomes stronger and starts to suppress the roots at temperatures higher than the WMM solution $T_\mathrm{sol}=910$~K. In the case $A_V=0.75$ (only shown in panel b) as red curve) we find that $E_\mathrm{tot}(T>910~\mathrm{K})<0$. Looking at Fig.~\ref{fig:V4-benchmark-model} b) shows that at 6000~K the energy is almost balanced with $E_\mathrm{tot}=-5.5\times 10^{-20}$~erg~s$^{-1}$~cm$^{-3}$. \footnote{Numeric root search algorithms may find a root here depending on their numerical parameters even though this is mathematically not a root. }
	
In \citet{roellig06} we discussed how the dominant heating and cooling at the surface of PDRs changes with gas density $n$ and with FUV intensity $\chi$.  The energetic behavior described above is  associated with the cooling (and heating) capabilities of molecular hydrogen H$_2$. However, we note, that this is not due to increased cooling in the high density case. The important point is that the cooling needs to be balanced by a sufficiently strong heating in order to produce a temperature solution ($E_\mathrm{tot}=0$). In case of low FUV or low density we don't have an efficient heating process in this regime and accordingly we will not find multiple solutions. Adding additional or stronger heating to the model can possibly alter this behavior.

Such phase transitions result from physical processes that become effectively inefficient once they cross an energetic threshold. The main difficulty is that even though such transitions are physically possible, their prescription in numerical models suffers from inherent uncertainties: 1)  numerical approximations of complex physics, 2) unknown or inaccurate material constants (e.g. collision rates, spectroscopic constants, binding energies), 3) unavoidable model simplification (e.g. geometry, numerical resolution), 4) numerical inaccuracies. Any of these can either lead to a phase transition or prevent it.

\begin{figure}[thb]
	\resizebox{\hsize}{!}{\includegraphics{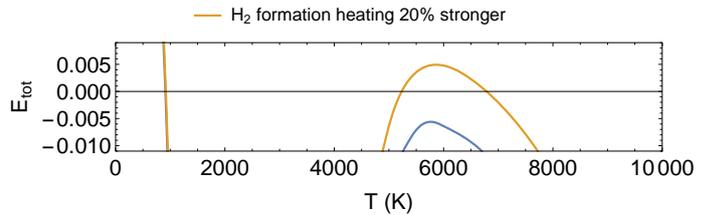}}
	\caption{Energy balance for the V4 model at $A_V=0.75$. The blue line shows the standard computation. The orange line was computed using a 20\% enhanced H$_2$ formation heating.}
	\label{fig:modH2form}
\end{figure}

In the model V4 at $A_V=0.75$ we noted in Fig.~\ref{fig:V4-benchmark-model} that $E_\mathrm{tot}(T)$ is smaller but close to zero between 5000 and 7000~K. This is because the total cooling is only slightly stronger in that temperature range than the total heating. In Fig.~\ref{fig:modH2form} we show how changing a single heating process can change the situation. The orange line shows $E_\mathrm{tot}$ in case of a 20\% enhanced H$_2$ formation heating. This would result in two additional temperature solutions, i.e. a possible temperature solution about 7 times higher than in the case of standard H$_2$ formation heating.

Predicting the effect of the two different temperature solutions on observable line intensities is not straight forward because excitation conditions and column density effects are mixed. We found up to 20\% higher CO emissivities for the upper lines and up to 70-80 \% higher \ce{CH+} line intensities for higher transitions when selecting the HAM solution instead of the WMM solution.

The final choice between WMM and HAM depends on the modeled physical scenario as already described by \citet{burton90}. An initially hot and fully atomic (or ionized) medium that becomes subject to significant lower FUV illumination will settle down at lower temperatures coming from a much higher temperature. This clearly prefers the HAM solution. In the opposite case, where the gas starts from cold and molecular conditions at the onset of FUV illumination, the WMM solution is favored because the gas is gradually transitioning from lower to higher temperatures. A  scenario for the second case is the onset of massive star formation in the vicinity of the model cloud while the first scenario could occur in case of some shielding due to cloud motions or sudden changes in the illuminating source due to supernova events.

\section{New model results}\label{sect:new_results}

The most important surface reaction in the ISM is the formation of molecular hydrogen \citep{cazaux2002,cazaux2004,lebourlot2012,bron2014,wakelam2017,thi2020} and all PDR models already account for it with various degrees of complexity. Other surface reactions were typically not included until sufficient computing power became available.  \citet{woitke2009} presented the disk PDR model ProDiMo with surface chemistry and
\citet{hollenbach2009} published a plane-parallel PDR model with a small surface network to study the chemistry of \ce{H2O} and \ce{O2} and \citet{guzman2011} studied the chemistry in the Horsehead using an updated version of the Meudon code \citep{lepetit2006}. 
Recently, \citet{esplugues2016,esplugues2019} presented PDR model results including (time-dependent) surface chemistry. In a more recent work, they expanded their chemical desorption scheme to include partial desorption of the products  \citep{esplugues2019}. 
We compare our model computations with their results and in addition we investigate how the revised description of CR induced desorption  affects the  structure of the PDR. 
As shown by \citet{sipilae2021} Eqs.~\ref{eq:tau_col} and \ref{eq:tau_heat} result in higher values of $f_{CR}$ compared to $3.16\times 10^{-19}$ as suggested by \citet{hasegawa1993} making the CR-induced desorption more efficient.
We use the label CR$_1$ and CR$_2$ to indicate CR desorption by \citet{hasegawa1993} and \citet{sipilae2021}, respectively.  
Model parameters for computations in Sect.~\ref{sect:new_results} are summarized in Table~\ref{tab:model_parameter}.
\begin{table*}[htb]
	\caption{Model parameter for  models discussed in Sect.~\ref{sect:new_results}.}
	\label{tab:model_parameter}
	\centering
	\begin{tabular}{lll}
		\hline\hline
		\vrule width 0pt height 2.6ex 
		parameter&value&comment\\
		\hline
		\vrule width 0pt height 2.6ex 
		$n_0$&$10^4$~cm$^{-3}$& surface density\\
		$M$&$50,1000\,\mathrm{M}_\odot$&clump mass\\
	    $\chi$&$1-10^6$& FUV in units of \citet{draine78}\\
	    $\alpha$&1.5&density power law index\\
	    $R_\mathrm{core}$&0.2&constant density core fraction\\
	    FWHM&1 km~s$^{-1}$& micro-turbulent line width\\
	    $\zeta_\mathrm{CR}$&$10^{-16}$~s$^{-1}$&CR ionization rate\\
	    dust composition& WD01-7&\citep[][ entry 7 in their Table 1]{wd01}\\
	    $\Gamma_8$& PEHWD01-4&photo electric heating rate \citep[][ entry 4 in their Table 2]{WD01PEH}\\
	    $R_\mathrm{H_2}$&&\ce{H2} formation described in \citep{roellig2013dust}, formation on PAHs larger than $a_\mathrm{carb}=29 \text{\AA}$  \\
	    $f_\mathrm{H_2,ss}$&&\ce{H2} self-shielding \citep{federman79}\\
	    $[\ce{He}]/[\ce{H}]$&0.851&\ce{He} elem. abundance \citep{simon-diaz2011}\\
	    $[\ce{C}]/[\ce{H}]$&$2.34\times 10^{-4}$&\ce{C} elem. abundance \citep{simon-diaz2011}\\
	    $[\ce{^{13}C}]/[\ce{H}]$&$3.52\times 10^{-6}$&\ce{^{13}C} elem. abundance \citep{simon-diaz2011}\\
	    $[\ce{O}]/[\ce{H}]$&$4.47\times 10^{-4}$&\ce{O} elem. abundance \citep{simon-diaz2011}\\
	    $[\ce{N}]/[\ce{H}]$&$8.32\times 10^{-5}$&\ce{N} elem. abundance \citep{simon-diaz2011}\\	    
	    $[\ce{Si}]/[\ce{H}]$&$3.17\times 10^{-6}$&\ce{Si} elem. abundance \citep{simon-diaz2011}\\
	    $[\ce{S}]/[\ce{H}]$&$7.41\times 10^{-6}$&\ce{S} elem. abundance \citep{simon-diaz2011}\\
	    \hline   
		
	\end{tabular}
\end{table*}

\subsection{Selective freeze-out}
The ice composition in a molecular cloud varies with extinction because it is mostly governed by the dust temperature, which decrease with $A_V$. The gas temperature does not always steadily decrease with cloud depth because cooling radiation may become trapped in the cloud for large optical depths, leading to increasing gas temperature for high values of $A_V$. Table~\ref{tab:snow_temps} lists the average condensations (dust) temperatures for the most abundant ice species, derived from the balance between accretion and desorption rates. It is important to note, that major carbon-bearing ice species condense below 50~K (with the exception of methanol). If $T_\mathrm{dust}$ is sufficiently high, this may selective favor oxygen-bearing ice components or lock up elements in certain ice species that are favored under current dust temperature conditions.

\begin{table}[thb]
	\centering
	\caption{Condensation temperatures for some relevant species in K derived from the balance between accretion and desorption.}
	\label{tab:snow_temps}
	\begin{tabular}{cc|cc}
		\hline \hline
		\vrule width 0pt height 2.2ex
		\ce{H} & 11 & \ce{O2} & 18\\
		\ce{CO} & 21 & \ce{CO2} & 48 \\
		\ce{H2O} & 85 & \ce{H2O2} & 100 \\
		\ce{CH4} & 25 & \ce{SO2} & 95 \\
		\ce{N2} & 14 & \ce{NO} & 29 \\
		\ce{HCN} & 36 & \ce{CH3OH} & 66 \\
		\hline
	\end{tabular}
\end{table}

\begin{figure}[htb]
	\resizebox{\hsize}{!}{\includegraphics{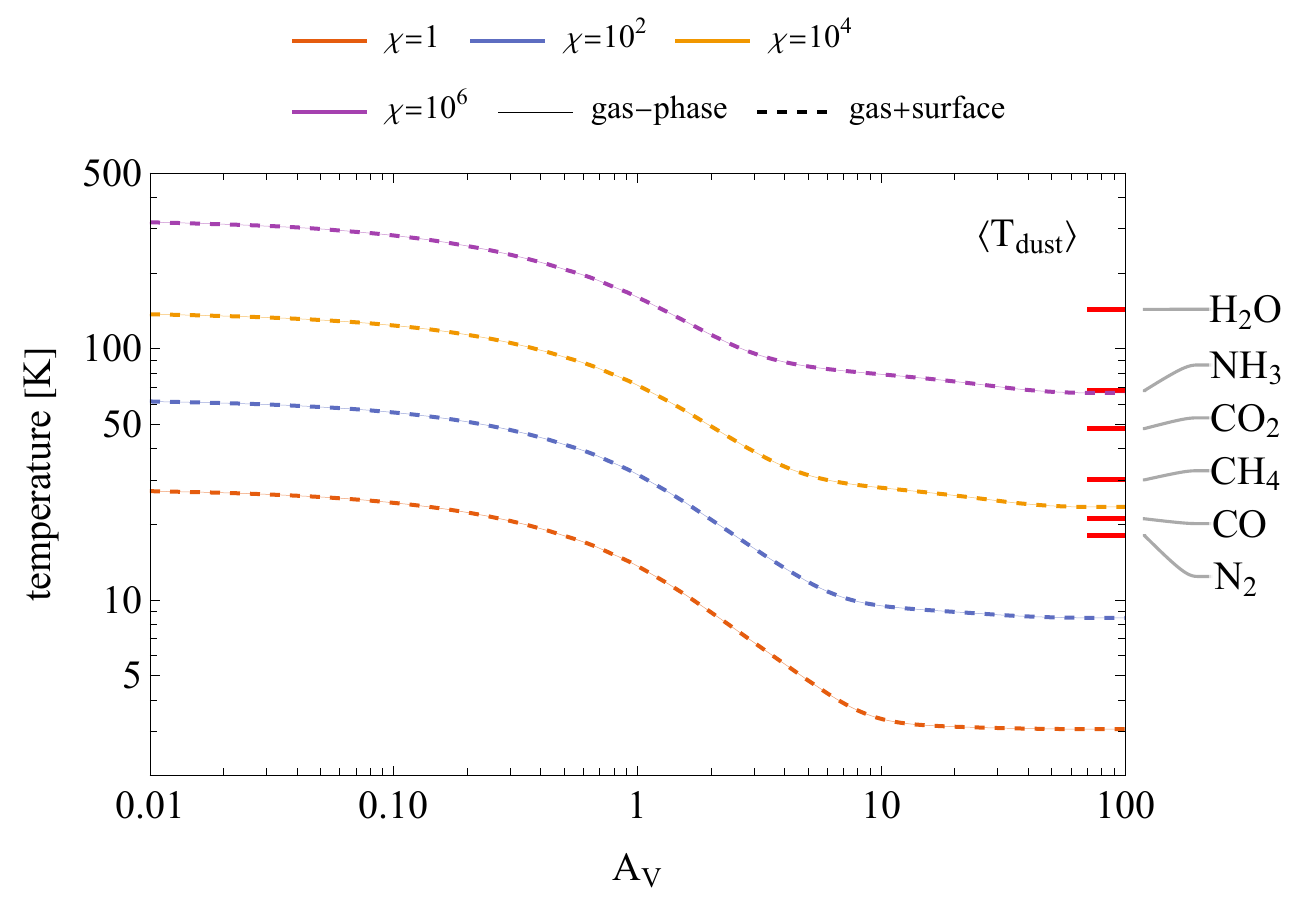}}
	\caption{ Dust temperature $\langle T_\mathrm{dust}\rangle$ profile for $n_0=10^4$~cm$^{-3}$ and different FUV fields. Sublimation temperatures of selected ice species are indicated by red marks on the right.}
	\label{fig:Tdust-vs-FUV}
\end{figure}

\begin{figure}[hbt]
	\resizebox{\hsize}{!}{\includegraphics{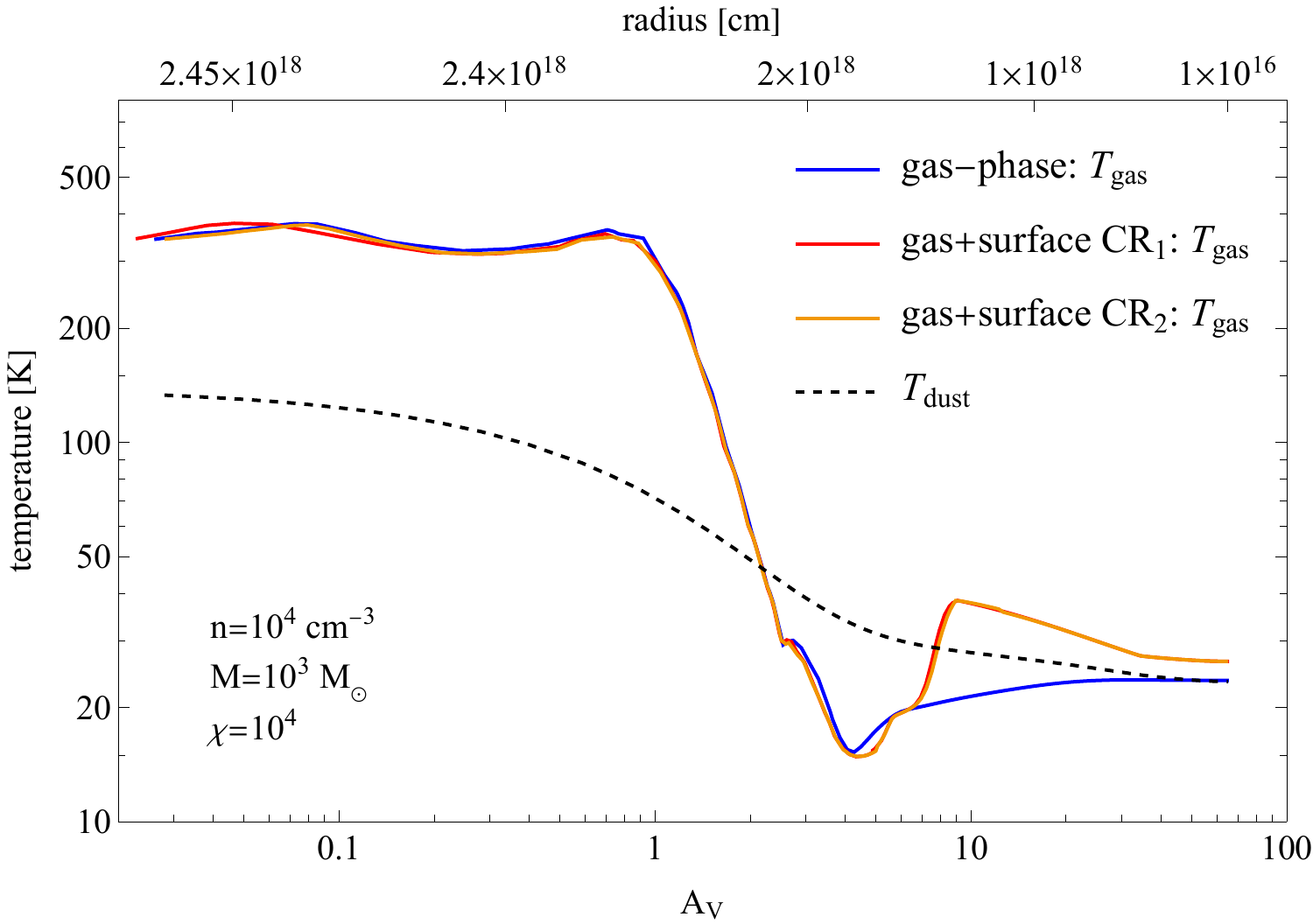}}
	\caption{Influence of the surface chemistry on the thermal structure of a PDR. The model parameters are: $n_0=10^4~\mathrm{cm}^{-3} (\alpha=1.5), M=10^3~\mathrm{M}_\odot, \chi=10^6$. CR$_1$ and CR$_2$ indicate the different models for the CR induced desorption.  }
	\label{fig:surface_Tgas}
\end{figure}

We compare how the structure of a strongly illuminated PDR ( $n=10^4~\mathrm{cm}^{-3}$, $M=10^3~\mathrm{M}_\odot$, $\chi=10^4$) depends on the surface chemistry. Fig.~\ref{fig:surface_Tgas} shows the gas and dust temperatures of a PDR model cloud as function of $A_V$. The dust temperature does not alter with adding surface chemistry. The gas temperature reaches a minimum at $A_V\approx 5$. At this depth, the FUV radiation is sufficiently attenuated and does not dominate the gas heating any more. Other processes, such as heating by gas-grain collisions start to dominate. Deeper in the cloud it becomes increasingly more difficult for cooling line photons to escape the model cloud due to optical thickness. This leads to an increase in gas temperature until it approaches $T_\mathrm{dust}$. This effect is enhanced by any process that further limits the cooling capacity of the gas, for instance freeze-out of relevant line cooling species. Figure~\ref{fig:surface_Tgas} shows how the gas temperature increase is stronger if surface chemistry is included and reaches $T_\mathrm{gas}\approx 40$~K, almost 20~K warmer compared to the model without freeze-out.  This effect also occurs in case of the more efficient CR induced desorption model CR$_2$.

\begin{figure*}[hbt]
	\resizebox{\hsize}{!}{\includegraphics{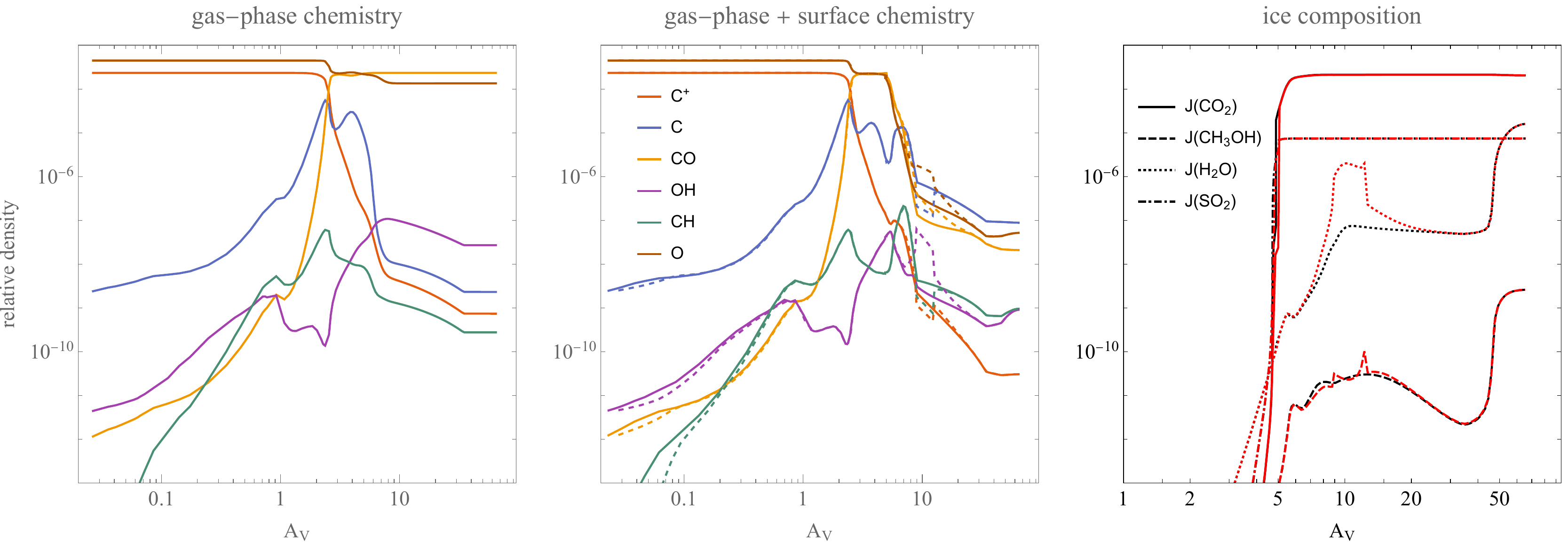}}
	\caption{Influence of the surface chemistry on the chemical structure of a PDR. The results are for the same model as in Fig.~\ref{fig:surface_Tgas}. Solid (black) lines in the central(right) panel correspond to CR$_1$ while dashed(red) lines are for CR$_2$.   }
	\label{fig:surface_chem1}
\end{figure*}

Fig.~\ref{fig:surface_chem1} shows the corresponding chemical structure  excluding (left panel) and including (two right panels) surface chemistry. The incident FUV field in this particular case is strong leading to the typical transition from \ce{C+ -> C -> CO} at $A_V\approx 3$. In the center of the  model cloud without surface chemistry the majority of the carbon is bound in \ce{CO} with a temperature slightly above 20~K. The atomic carbon density peaks at $A_V=2-5$. In the models with surface chemistry we find the same behavior up to an $A_V\approx 7$. 
At higher visual extinction, the \ce{CO} abundance  decreases while other gas phase species have higher abundances, e.g. \ce{C} and \ce{CH}. 


Note, that the dust is too warm to host a significant \ce{J(CO)} population. Under these conditions the ice consists mainly of \ce{J(CO2)} which locks up most of the available \ce{C} and \ce{O} atoms in the ice mantle. This renders the main destruction reactions of atomic carbon \ce{C}, e.g. by collisions with \ce{O2} and \ce{SO2}, inefficient. These reactions would otherwise replenish the \ce{CO} population after destruction by \ce{H3+} and \ce{He+}. Instead, the carbon remains locked in atomic form and the oxygen is converted to ice species. That affects related species like \ce{HCO+} and \ce{OH}. The higher \ce{C} abundance leads to an enhanced abundance of light carbon hydrides \ce{CH_n}.  As a consequence \ce{CO} is not available as coolant any more and the gas temperature rises by 15-20~K compared to the pure gas-phase model as shown in Fig.~\ref{fig:surface_Tgas}. This effect is more pronounced in models with lower FUV illumination where the  additional warm \ce{C} core is visible through $50-100\%$ stronger {\ci} emission at 610 and 370 {\textmu}m.  

For $\chi=10^4$ the relatively high $T_\mathrm{dust}$ prevents most ices to remain on the surface. 
 Comparing the frost temperatures from Table~\ref{tab:snow_temps} with $T_\mathrm{dust}$ from Fig.~\ref{fig:Tdust-vs-FUV} shows that \ce{J(CO2)} is the only viable carbon reservoir in the ice mantle that could survive temperatures above 40~K. \ce{CH3OH} condensation temperatures would be sufficiently high, but the formation in the ice is prevented because the relevant precursor species do not survive long enough in the solid phase. There are no efficient chemical formation routes for \ce{CO2 -> H2O} available, which prevents the formation of significant amounts of \ce{H2O} ice.

 The influence of surface chemistry on the \ce{CO} abundance has already been described by \citet{hollenbach2009}. They modeled a significantly simpler chemistry and are mainly focused on the predictions for the \ce{H2O} and  \ce{O2} gas-phase abundance but we find a similar formation and destruction behavior for \ce{CO} when we compare with our $\chi=10^3$ results. For the higher FUV field shown in Fig.\ref{fig:surface_chem1} we find that for $ A_V\gtrsim 5$ \ce{CO} is mostly destroyed in gas-phase models by \ce{H3+ + CO  ->  HCO+ + H2} and \ce{He+ + CO  ->  O + C+ + He}. Adding surface chemistry does not introduce fundamental different channels with the exception of freeze-out.   

The increasing \ce{C} density at high visual extinction can also be understood by looking at the respective formation/destruction rates. Surface chemistry suppresses the dominant destruction channel of \ce{C} at high $A_V$: \ce{C + O2 -> CO + O} because most of the oxygen is locked up in \ce{J(CO2)} and \ce{J(SO2)} ice. This reaction can no longer replenish the \ce{CO} population leading to a strong increase in \ce{C} abundance and a decrease in \ce{CO}  in the dark cloud portion of the model cloud.

\subsection{Illumination effects on the surface chemistry}

\begin{figure*}[t]
	\centering
	\begin{subfigure}[t]{.48\textwidth}
		\centering
		\includegraphics[width=\linewidth]{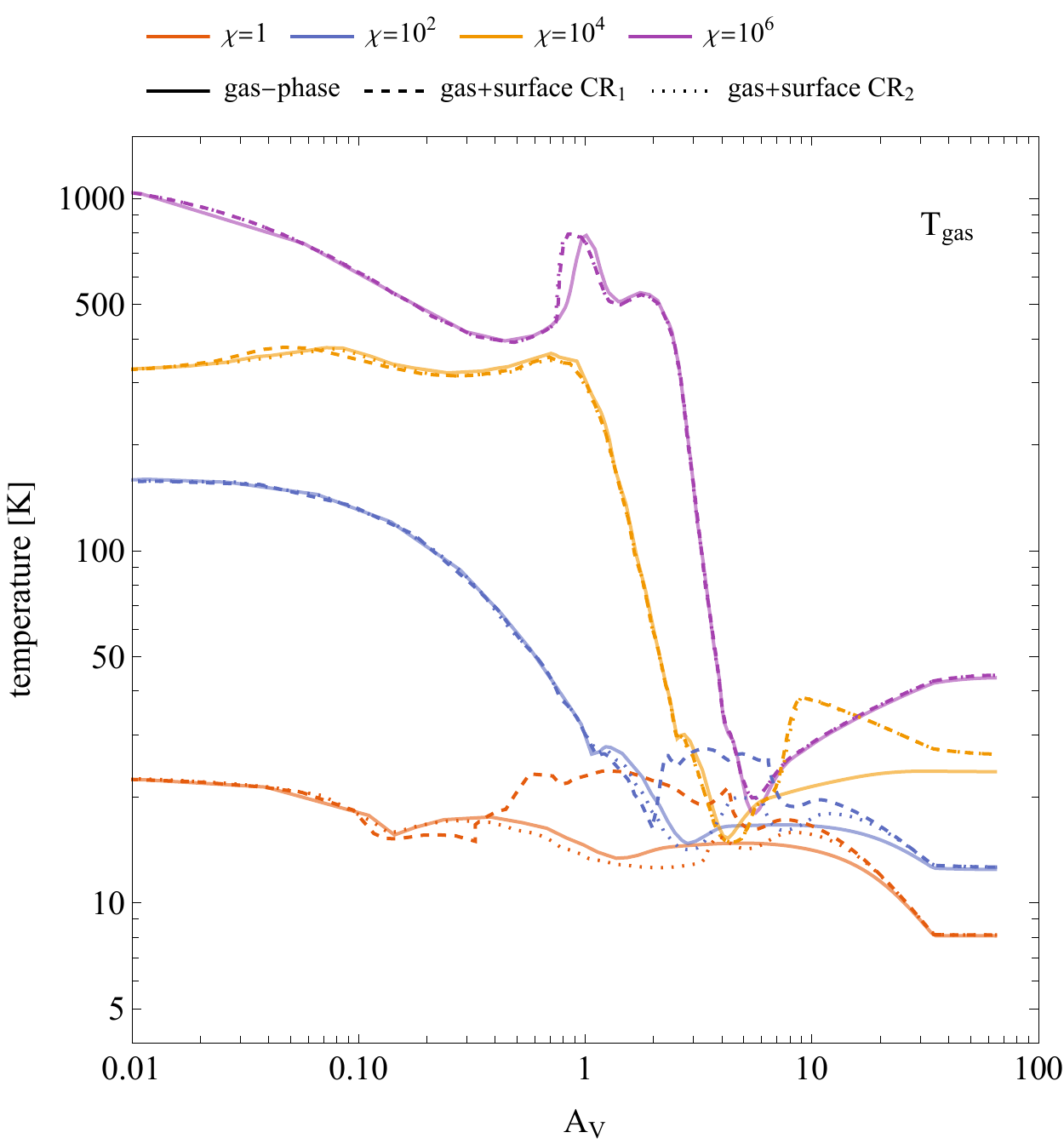}
		\caption{$T_\mathrm{gas}$ profile.}
		\label{fig:Tgas-vs-FUV}
	\end{subfigure}
	\begin{subfigure}[t]{.48\textwidth}
		\centering
		\includegraphics[width=\linewidth]{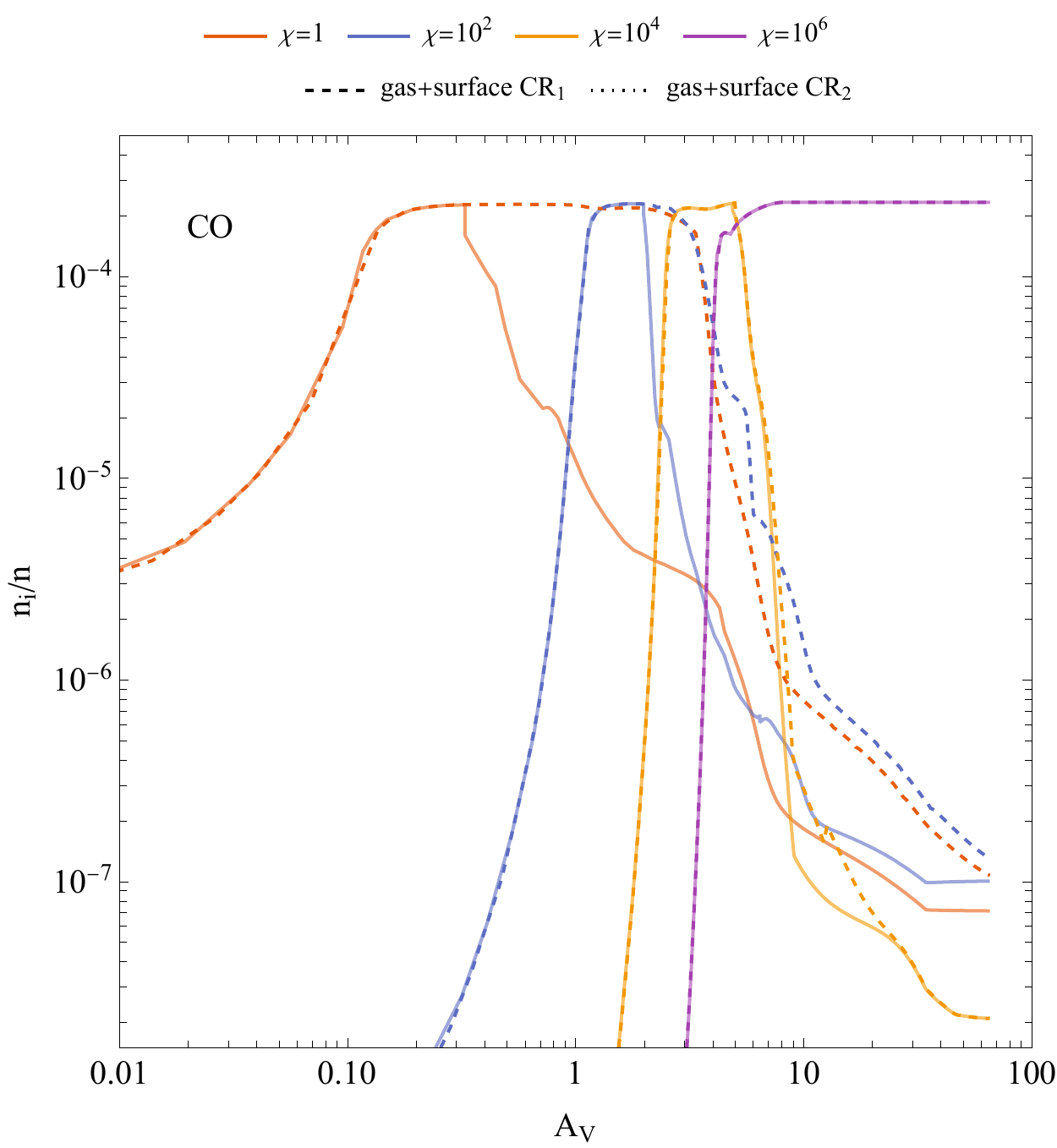}
		\caption{\ce{CO} gas-phase abundance.}
		\label{fig:CO-CR-effect}
	\end{subfigure}
	\caption{\textbf{Left:}Temperature profile changes with FUV strength ( $n=10^4~\mathrm{cm}^{-3}$). Solid lines show pure gas-phase results, dashed lines correspond to  gas+surface chemistry. 
	\textbf{Right:} Influence of the surface chemistry on the CO gas-phase abundance for the same parameters as in Fig.~\ref{fig:Temperature-vs-FUV}. CR$_1$ and CR$_2$ indicate the different models for the CR induced desorption.}
	\label{fig:Temperature-vs-FUV}
\end{figure*}

For a more systematic analysis, we investigate how the additional surface chemistry affects individual species depending on the FUV field strength $\chi$ and whether the chemical desorption reactions discussed in Sect.~\ref{sect:chem_desorption} changes the overall chemistry.

\begin{figure*}
	\begin{subfigure}[t]{.31\textwidth}
		\centering
		\includegraphics[width=\linewidth]{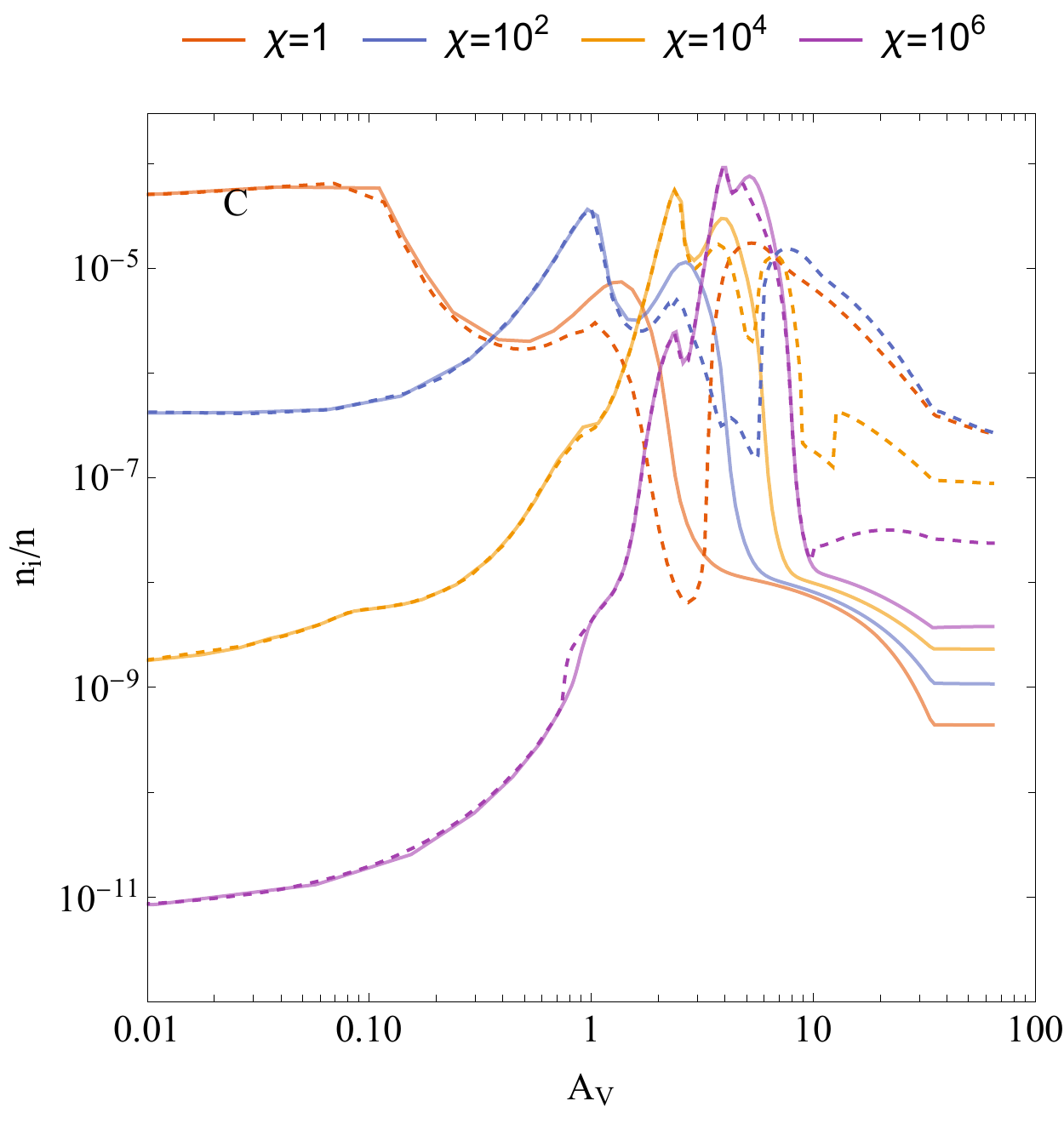}
		\caption{\ce{C} density profile.}
		\label{fig:structure_fuv-C}
	\end{subfigure}
		\hfill
	\begin{subfigure}[t]{.31\textwidth}
		\centering
		\includegraphics[width=\linewidth]{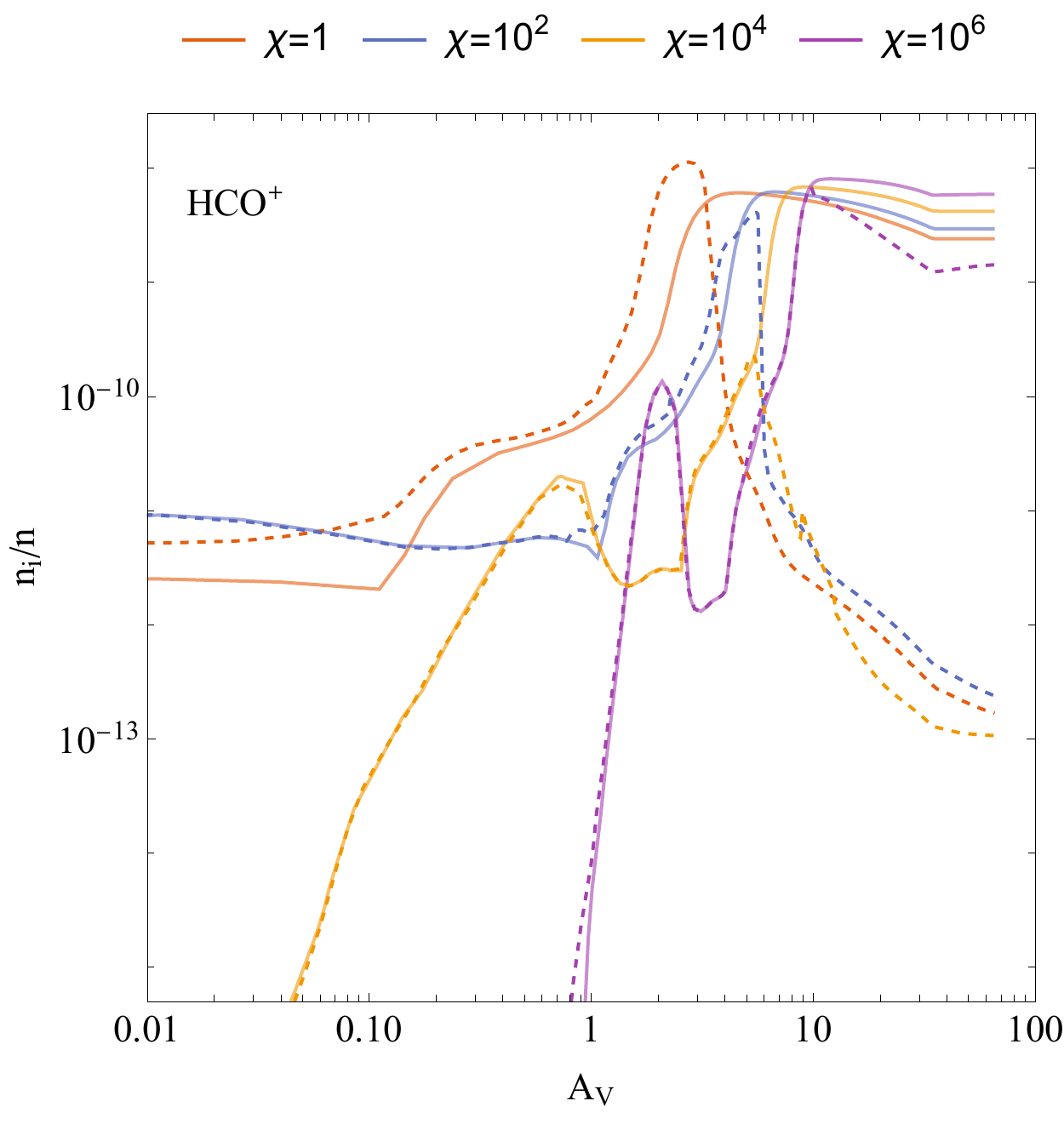}
		\caption{\ce{HCO+} density profile.}
	\end{subfigure}
	\hfill
	\begin{subfigure}[t]{.31\textwidth}
		\centering
		\includegraphics[width=\linewidth]{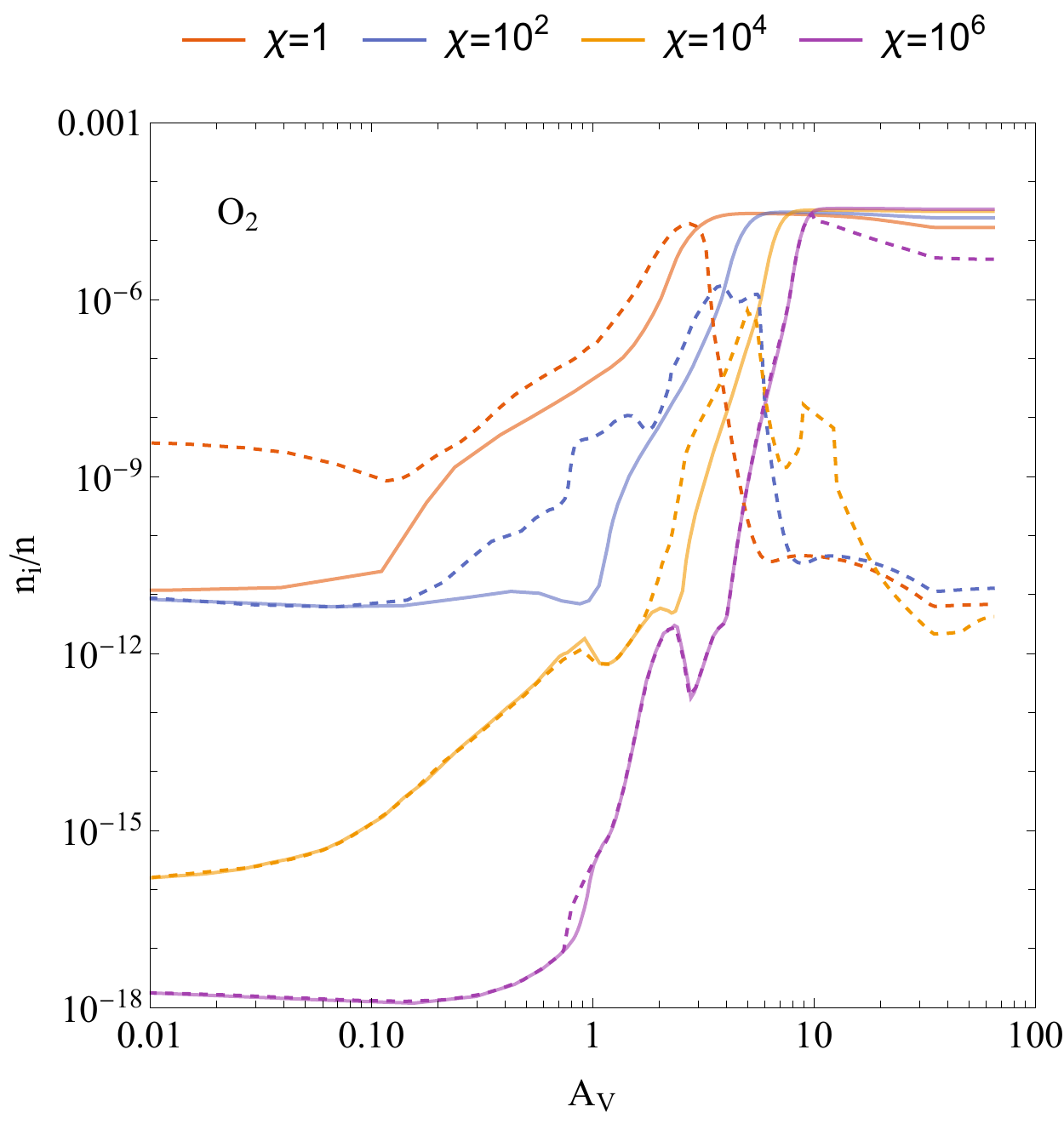}
		\caption{\ce{O2} density profile.}
		\label{fig:structure_fuv-O2}
	\end{subfigure}
	\caption{Chemical structure changes with FUV strength.  $n=10^4~\mathrm{cm}^{-3}$. Solid lines indicate pure gas-phase chemistry, dashed lines correspond to the gas+surface chemistry using the CR$_2$ model.}\label{fig:structure_fuv} 
\end{figure*}

The gas temperature remains mostly unaffected for $\chi \gg 10^2$ as shown in Fig.~\ref{fig:Tgas-vs-FUV}. In Fig.~\ref{fig:structure_fuv} we show how different values of $\chi=1, 10^2, 10^4, 10^6$ affect the chemistry for \ce{C}, \ce{HCO+}, and \ce{O2}. 
The high FUV case $\chi=10^6$ shows no significant effect of the surface chemistry. All relevant coolants remain abundant in the gas-phase due to the warm grain surfaces. For lower FUV field strengths we find that the gas temperature increases above the pure gas-phase case once the dominant coolant \ce{CO} starts to deplete from gas-phase (see Fig.~\ref{fig:CO-CR-effect}). A dominant \ce{CO} gas-phase population survives until the dust temperature falls below $\langle T_\mathrm{dust}\rangle \lesssim 20$~K (see Fig.~\ref{fig:Tdust-vs-FUV}) and desorption becomes too weak to replenish gas-phase \ce{CO}. At lower $\chi=1, 10^2$ this occurs at: $A_V\approx0.2, 1$, respectively. At higher FUV this requires significantly higher values of $A_V$. Any effect that desorbs \ce{J(CO)} more efficiently will thus result in a lower gas temperature. This can be seen in Fig.~\ref{fig:CO-CR-effect} which compares the  \ce{CO} gas-phase abundance between the CR$_1$ and CR$_2$ models. In the CR$_2$ model, the energy injected per grain as well as the grain cooling time are enhanced which effectively corresponds to a higher $\langle T_\mathrm{dust}\rangle$. As a result we find that \ce{CO} freeze-out sets deeper in the cloud if $\chi\le 10^2$ and consequently we see a somewhat lower $T_\mathrm{gas}$.

A general trend is that most carbon bearing species show enhanced abundances at high $A_V$ while most oxygen-bearing species have lower densities compared to the pure gas-phase chemistry. This is because ice composition is dominated by oxygen-bearing species, which locks-up a significant fraction of the available oxygen in the ice mantles.  For models with significant \ce{CO} freeze-out we find \ce{C}  to be the main carbon reservoir in the gas-phase. As a consequence, the CR induced ionization of atomic carbon can significantly contribute to the electron formation. Thus, we observe the same enhancement for the electron density. This strongly affects all species that form via dissociative recombination, such as for instance \ce{CH}. We note similar inherited effects in Fig.~\ref{fig:structure_fuv}, where \ce{HCO+} follows the abundance change of \ce{CO}. The selective freeze-out of oxygen-bearing ice species also leads to a significantly decrease gas-phase abundance of \ce{O2}. This is in agreement with observations having difficulties to confirm high molecular oxygen abundance predictions from pure gas-phase models.

Enhanced cosmic ray induced desorption in the CR$_2$ models allows for a higher gas-phase abundance of certain species with lower binding energies, such as \ce{CO} and \ce{CH4}. The corresponding ice mantle composition is shifted towards tighter bound ice species, e.g. \ce{J(CO2)} and \ce{J(H2O)}. A weaker dominance of overall depletion over gas-phase abundances seems to be in agreement with observations of e.g. S bearing species in PDRs that do not show signs of significant freeze out \citep{riviere2019}.   However, our current implementation of CR induced desorption (CR$_1$ vs. CR$_2$) remains a crude approximation with large uncertainties. Nevertheless, it might be feasible to use observations of low FUV source as calibrators for more detailed models of CR induced desorption.

Chemical desorption does affect the formation and destruction of some species at various cloud depths. For instance, \ce{J(O) + J(O) -> O2} takes over as the dominant \ce{O2} formation channel for $2\lesssim A_V \lesssim 5$ so that we observe an enhanced \ce{O2} density before oxygen freeze-out starts to become important (Fig.~\ref{fig:structure_fuv-O2}). Another example is the reaction \ce{J(O) + J(H) -> OH} which contributes approximately 10\% to the total \ce{OH} formation rate for $A_V>30$.

\begin{figure*}
	
	\begin{subfigure}[t]{.49\textwidth}
		\centering
		\includegraphics[width=\linewidth]{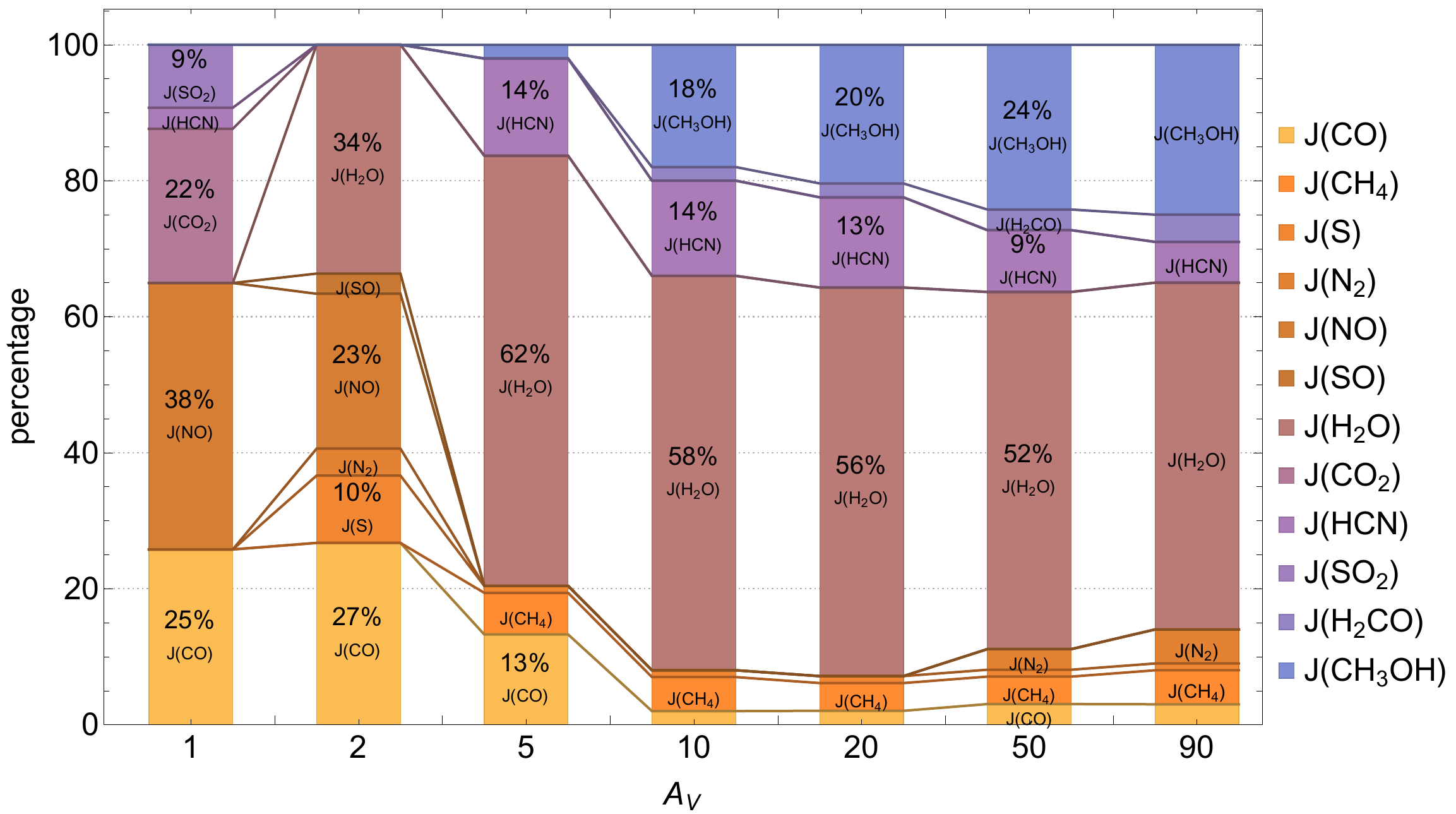}
		\caption{$\chi=1$}
		\label{fig:ice_structure_0}
	\end{subfigure}
	\hfill
	\begin{subfigure}[t]{.49\textwidth}
		\centering
		\includegraphics[width=\linewidth]{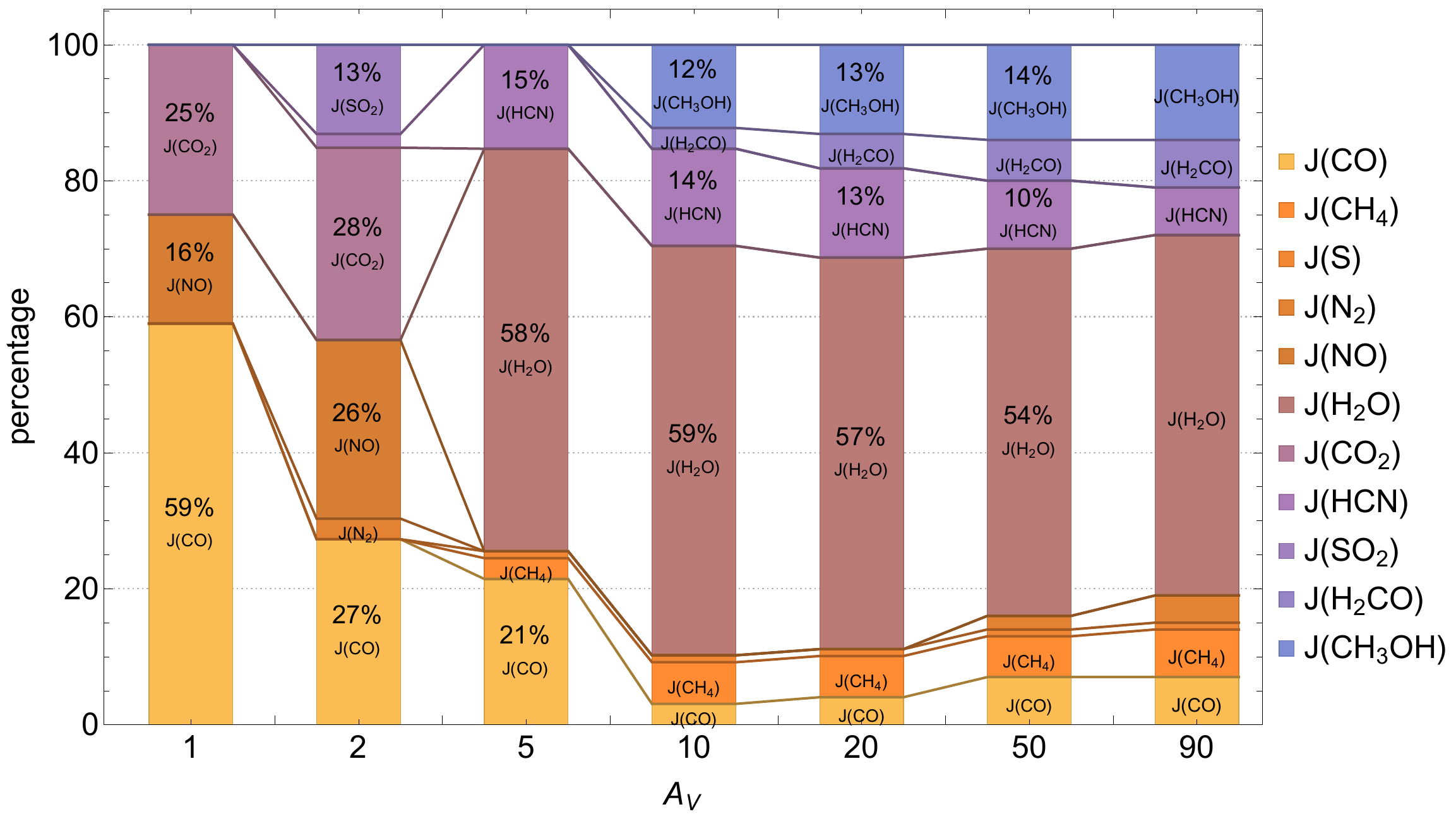}
		\caption{$\chi=10$}
		\label{fig:ice_structure_1}
	\end{subfigure}
	\medskip
	\begin{subfigure}[t]{.49\textwidth}
		\centering
		\includegraphics[width=\linewidth]{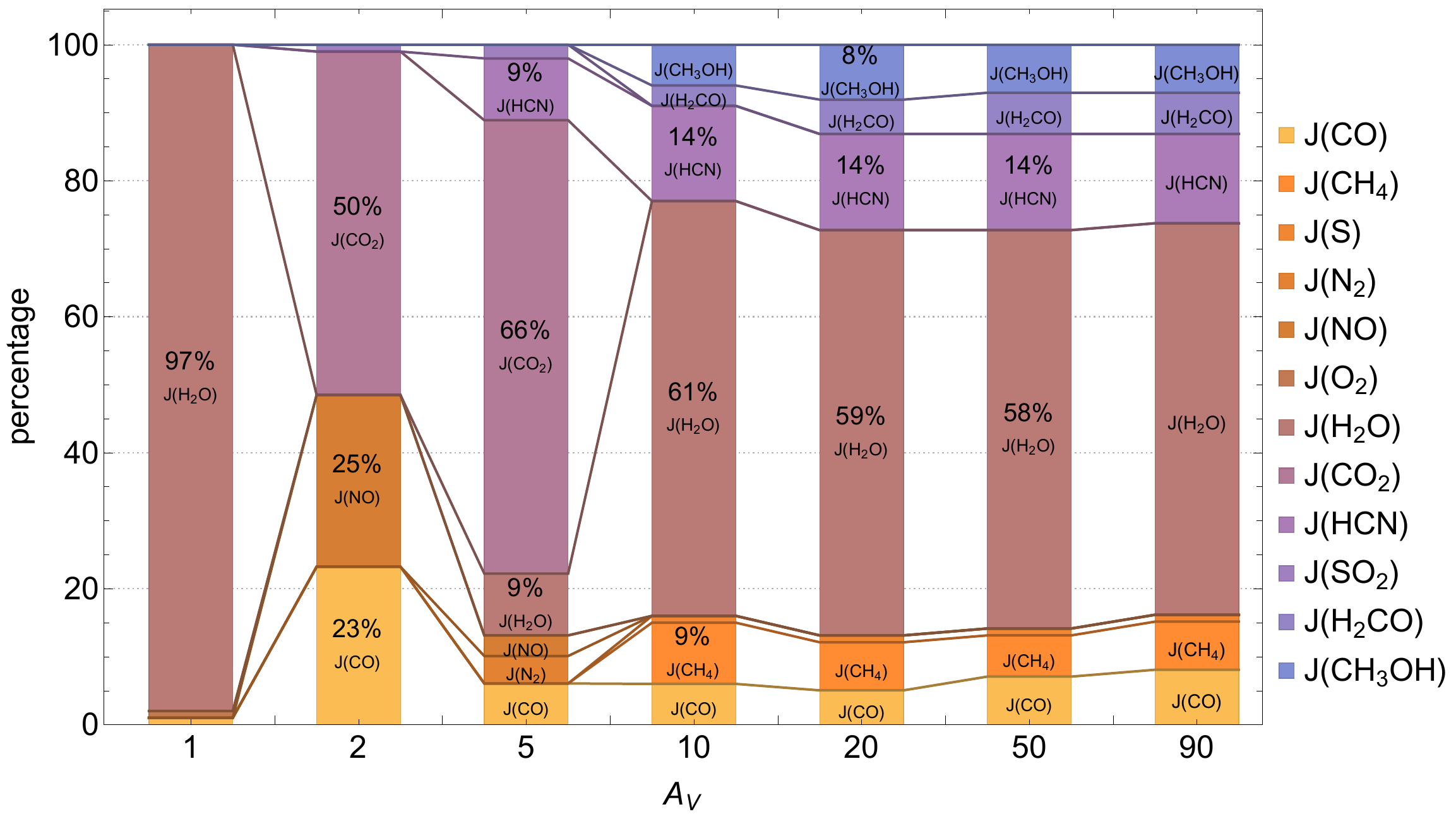}
		\caption{$\chi=10^2$}
		\label{fig:ice_structure_2}
	\end{subfigure}
	\hfill
	\begin{subfigure}[t]{.49\textwidth}
		\centering
		\includegraphics[width=\linewidth]{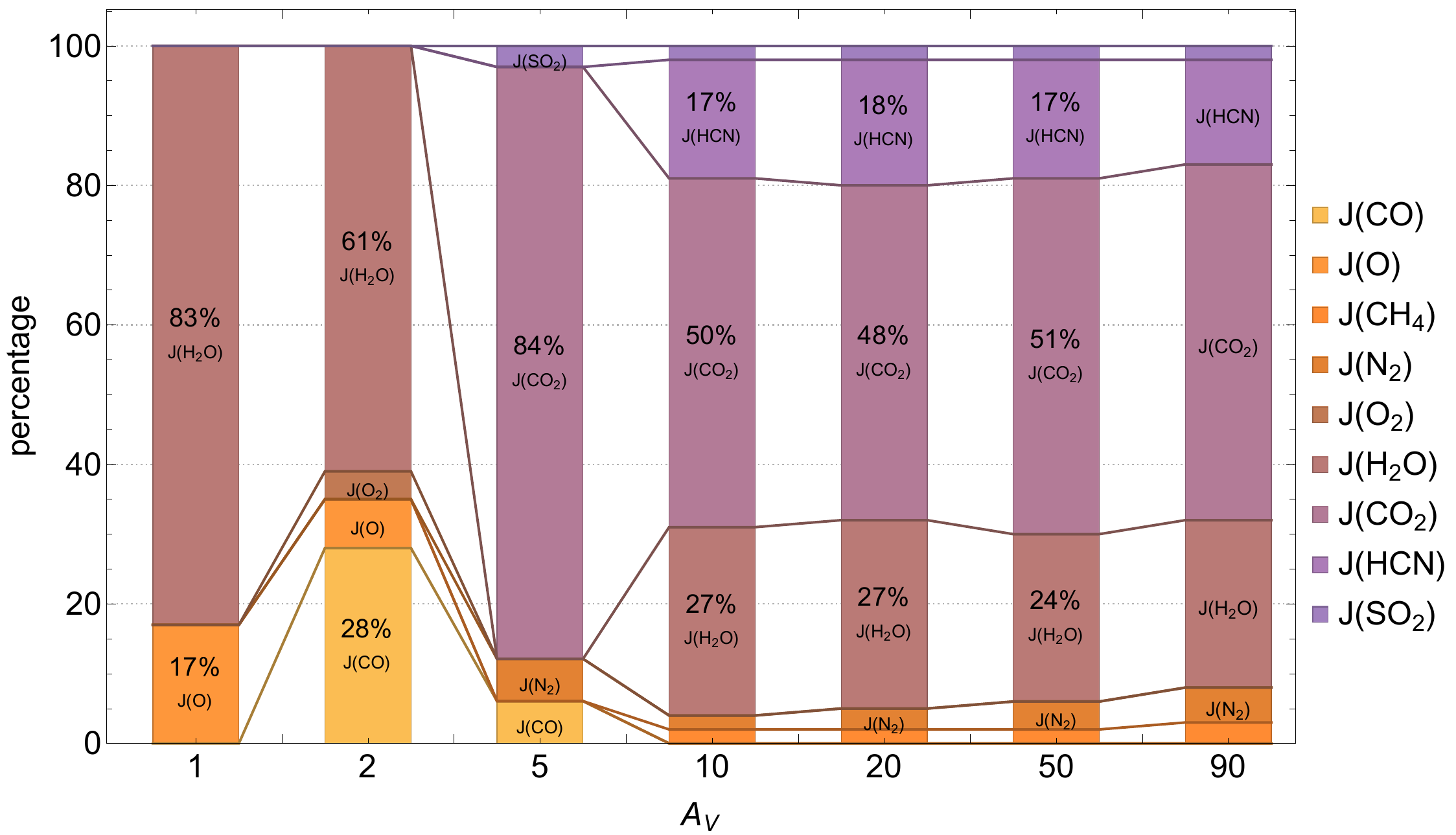}
		\caption{$\chi=10^3$}
		\label{fig:ice_structure_3}
	\end{subfigure}
	\medskip
	\begin{subfigure}[t]{.49\textwidth}
		\centering
		\includegraphics[width=\linewidth]{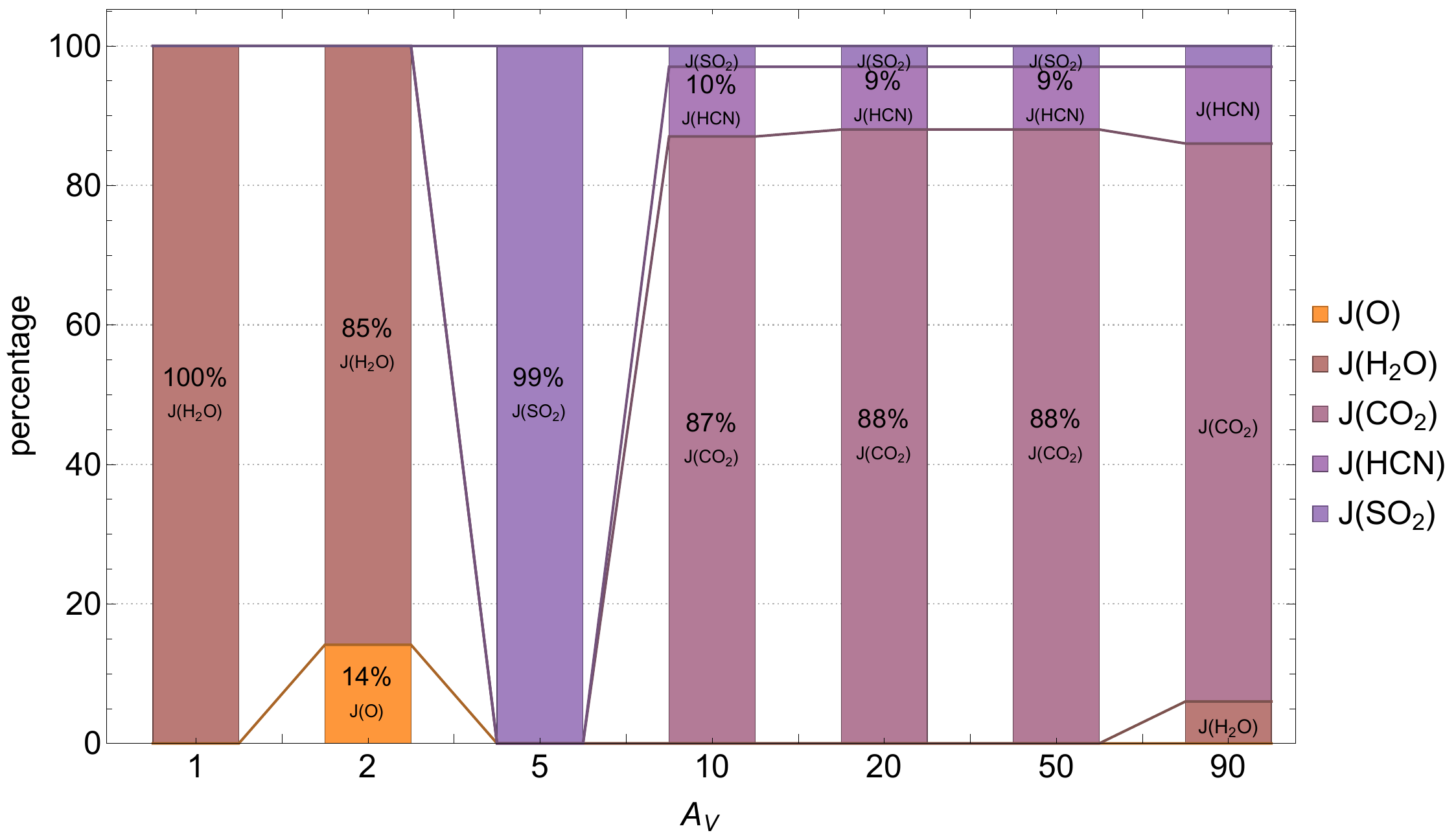}
		\caption{$\chi=10^4$}
		\label{fig:ice_structure_4}
	\end{subfigure}
	\hfill
	\begin{subfigure}[t]{.49\textwidth}
		\centering
		\includegraphics[width=\linewidth]{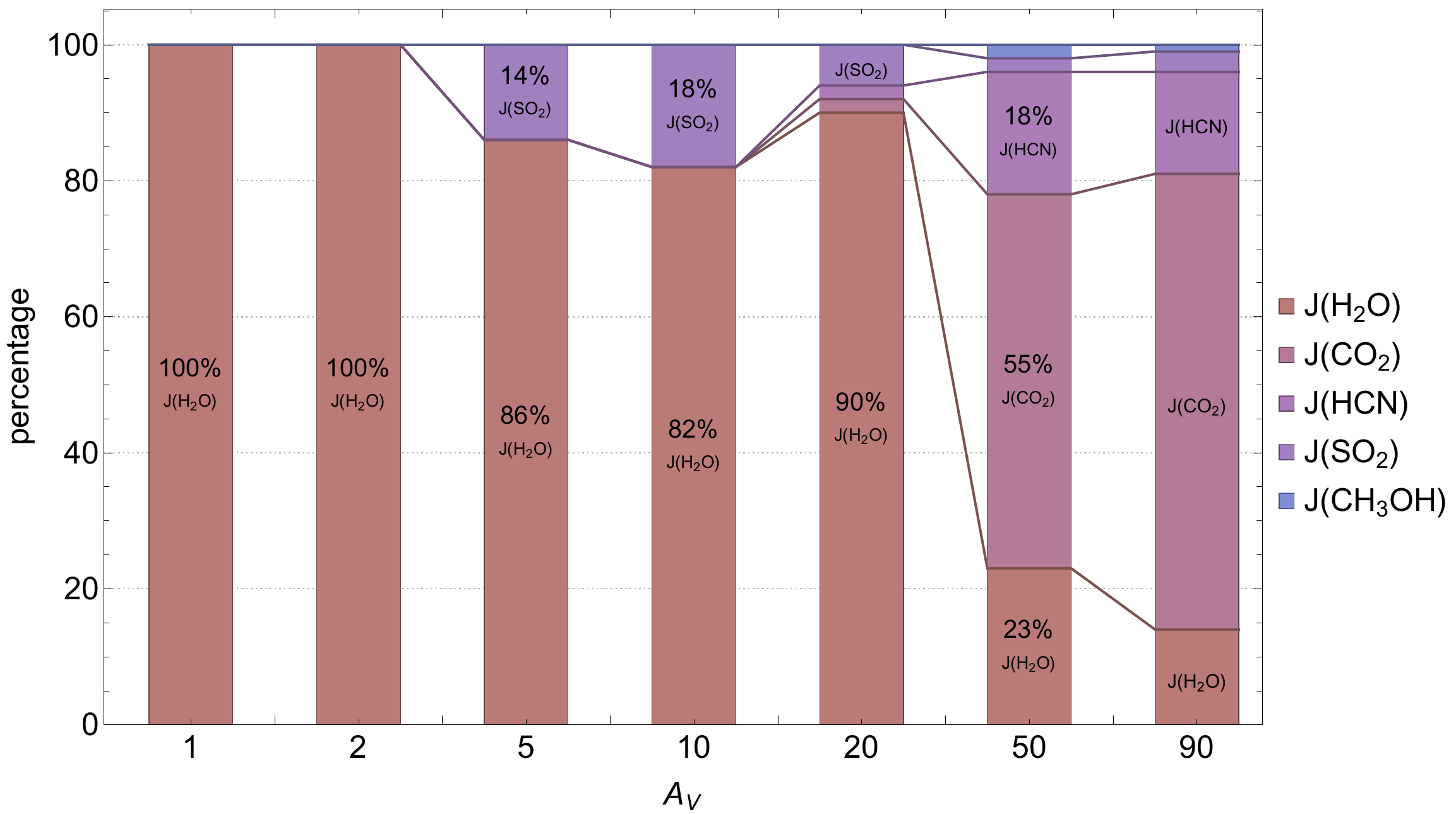}
		\caption{$\chi=10^5$}
		\label{fig:ice_structure_5}
	\end{subfigure}
	\medskip
	\begin{subfigure}[t]{.49\textwidth}
		\centering
		\includegraphics[width=\linewidth]{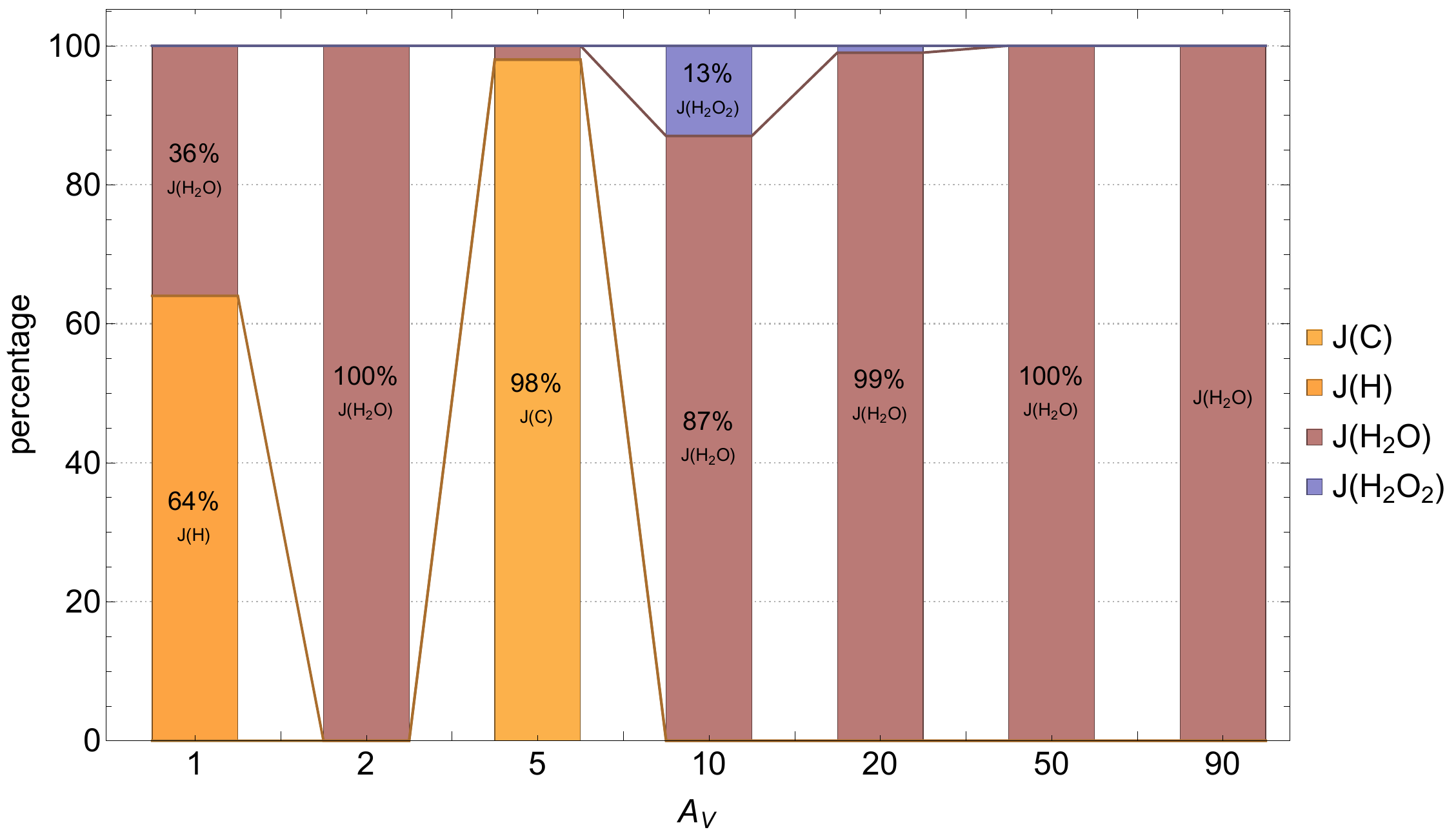}
		\caption{$\chi=10^6$}
		\label{fig:ice_structure_6}
	\end{subfigure}
	\hfill
	\begin{subfigure}[t]{.49\textwidth}
		\centering
	\end{subfigure}
	\caption{Percentile ice composition as function of $A_V$ for $n=10^4$~cm$^{-3}$, $M=10^3$~M$_\odot$ assuming CR$_2$.}
	\label{fig:ice_structure}
\end{figure*}

Figure~\ref{fig:ice_structure} shows how the ice composition changes with cloud depth.
Each panel in the Fig. correspond to a different FUV illumination (assuming CR$_2$). In the low FUV cases $\chi\le10$ the ice forming closest to the cloud surface consist mainly of \ce{J(CO)} (60-80\%) up to a few $A_V$. Deeper into the cloud the  ice is converted to water ice ($\sim50\%$) as well as methanol ice \ce{J(CH3OH)} ($\sim20-25\%$). Under increasing FUV conditions the ice composition shows some significant changes. For $\chi=10^2,10^3$ we find \ce{J(H2O)} ice closer to the cloud surface. \ce{J(CO)} becomes less abundant (10-25\%) and forms only deeper in the clump ($A_V>2$).
For  FUV strengths  $\chi\ge10^3$  the dust temperatures prohibit large amounts of \ce{J(CH4)} and \ce{CO} ice and the carbon ice consists to 50-90\% of \ce{J(CO2)}, \ce{J(SO2)}  and water ice. For $\chi=10^5$ the ice is dominated by \ce{H2O} ice and \ce{J(CO2)} at $A_V>50$. The deeply embedded \ce{J(CO2)} population vanishes at $\chi=10^6$ and water ice remains the dominant ice component.   

\begin{figure}[hbt]
	\resizebox{\hsize}{!}{\includegraphics{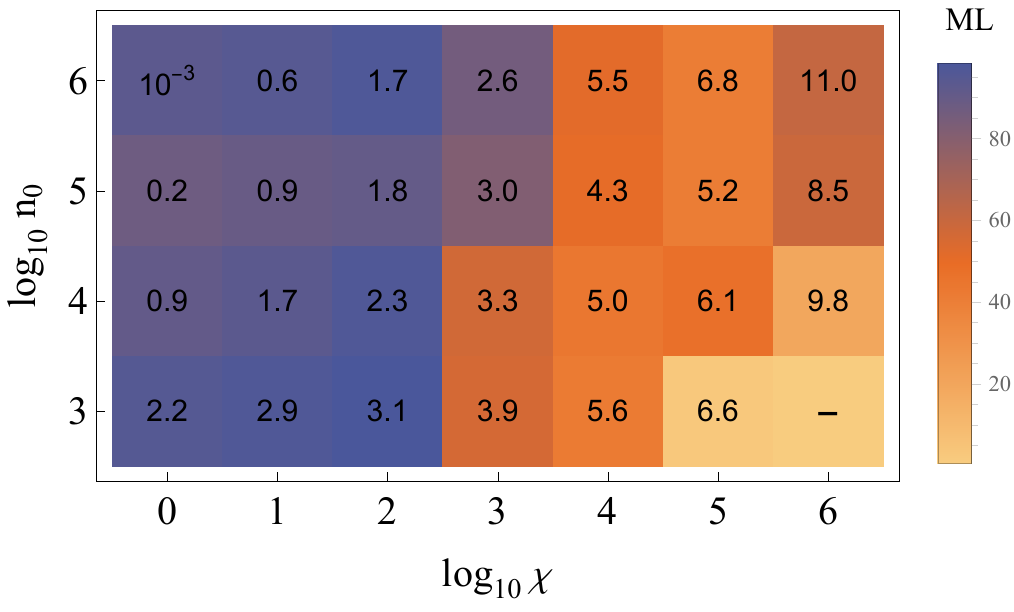}}
	\caption{ Ice thickness in ML for the model clump center as a function of total gas density $n_0$ and FUV strength $\chi$, using CR$_2$. The numbers give the threshold for the formation of one monolayer of ice in $A_V$.}
	\label{fig:monolayers}
\end{figure}

The most obvious conclusion is that adding or removing some surface species from the chemical network may result in significantly different ice structures. The same is of course true for different sets of binding energies in use. Even so, another main effect of (any) existing surface chemistry is to open up new formation \& destruction channels that become active below certain grain temperatures and will therefore significantly alter the gas-phase abundances. The freeze-out of \ce{CO} is a good example for this effect. In terms of typical PDR tracers it is important that they are removed from the gas-phase and not so much whether they end up forming \ce{J(CO)} or \ce{J(H2O)} ice.

The effective dust temperature is the most important factor in determining the ice structure. Fig.~\ref{fig:Tdust-vs-FUV} shows that assuming very low dust temperature for deeply embedded parts of the model PDR is not always justified. High FUV models may show enhanced dust temperatures and selectively prevent certain ice species to form. This modifies the overall ice composition on the grain surfaces also affecting the gas-phase abundances. As an example we showed how gas-phase \ce{CO} is diminished in the dark cloud even though no explicit \ce{J(CO)} ice is formed.

\subsection{Coupling between chemistry and line excitation}

\begin{table*}[htb]
	\caption{Comparison of model fine-structure emission (in units of \Kkms).}
	\label{tab:emission_comparison}
	\resizebox{\textwidth}{!}{%
		\begin{tabular}{lccccccc}
			\hline\hline
			\vrule width 0pt height 2.2ex
			transition	& $\chi=1$ & $\chi=10$&$\chi=10^2$&$\chi=10^3$&$\chi=10^4$&$\chi=10^5$&$\chi=10^6$\\
			\hline
			\vrule width 0pt height 2.2ex

                \ciitrans &0.559/0.562(0.622)& 9.79/10.2(10.5)& 61.3/62.6(63.5)&               139/140 (142)& 226/227 (229)& 321/323 (316)& 441/438 (418)\\
                \cilo &8.55/11.1 (6.07)& 10.7/13 (8.06)& 11.4/15.7 (9.82)& 15./14.1 (11.5)& 16.5/16.7 (15.7)& 18.6/18.4 (20)& 24.1/28.5 (24)\\
                \ciup & 2.16/2.78 (1.42)& 3.9/4.46 (2.99)& 5.39/7.03 (4.65)& 7.94/7.29 (5.97)& 9.32/9.52 (8.66)& 11.5/11.4 (12.1)& 16.8/20.3 (15.8)\\
                \oilo & 0.018/0.005 (0.006)& 1.42/1.45 (1.46)& 78./80.4 (81.2)&   283/280 (279)& 500/495 (500)& 692/702 (699)& 920/904 (861)\\
                \oiup & 0/0 (0)& 0.128/0.134 (0.135)& 32.7/34 (35)& 160/158 (158)& 317/316 (317)& 485/503 (491)& 758/756 (697)\\ \hline
	\end{tabular}}
	\tablefoot{CR$_1$ and CR$_2$ model results are shown as CR$_1$/CR$_2$; models without surface chemistry are given in parenthesis. The columns give results for models with different FUV strength $\chi$ assuming  $n=10^4$~cm$^{-3}$, $M=10^3$~M$_\odot$.}
\end{table*}

\begin{figure*}[hbt]
	\resizebox{\hsize}{!}{\includegraphics{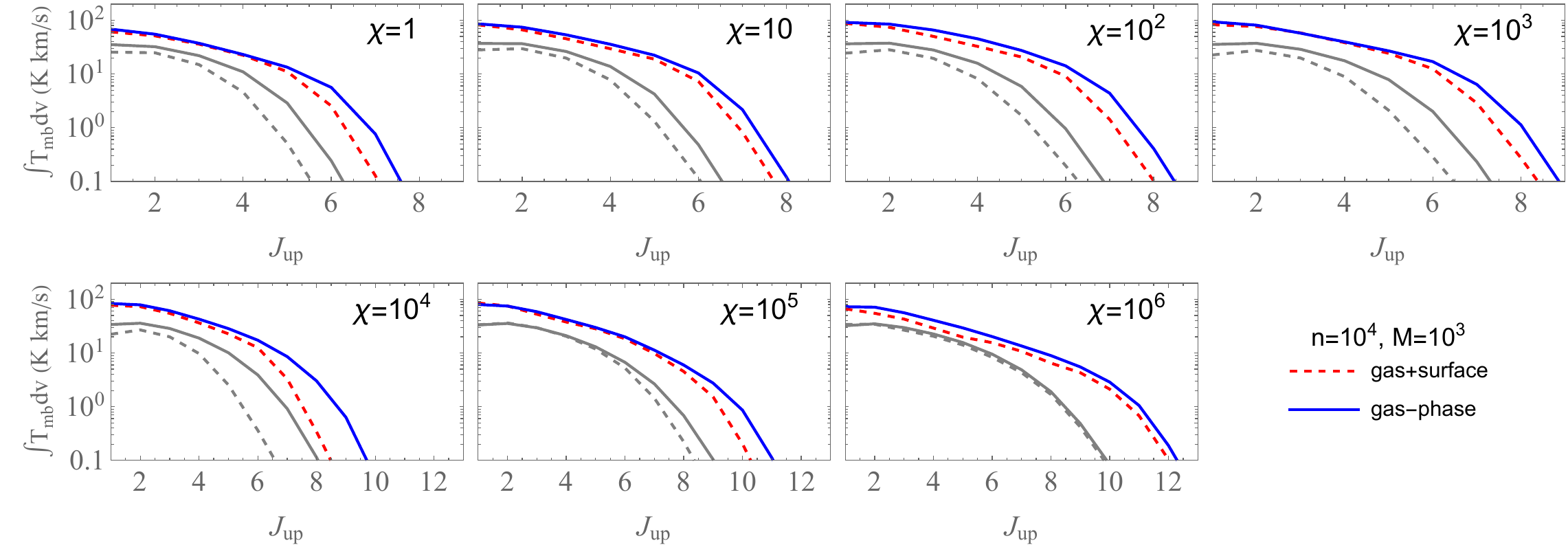}}
	\caption{Effect of surface chemistry on the CO spectral line emission distribution (in \Kkms{}) as function of $J_\mathrm{up}$ . Each panel corresponds to a different FUV field strength $\chi$. \ce{^{12}CO} lines are in color, while \ce{^{13}CO} lines are plotted in gray-level.  }
	\label{fig:CO-SLED-FUV}
\end{figure*}

The previous section showed that the removal of \ce{CO} from the gas-phase results in a reduced cooling capacity of the gas and an increase in gas temperature. In terms of observable line intensities the higher gas temperatures may partially compensate the reduced abundance of \ce{CO}. Nevertheless, the general behavior is that {\twco} and {\thco} line intensities are lower in models with surface chemistry due to the freeze-out of \ce{CO}.  This is shown in Fig.~\ref{fig:CO-SLED-FUV} where we plot the \ce{CO} line emission of the transition $\mathrm{J_{up}}\rightarrow(\mathrm{J_{up}}-1)$  as function of the angular momentum quantum number $\mathrm{J_{up}}$ for pure gas-phase and surface chemistry models.
We find the same behavior across the full density range.

Table~\ref{tab:emission_comparison} lists the clump averaged emission of the fine-structure lines. The selective freeze-out of \ce{O} bearing species that leads to a dominant gas-phase population of \ce{C} at large $A_V$ results in enhanced fine-structure emission for models with low to intermediate values of $\chi$. This effect is $\sim20\%$ stronger in the CR$_2$ models due to the slightly enhanced gas-phase abundance of atomic carbon. The enhancement of the  \ci\ intensities due to the surface chemistry occurs for the \cilo and the \ciup lines. The same effect does not occur for the \cii\ line because of the higher energy of 92~K required to excite the \ciitrans\ transition. The same holds for the \oi\  lines at 63 and 145{\textmu}m.

Altogether, the surface chemistry in \Kt{} has a small effect on the total \ce{CO} cooling budget, raising the gas temperatures by typically less than a few K (with some exceptions visible in Fig.~\ref{fig:Tgas-vs-FUV}) .  The \cii\ and \oi\ cooling lines are rather insensitive to the modified chemical and temperature profile while the \ci\ fine-structure lines
may be affected depending on the effectiveness of overall desorption, which makes the atomic carbon an interesting tracer of surface chemistry in PDRs. Note, that in diffuse and translucent clouds gas-phase PDR models notoriously over-predict \ci\ abundances (the so-called '\ci\ problem', \citet{gong2017}).  However, column densities are not a direct observable and therefore comparing model column densities to column densities derived from observations introduces additional uncertainties. Here, we are more concerned with denser clouds where PDR model prediction can over- and under-predict  \ci\ intensities depending on the detailed modeling approach. Standard chemical models are not able to reproduce the described \ci\ enhancement because they do not solve the temperature self-consistently with the chemistry and the non-local radiative transfer through the model cloud. 

The line intensities in Table~\ref{tab:emission_comparison} and Fig.~\ref{fig:CO-SLED-FUV} were computed for a model with $n=10^4$~cm$^{-3}$, $M=10^3$~M$_\odot$. Many observations of PDRs show significantly brighter emission lines. Examples are the Orion Bar and NGC~7023 \citep{joblin2018} and the Carina Nebula \citep{Wu2018}. They require significantly higher gas densities ($n=10^5-10^6$~cm$^{-3}$) to explain the observed intensities . At those high densities rotational lines of CO up to $J>20$ can be excited;  this is not possible for $n=10^4$~cm$^{-3}$. For comparison we provide plots of the CO SLED for higher densities in the appendix.

\subsection{Comparison with other models}
Even though inter-model comparison tends to be difficult \citep[compare for e.g.][]{comparison07} we try to compare our results to other computations. 

\citet{hollenbach2009} added surface chemistry to their PDR model \citep{kaufman99,kaufman06} to study \ce{H2O} and \ce{O2} abundances observed in the gas-phase. They added only a small surface network to their chemistry and used significantly different desorption energies. In particular the values for \ce{C} and \ce{OH} are very different significantly affecting the ice composition. Nevertheless we find that \ce{H2O} peaks at $A_V\approx5$ with $n(\ce{H2O})\approx 10^{-7}$~cm$^{-3}$ and drops to $10^{-9}-10^{-10}$~cm$^{-3}$ deeper in the clump, comparable to their result. We also see a corresponding increase of the water ice density at $A_V\approx3-5$ locking-up most of the oxygen atoms in the ice mantle. On the other hand, they find a significantly different carbon ice structure, except for \ce{J(CO)}. For $A_V>6$ they find the carbon atoms locked-up in \ce{J(CH4)} which does not occur in our models. They do not report details of their carbon surface network but the most likely reason for this is the assumed binding energies. Using comparable values we also find a dominant \ce{J(CH4)} population. They discuss the threshold $A_V$ for ice formation (ML=1) and we provide the corresponding numbers in Fig.~\ref{fig:monolayers}. For conditions applicable to the Taurus cloud ($n_0=10^3$~cm$^{-3}$ and $\chi\sim 1$) they find a value of 2 almost identical to our result of 2.1. For $n_0=10^4$~cm$^{-3}$ and $\chi\sim 100$  they find approximately 3 where we have about 2.3 which can be explained by our increasing total gas density profile, which shifts the ML=1 threshold closer to the surface.
 
 \citet{esplugues2016} presented a significantly improved version of the \textit{Meijerink}  PDR code \citep{meijerink2005} including an up-to-date surface chemistry. Unfortunately, they computed their models only up to $A_V=10$. More importantly, they used a simple approximation for their dust temperature that produces much too low $T_\mathrm{dust}$ (compared to detailed computations) and is not valid for $A_V\gtrsim 10$. Therefore, their model is not able to selectively freeze-out particular ices only and thus could not produce the dark cloud \ce{C} population we find.  
Their \ce{C} density profile for $n=10^4\,\mathrm{cm}^{-3},\,G_0=10^4$ at $A_V < 10$ \citep{esplugues2016} was very similar to our results (see Fig.~\ref{fig:structure_fuv-C}). Note, that typical plane-parallel PDR computations often do not cover deeply embedded regions with $A_V\gg 10$ and therefore miss out the strong peaks in, e.g. \ce{C} and \ce{CH}. The common argument that beyond a certain visual extinction no significant chemical variation takes place is not necessarily valid any more once we include surface reactions to the chemical network and account for higher dust temperatures due to diminishing cooling efficiencies.
Looking at their Model~1 results ($n=10^4$~cm$^{-3}$, $G_0=10^4$) \citep{esplugues2016} we note that their ice composition is dominated by \ce{J(CO2)} for $A_V>3$. Our models show  \ce{J(CO2)} as major ice component for slightly larger values of $A_V$. They find water ice  as second most abundant ice component for $A_V>3$. Our results indicate that water ice is dominating the ice composition at $A_V$ of a few with a \ce{J(SO2)} peak around $A_V\approx5$  which is not included in the model chemistry of \citet{esplugues2016}. \ce{J(CO2)} is dominating the ice composition at $A_V>5$.
Fig.~\ref{fig:monolayers} shows how the ice thickness (in MLs) in the center of our model clumps depends on the gas density and the FUV illumination. The number of MLs decreases with $\chi$ as well as with $n_0$. The same is true for the depth where $\mathrm{ML}=1$. For $n_0=10^4$~cm$^{-3}$ and $\chi=10^4$ we find that the first ML builds at $AV=5.2$ roughly consistent with  \citet{esplugues2016} who find 0.6~ML at that extinction. However, our density increases with depth and therefore we have a slightly earlier onset of ice formation.

In a recent update \citet{esplugues2019} present computations with a modified dust temperature formulation \citep{hocuk2017} allowing for higher central dust temperatures. Comparing their dust temperatures with the detailed results from \texttt{MCDRT} shows a factor two lower values at the edge of the cloud but a factor $\sim 2$ higher central dust temperatures (except for $\chi=10^6$). At $A_V=10$ their $T_\mathrm{dust}$ is consistently $\sim10$~K hotter except for $\chi=1$ where they have 8.8~K compared to our 3.6~K.
As a result they also see a preferentially oxygen-bearing ice composition but with a different spatial behavior and a different ice composition because of their dust temperature exceeding the sublimation temperatures of \ce{CO} and \ce{CH4}. They present three models with different values for density and FUV strengths compared to \citet{esplugues2016}. A comparison with their results for the \ce{OH} density profiles shows a similar behavior for $A_V<5-6$. The density peak shifts closer to the surface for increasing $n$ and $G_0$ and we observe the same trends. For $n=10^5$~cm$^{-3}$ they show a peak density of $n_i/n_\mathrm{H}\sim 10^{-8}-10^{-7}$. We find similar densities for $\chi=10^4$. For the lower FUV illumination we find peak \ce{OH} abundances of $n_i/n_\mathrm{H}\sim 10^{-6}$. Their model 3 ($n=10^6, \, G_0=10^4$) shows a much higher peak density for $A_V<1$ of a few $10^{-5}$ and we find similar values at $A_V\sim0.2$. However, all their models show a high \ce{OH} density at their maximum $A_V=10$ of $2-3 \times10^{-8}$. Our models show a significantly lower central \ce{OH} density of a few $10^{-10}$ in for $n_0=10^5$ models. We find comparably high central densities only for  $\chi=10^6$ which is consistent with our lower dust temperatures for $\chi<10^6$. Our results for their model 3 configurations shows similar densities at $A_V=10$ but differs significantly in shape and drops to $10^{-10}$ in the center of our cloud. For \ce{O2} we find significantly different results. At $\chi=10^2,10^4$ we find peak relative densities of a few $10^{-6}$ at $A_V=3-5$. Our central \ce{O2} density drops with $n_0$ to $n_i/n_\mathrm{H}<10^{-12}$ for $n_0=10^5$~cm$^{-3}$ compared to values above $10^{-6}-10^{-5}$ for their models 1 and 2. Such high values seem to be in conflict with observations \citep[e.g.][and references therein]{goldsmith2011b,wirstroem2016}.  Only for $\chi=10^6$ are we finding a strong central \ce{O2} population of larger than $10^6$. A similar difference is also visible for \ce{H2O} that has lower central and higher peak values in our results. \citet{nagy2017} gives relative \ce{H2O} densities for the Orion Bar of $~1-5\times 10^{-12}$. The closest parameter set from \citet{esplugues2019} is their model 3 with peak/center abundances of $\sim10^{-7}$. Within the same $A_V$ range we find factor 10 lower values with central densities of $3\times 10^{-11}$ for $n_0=10^6$~cm$^{-3}$ and $\chi=10^4$. Our total  \ce{H2O} column densities at $A_V=5$ are $2\times10^{13}$~cm$^2$ for $n_0=10^5$~cm$^{-3}$ and $7\times10^{13}$~cm$^2$ for $n_0=10^6$~cm$^{-3}$, which is of the same order as the observed values of $2\times 10^{12}$ to $2\times 10^{13}$~cm$^2$ \citep{nagy2017}. Within their computations, changing the dust temperature prescription gave \ce{J(CH4)} abundances at $A_V=10$ that differed by a factor $10^{16}$! Given their different dust temperature behavior and a significant different set of assumed $E_D$ they find a very different ice composition and we conclude that a detailed comparison with our results is difficult. Qualitatively, we find comparable amounts of \ce{J(CH4)} and \ce{J(CO)} ice for low values of $\chi$ but significantly lower abundances for higher FUV fields. 

\citet{guzman2011} presented observations of \ce{H2CO} emission from the Horsehead together with surface chemistry computations from the Meudon PDR code \citep{lepetit2006}. \ce{H2CO} is interesting because it can be efficiently formed in the gas-phase and on the surface of grains. They showed that adding the surface network leads to an increase of \ce{HCO} densities by 1-2 orders and a  pronounced density peak of \ce{H2CO} at a few $A_V$ that is not produced in their gas-phase results. We find a similar qualitative behavior and can reproduce their \ce{H2CO} peak densities (relative to $n_\mathrm{tot}$) of $10^{-8}$ for comparable PDR parameters ($n_0=10^5$~cm$^{-3}$, $\chi=10^2$ ).

\subsection{Comparison with observations}

\begin{figure}[hbt]
	\resizebox{\hsize}{!}{\includegraphics{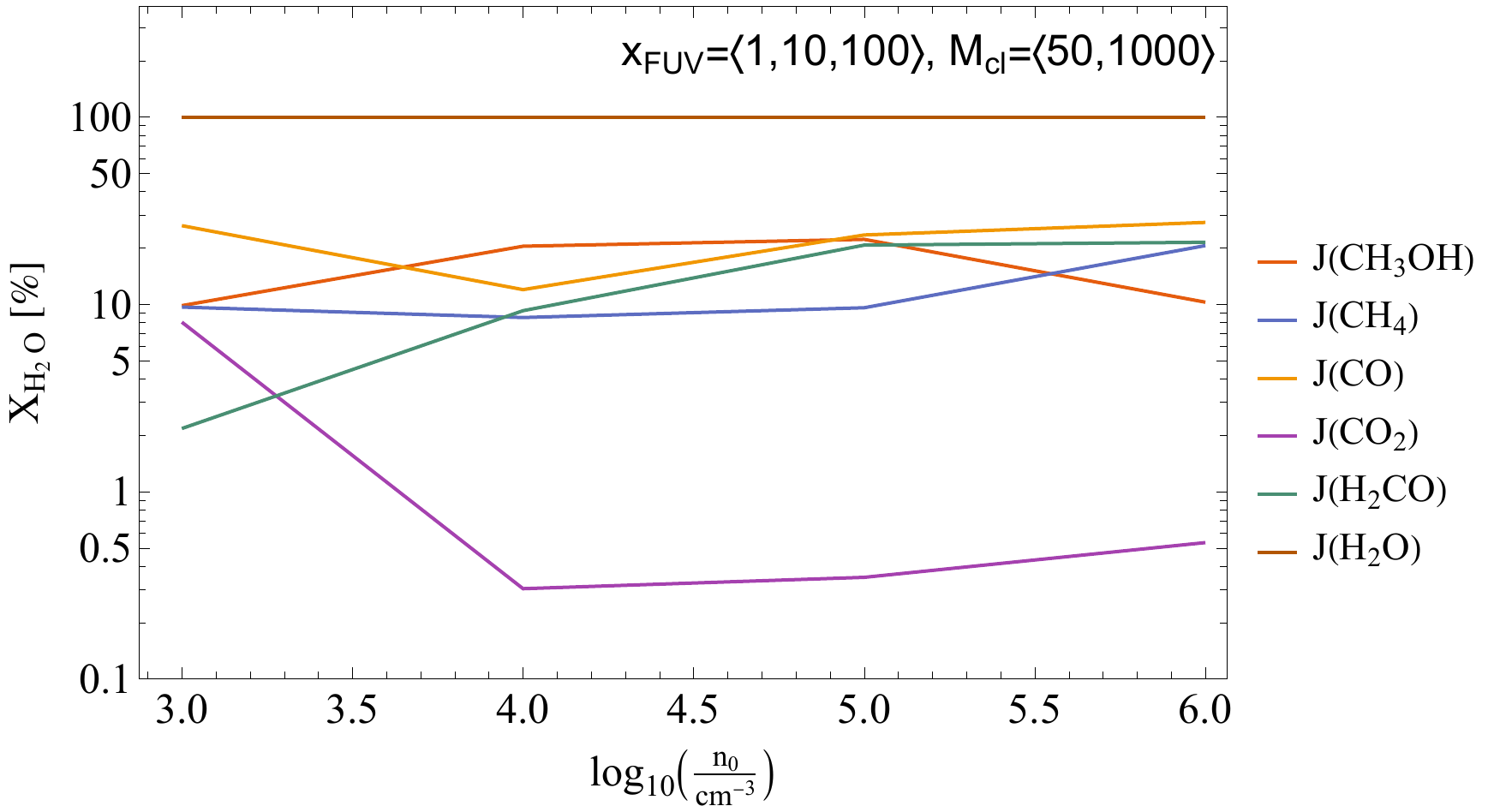}}
	\caption{Ice abundances relative to the \ce{H2O} ice column density. The median of all models with $\chi=1-10^3$ and $M=50,1000$~M$_\odot$ is shown.}
	\label{fig:ice-abundance}
\end{figure}
citet{boogert2015} summarizes observed ice abundances based on column density derivations. The dust temperature variations along the observed lines of sight have a major effect on the ice but remain unknown which makes a direct comparison of the model ice composition with observations not trivial. Secondly, the limited number of chemical species included in the presented computations limits the predictive power for some of the ice species, e.g. for \ce{J(NH3)}. \citet{boogert2015} give ice abundances for different environments: massive young stellar objects (MYSOs), low mass young stellar objects (LYSOs) and BG stars. Naturally, they correspond to different local physical conditions, but we will focus on general trends only. All abundances in this section will be relative to \ce{J(H2O)} column densities $X_\mathrm{H_2O} [\%]$ unless stated otherwise. 

The securely observed ices are in descending order: \ce{J(CO2)} and \ce{J(CO)} with 20-30\% abundance each, \ce{J(CH3OH)} with 6-10\%, \ce{J(NH3)} with $\sim6$\% and \ce{J(CH4)} with $\sim5$\%. \ce{J(H2CO)} is likely identified with a few \% abundance. In Fig.~\ref{fig:ice-abundance} we show our model ice predictions where we average over a range of FUV strengths $\chi=1-10^3$ and clump mass $M=50,1000$~M$_\odot$ and plot  $X_\mathrm{H_2O}(i)$ as function of the gas density $n_0$. \ce{J(H2O)} is the most common ice across the whole density range and \ce{J(CO)} varies between 20-40\%. Both results are in agreement with observed numbers. We also find that \ce{J(CH3OH)} is predicted with 3-20\% relative abundance which is also consistent with observations. \ce{J(CH4)} model abundances are higher than the observed ranges by a factor 2-3 and model \ce{J(CO2)} is consistent with the observed 20-30\% only for $n_0\sim10^3$~cm$^{-3}$. Our \ce{J(H2CO)} predictions recover the observed values in the lower density range of Fig.~\ref{fig:ice-abundance}. Given the crude averaging over the physical parameters and our small chemical network we find our model predictions to be in reasonable agreement with observations.

\section{Clumpy ensemble model}

Interstellar clouds are neither a plan-parallel slab nor of perfect spherical shape and results from these kind of models will always be a rough approximation to reality, a reality where the ISM is clumpy/fractal, turbulent, organized in filaments or fibers, and most importantly not  in equilibrium. PDR models with more complex geometries have been designed to address this deficiency, but the higher complexity always comes with the price of much higher computation costs \citep{bisbas2012,levrier2012,grassi2014,girichidis2016,bisbas2021}. The spherical setup of \Kt{} offers the attractive option of modeling clumpy clouds as a superposition of differently sized clumps following a well-defined clump-mass spectrum. This has been described in detail in \citet{zielinsky2000}, \citet{cubick08} and \citet{andree2017}. For details see Appendix~\ref{sect:clumpy_appendix}.

Clumpiness has frequently been invoked to explain certain emission characteristics of the ISM in spatially unresolved observations. The main driver was always that clumpy gas has a higher surface to volume ratio and therefore shows an excess of emission primarily produced in the PDR surface regions of molecular clouds. A typical example is the {\cii} emission, which is produced by strong FUV illumination and a good tracer of PDRs, this means a surface tracer. Conversely, rotational \ce{CO} line emission is a good volume tracer because \ce{CO} requires shielding from intense FUV illumination, which is effective only for extinctions $A_V>1$. Non-clumpy PDR models were not able to explain the observed excess in e.g. the \cii/\ce{CO}(1-0) line ratio in active star forming regions \citep{stutzki1988,spaans1997,dedes2010,graf2012} but the observed ratios asked for models with a larger surface-to-volume ratio. A similar explanation has also been presented by \citet{MeixnerTielens1993,hogerheijde1995,zielinsky2000} to explain variations in observed line ratios of several different PDR and molecular cloud tracers through clumpy ensembles of PDRs. \citet{cubick08} showed that the global far-infrared (FIR) emission of the Milky Ways can  be explained in terms of clumpy PDR emission. 

This has been supported by observations of clumps spectra in molecular clouds. Based on molecular line observations \citet{heithausen1998} measured the scaling relations for one source over several orders of magnitude and \citet{kramer1998} measured the clump-mass distribution in various sources confirming a common power-law across all sources with power-law index 1.6 to 1.8, extending down to the resolution limit and clump masses as low as $10^{-3}~M_\odot$ for at least two of the sources. These findings have been put at question by models that managed to explain the observed line ratios from high-pressure PDR models \citep[e.g.][]{Marconi1998}. \citet{joblin2018} showed that the pure line intensities in the Orion Bar can be explained from a plane-parallel PDR model. Their high-pressure models however, do not reproduce the observed spatial stratification of the different tracers but predict a very thin PDR layering. The key information on the clumpiness comes from the spatially resolved structures. \citet{andree2017} showed that the observed spatial stratification within the Orion Bar is in disagreement with any  simple plane-parallel model but that some kind of clumpiness had to be invoked. \citet{andree2017}  presented \Kt-3D, where individual volumetric elements (voxel) are populated with unresolved clumpy PDR ensembles. They modeled the 3-dimensional structure of the Orion Bar and found a good agreement of their results with a multi-line data set from \textit{Herschel} and Caltech Submillimetre Observatory (CSO) observations.

Velocity-resolved observations from Herschel, SOFIA and ALMA confirm the dynamical nature of PDRs \citep{goicoechea2016,goicoechea2017,joblin2018,Wu2018,luisi2021,kabanovic2022}. Photo-evaporation flows from globules and other dense clumps are ubiquitous \citep{mookerjea2012,bron2018}. Theoretical papers have predicted for a long time that dense clumps are carved out from their parental cloud by UV irradiation \citep{Lefloch1994,Henney2009,Bisbas2011}. The inter-clump medium is then fed by the photo-evaporation flows from the PDR surfaces. The observations confirm the theoretical predictions that the inter-clump medium density is lower than the clump density by at least two orders of magnitude \citep{Arkhipova2013,Mookerjea2019,schneider2021}. The photo-evaporation from the clumps provides a continuous low-density mass flow from the surfaces with velocities of 1--2~\kms{} \citep{Makai2015, mookerjea2012,Mookerjea2019,Sandell2015,Goicoechea2020}. The inter-clump medium inherits the chemical properties of the PDR surfaces. Despite of the strong density difference their chemistry is rather similar because it is dominated by the FUV radiation rather than collisions. However, in principle this constitutes a non-stationary scenario due to the constant mass loss to the inter-clump medium \citep{bertoldi1996,stoerzer1997,stoerzer1998,stoerzer1999,maillard2021}. It violates the assumption of a constant mass in the \Kt{} framework primarily affecting low-mass clumps ($M_{cl}\la10^{-2}$~M$_\odot$ for $n\ga10^6$~cm$^{-3}$) \citep{Decataldo2017,Decataldo2019}. As a dynamic effect it is currently also not contained in isobaric PDR models where the low density gas only occurs as a very thin and hot surface layer.   

The picture of the mass flow constitutes a recipe for us to represent the clump/inter-clump structure within the clumpy model framework. In \Kt{} we can model the dense PDR clumps by \Kt{} clump masses $M_{cl}>10^{-2}$~M$_\odot$ while the inter-clump gas is represented by UV dominated conditions that are well modeled by low mass clumps with small $A_V$.

With the continuously improving spatial resolution of modern observatories, such as ALMA and IRAM/NOEMA, it  becomes possible to further test the clumpy ensemble picture by new observational data.
However, even with today's interferometric observations it remains difficult to resolve clumps with masses below 0.1~M$_\odot$ in continuum observations. The completeness limit of the clump analysis is often reached at a few 0.1~M$_\odot$ so that the slope of the mass distribution below one solar mass is uncertain and highly debated \citep[see e.g.][]{Pineda2009,MivilleDeschenes2017,Cheng2018,Kong2019,koenyves2020}. For an overview of how different clump finding strategies affect the derived size and mass distribution see \citet{schneider2004} or \citet{Li2020}. 
Moreover, the photo-evaporation modifies the clump distribution in molecular clouds under the irradiation forming PDRs. Hence, it is an important question to study whether the observational data provide further constraints on the slope in the clump size spectra below the scale resolved in the continuum observations.

\subsection{Example: Orion Bar - spatial structure}

\begin{figure}
	\resizebox{\hsize}{!}{\includegraphics{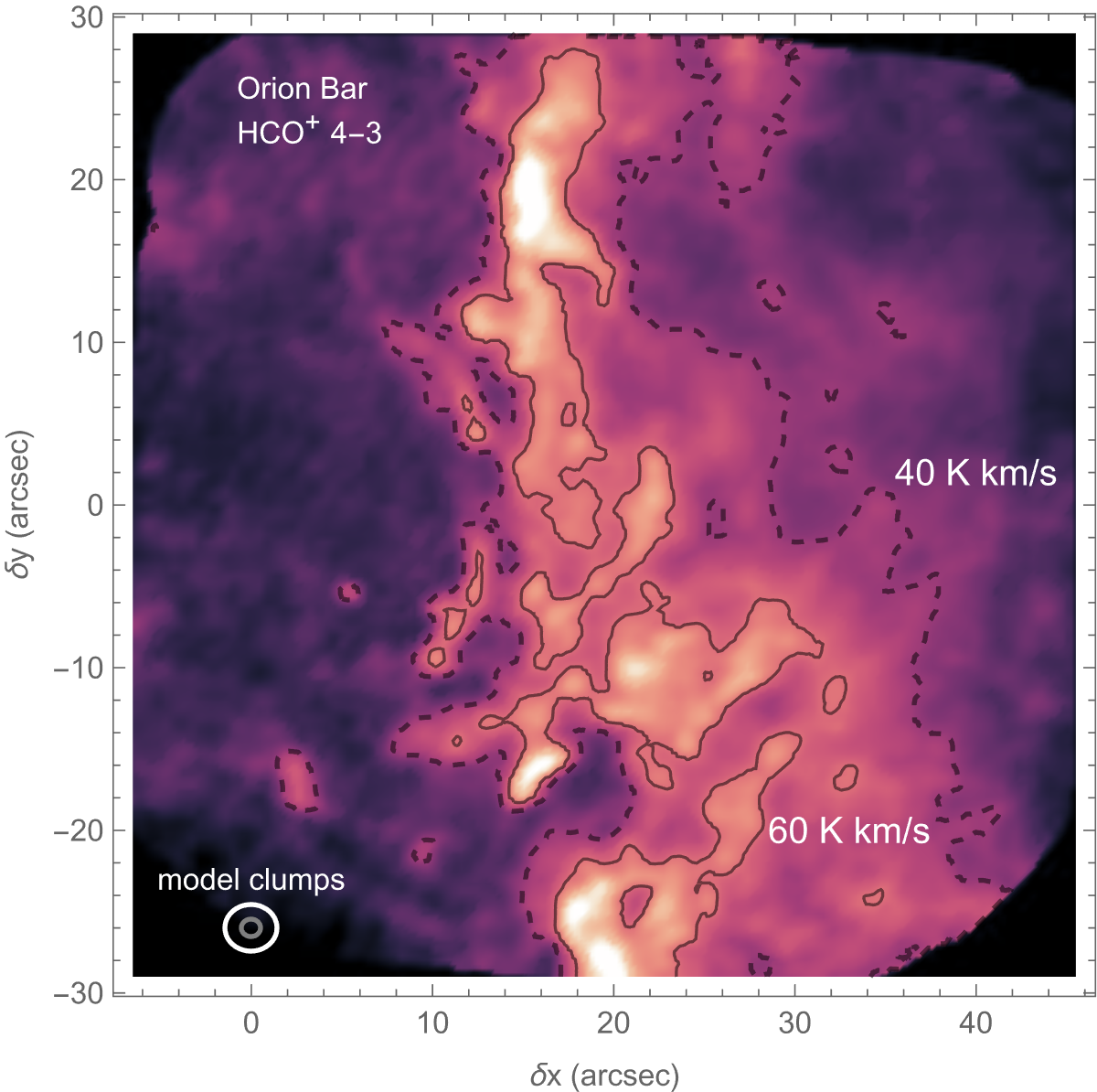}}
	\caption{Integrated intensity map of the \ce{HCO+} 4-3 line emission in the Orion Bar \citep[see][for details on the mapped area]{goicoechea2016}. The contours corresponds to integrated intensities of $40\, \mathrm{K\, km\, s^{-1}}$ (dashed) and $60\, \mathrm{K\, km\, s^{-1}}$ (solid). The colors indicate intensities between 0 and 100 \Kkms{}. The two circles in the lower left show the area of clumps with $M=0.01\,\mathrm{M_\odot}$ (white) and $M=0.001\,\mathrm{M_\odot}$ (gray).}
	\label{fig:orionbar}
\end{figure}

\begin{table*}[ht]

	\caption{Properties of a simple discrete Orion Bar ensemble.}
	\label{tab:orionbar}
		\centering
	\begin{tabular}{rcrrrrrrc}
		\hline \hline %
		\vrule width 0pt height 2.6ex %
		$M_i$ & $N_i$\tablefootmark{(1)} & $n_i$\tablefootmark{(2)} & $R_i$ & $R_i$\tablefootmark{(3)} & $\Omega_i$\tablefootmark{(3)} & $I_i(\ce{HCO+})$\tablefootmark{(4)} & $\Omega_{\ce{HCO+}} $\tablefootmark{(5)} & $N_i\Omega_i/\Omega_{\ce{HCO+}}$  \\
		$(\mathrm{M_\odot})$ & & $(\mathrm{cm^{-3}})$ & $(\mathrm{pc})$ & $(\arcsec)$  & $(\arcsec^2)$ & $(\mathrm{K\,km\,s^{-1}})$& $(\arcsec^2)$ &  \\
		\hline%
		\vrule width 0pt height 2.6ex %
	    &&
		\multicolumn{6}{c}{$M_\mathrm{ens}=1\,\mathrm{M_\odot}$, $\langle n_\mathrm{ens}\rangle=5\times 10^6\,\mathrm{m^{-3}}$, $\chi=10^4$,  $\alpha=1.8$, $\gamma=2.3$}\\
		\hline
		\vrule width 0pt height 2.6ex 		
		0.001 & 387(794) & $8.1\times 10^6$ & 0.0011 & 0.53 & 0.88 & 39.4& 1097 & 0.48(1.0)  \\
		0.01 & 61 & $4.0\times 10^6$ & 0.0029 & 1.44 & 6.52 & 57.2& 385 & 1.0  \\
		\hline
		\vrule width 0pt height 2.6ex 
	\end{tabular}
	\tablefoot{
		\tablefoottext{1}{$N_I$: clump number in the ensemble. The numbers in parenthesis give the clump number to completely fill the intensity contour area for 40~\Kkms $\le I\le$ 60~\Kkms.}
		\tablefoottext{2}{$n_i$: average clump density.}
		\tablefoottext{3}{$R_i$ and $\Omega_i$ denote radius and solid angle of a single clump. $I_i$ corresponds to the clump-averaged intensity.}
		\tablefoottext{4}{$I_i$: clump averaged intensity.}
		\tablefoottext{5}{$\Omega_{I}$ denotes the solid angle enclosed by the respective intensity contour including the solid area of all inner contours. The area of the 39~$\mathrm{K\,km\,s^{-1}}$ contour is exclusive of the 57~$\mathrm{K\,km\,s^{-1}}$ contour listed in the second line. In the last column $N_i\Omega_i/\Omega_{\ce{HCO+}}$ gives the ratio between the solid angle covered by the model clumps with the matching intensity and the corresponding observed solid angle in the map (excluding inner contours).}
	}
\end{table*}

High-resolution ALMA data of the Orion Bar directly resolve the larger
clumps in our description. \citet{goicoechea2016} presented ALMA observations of \ce{HCO+} 4-3 emission lines of the Orion Bar resolving the structure of the ionization and dissociation front with $\sim 1\arcsec$. Their data show
fragmented clumps with sizes in the order of 2\arcsec (see Fig.~\ref{fig:orionbar}). This corresponds roughly to a clump mass of about $0.01\,\mathrm{M_\odot}$ at densities of $n\approx 4\times 10^6\,\mathrm{cm^{-3}}$ and $\oslash=1.44$\arcsec. \Kt{} computes $I_{cl}(\ce{HCO+} 4-3) = 57\, \mathrm{K\,km\,s^{-1}}$ clump averaged emission for such a clump assuming a FUV field of $\chi= 10^4$. For a detailed discussion on suitable PDR model parameters for the Orion Bar we refer to \citet{andree2017}. For a qualitative discussion we overlay in Fig.~\ref{fig:orionbar} contours for 60~$\mathrm{K\,km\,s^{-1}}$ to the \ce{HCO+} 4-3 data presented by \citet{goicoechea2016}. The comparison with the size of a $0.01\,\mathrm{M_\odot}$ clump in the lower left corner (radius $1.44\arcsec$) shows a match of the typical structure size. Most of the emission at lower levels is not spatially resolved but forms an extended structure, mainly behind the Orion Bar. If we assume that this is due to smaller clumps, well below the beam size and not removed by photo-evaporation yet, we can also model them through \Kt{}. When following the original clump mass spectrum clumps with a mass of 0.001~M$_\odot$ have a radius of  $0.53\arcsec$ and a clump averaged \ce{HCO+} 4-3 intensity of 39 ~$\mathrm{K\,km\,s^{-1}}$. For the comparison, we also include dashed contours for 40~$\mathrm{K\,km\,s^{-1}}$ in Fig.~\ref{fig:orionbar}. Comparing the areas within the two contours we can count the required clumps to produce the observed emission and compare the ratio with our standard clump ensemble scaling.

Our standard clump ensemble setup approximately reproduces the properties of this \ce{HCO+} intensity map without any further fitting. Table~\ref{tab:orionbar} lists the parameters of the corresponding discrete clump ensemble. Here, we use just the two clump masses of $m_l=0.001$ and $m_u=0.01\,\mathrm{M_\odot}$ depicted in Fig.~\ref{fig:orionbar} and assume the standard scaling laws for the clump mass distribution (power law index $\alpha=1.8$) and the mass-size relation (power law index $\gamma=2.3$) \citep{heithausen1998}. The flux of the 0.01~$\mathrm{M_\odot}$ clumps fully explains the 57~$\mathrm{K\,km\,s^{-1}}$ contour if we ignore higher intensities within the area. This is an obvious oversimplification since the map shows few smaller condensations with $I_{\ce{HCO+}}>80$~$\mathrm{K\,km\,s^{-1}}$. The low mass clumps explain the outer contours to about 50\%. This is a remarkably good match given the fact that we did not perform any numerical fitting.

 To fully explain the intensity area it would take about twice as many clumps with $M=0.001$~M$_\odot$ (numbers in parenthesis in Table~\ref{tab:orionbar}). Consequently, this results in a steeper clump-mass distribution with a power-law index of $\alpha\approx 2.1$ or contributions from more even smaller clumps that were ignored in this simple picture. A steeper clump-mass distribution would agree with previous studies showing that denser regions, in particular in Orion A, tend to have steeper clump-mass indices $\alpha$ comparable to what we find \citep{bally1987,maddalena1986,nagahama1998,schneider2004}.

We can also compare the model column densities. For the model clumps from Table~\ref{tab:orionbar} we compute the mean column density of a species averaged over the projected clump area. The contour area filling factor of approximately unity allows to directly compare the column densities. Table~\ref{tab:columns} gives the observed values and our clumpy results. The observed column densities are mostly consistent with a clumpy PDR ensemble. The \ce{CO} column density predicted by the model is higher than the value derived from 3-2 observations by \citet{goicoechea2016}. This is easily explained by the fact that the derivation used there is not sensitive to CO at temperatures below 120~K so that a large fraction of cool CO was not accounted for. 
The major discrepancy in the \ce{SO} abundance may be attributed to some large uncertainties in the sulfur chemistry due to the unknown roles of vibrationally excited \ce{H2} \citep{goicoechea2021} and direct depletion \citep{fuente2016}.
The total column densities are also higher than the values given by \citet{goicoechea2016} but closer to estimates by \citet{hogerheijde1995} who derived $N_{\ce{H2}}=6.5\times 10^{22}$~cm$^{-2}$ at the peak of molecular emission. 

\begin{table}
	\centering
	\caption{Observed column densities vs. model results.
	}
	\label{tab:columns}
	\begin{tabular}{lrr}
		\hline \hline
		\vrule width 0pt height 2.2ex
		&$N_{obs}$~(cm$^{-2}$)\tablefootmark{(1)}&$N_{ens}$~(cm$^{-2}$)\tablefootmark{(2)}\\
		\hline
		\vrule width 0pt height 2.6ex %
					\vrule width 0pt height 2.6ex %
		\ce{CO}&$1\times 10^{18}$&$(8-10)\times 10^{18}$\\
		\ce{HCO+}&$5\times 10^{13}$&$(3-5)\times 10^{13}$\\
		\ce{H^{13}CO+}&$(5-20)\times 10^{11}$&$(6-8)\times 10^{11}$\\
		\ce{HOC+}&$(3-9)\times 10^{11}$&$(3-6)\times 10^{12}$\\
		\ce{SO+}&$(2-4)\times 10^{12}$&$(2-3)\times 10^{12}$\\
		\ce{SO}&$(5-10)\times 10^{13}$&$(8-14)\times 10^{16}$\\
		\ce{H + 2  H2}&$2\times 10^{22}$&$(4-5)\times 10^{22}$\\
		\hline
	\end{tabular}
\tablefoot{
	\tablefoottext{1}{Reference: \citet{goicoechea2016,goicoechea2017}},
    \tablefoottext{2}{Column densities are given for the ensemble clumps from Table~\ref{tab:orionbar}}.}
\end{table}

The simple numerical experiment shows that 1) the \ce{HCO+} 4-3 intensity levels predicted by \Kt{} clumps are consistent with the observations of the Orion Bar. 2) The total flux predicted by a clumpy ensemble covers the observed flux values to a significant degree. 3) The column densities from a clumpy PDR ensemble are consistent with observed values. 4) The fact that we do not resolve the smallest ensemble clumps is consistent with their filling factor which leads to roughly homogeneous intensity distribution. 

\subsection{Non-stationary clumpy PDR evolution}

The structure in Fig.~\ref{fig:orionbar} is the result of a high-pressure zone moving through the molecular cloud  \citep{goicoechea2016} forming a fragmented and dynamically moving PDR surface. Compression by this wave and photo-evaporation lead to enhanced density contrasts with dense clumps, subject to erosion, and a thin interclump medium \citep{gorti02}. \citet{goicoechea2016} find a dynamical crossing time for the wave front of a few $10^4$~yr. $10^4$~yr corresponds to the photo-evaporation destruction time for clumps with $M< 0.01$~M$_\odot$ \citep{Decataldo2017} and the  chemical time-scale for \ce{HCO+} formation or destruction in the clumps from Table~\ref{tab:orionbar}. Consequently, we do not expect the survival of any smaller clumps after the passage of the PDR zone. This is consistent with the fact that Fig.~\ref{fig:orionbar} shows almost no 40~\Kkms{} emission on the side facing the ionization front ($\delta x\la 15 \arcsec$). Clumps with $M<0.01$~M$_\odot$ did not survive the passage of the pressure front. The regions in  Fig.~\ref{fig:orionbar} with $\delta x\ga 20 \arcsec$ shows volumes which have not yet been affected by the front and where the lower mass clumps are still surviving. They may already be subject to some evaporation but their embedded \ce{HCO+}  population is not yet diminished. This picture also explains the emission peaks with levels above 60~{\Kkms} at the PDR front in Fig.~\ref{fig:orionbar}. Clumps with original masses above 0.01~M$_\odot$  are brighter than 70~{\Kkms} but may have been partially eroded so that we only see their larger fragments. We conclude, that the findings by \citet{goicoechea2016} are consistent with a clumpy medium with a mass spectrum that changes from the original distribution by photo-evaporation during the passage of the PDR. The observed gas dynamics and the related life times of such clumps explain the spatial distribution of the observed emission.

\section{Conclusion} 
We present the current status of the numerical PDR model code {\Kt}, which solves the coupled chemical and physical state of the ISM in a spherical model cloud under isotropic FUV illumination. \Kt{} has been thoroughly tested in the last 20 years and has been subject to a number of significant updates and improvements. In this paper, we provide a detailed description of the geometric and numerical setup of the code as a reference for comparison with other model codes. Details of adaptive stepping and numerical convergence are discussed and recommendations for application in other codes are derived. We discuss a series of numerical modifications of the code to improve convergence stability and overall performance. The implementation of a time-dependent chemical solver allows for an efficient fall-back algorithm in case regular steady-state solvers do not converge. Finally we present a publicly available sandbox PDR model realized in the programming language Mathematica by Wolfram Research.

\Kt{} now includes a complete surface chemistry network fully coupled to the gas-phase chemistry via all relevant accretion and desorption processes. In particular chemical desorption has been added to \Kt. We extended the prescription presented by \citet{minissale2016} and included all possible reaction branches also including partial desorption. 
We implemented the surface chemistry in a fully modular way analog to the gas-phase chemistry, allowing for simple addition or removal of chemical species. Adding surface chemistry to a PDR model can produce unexpected results. In particular the detailed treatment of the dust temperature is important because PDRs can sustain large amounts of warm dust particles at very large extinctions strongly affecting the ice chemistry. We discuss how higher dust temperatures lead to a selective freeze-out of oxygen bearing species compared to carbon-bearing species due to their higher condensation temperature. For high FUV models the reduction in \ce{CO} column density due to selective freeze-out of oxygen-bearing species onto grain surfaces produces a surplus of atomic carbon leading to a significantly enhanced \ci\ line emission that could explain the difficulties of previous generations of PDR models in fitting observed levels of \ci\ emission. The composition of the ice mantles changes in a complicated, not always intuitive way as a function of the FUV irradiation. This applies e.g. to the non-monotonous production of methanol that is suppressed at elevated FUV fields. We discuss the ice composition under different FUV conditions and compare our results with observations and with predictions by other models.
We use recent ALMA observations of the Orion Bar PDR to test the clumpy PDR model assumptions. Performing a simple numerical experiment shows that the PDR structure as well as the observed flux values can consistently described with the assumption of a clumpy ensemble of PDR clumps. Clumps with masses below 0.01~M$_\odot$ are eroded in the dynamical evolution of the the PDR explaining the asymmetry of the observed profile around the high-pressure zone when taking the observed time scales into account.

\acknowledgements{
We wish to thank J.R. Goicoechea for providing us with the ALMA data. We also would like to thank the anonymous referee for thoughtful suggestions and constructive feedback that helped to significantly improve the paper. 
The research presented here was supported by the Collaborative Research Centre  956, sub-project  C1, funded  by the  Deutsche  Forschungsgemeinschaft  (DFG), project ID 184018867.
}

\bibliographystyle{aa}
\setlength{\bibsep}{-2.1pt}
\bibliography{ref}
\clearpage
\begin{appendix}
\section{New model results}\label{sect:results_appendix}

\begin{figure*}[hbt]
	\resizebox{\hsize}{!}{\includegraphics{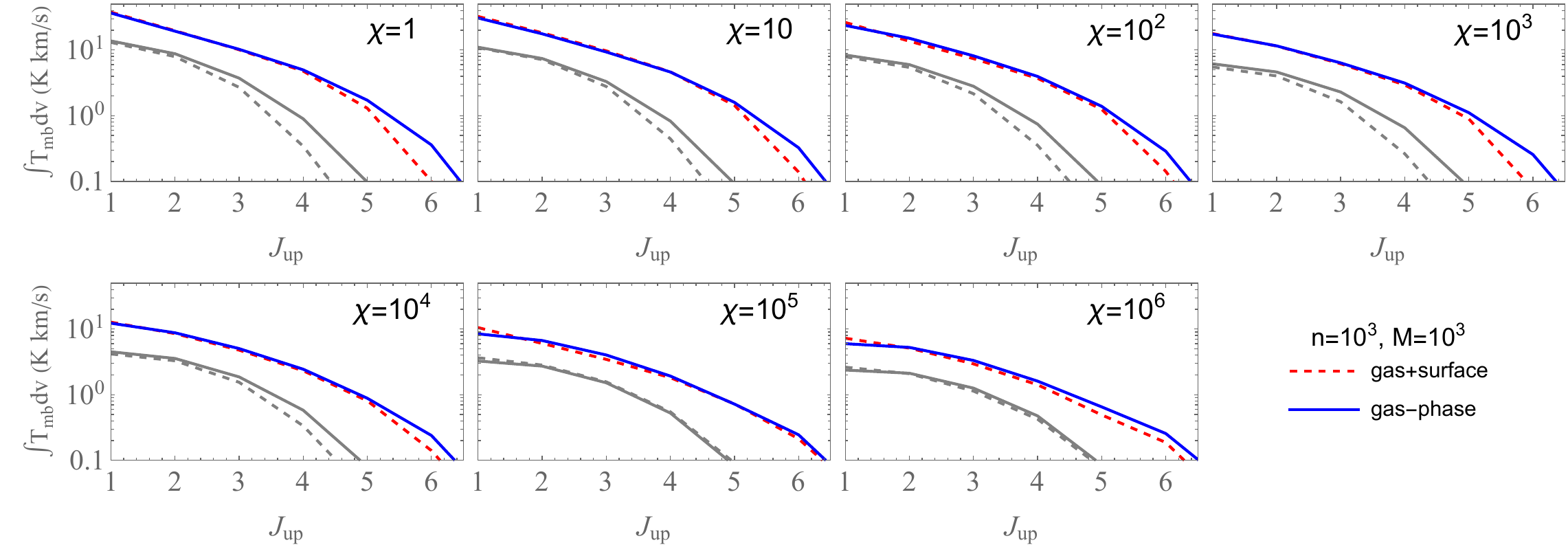}}
	\caption{Effect of surface chemistry on the CO spectral line emission distribution (in \Kkms{}) as function of $J_\mathrm{up}$ for $n_0=10^3$~cm$^{-3}$. Each panel corresponds to a different FUV field strength $\chi$. \ce{^{12}CO} lines are in color, while \ce{^{13}CO} lines are plotted in gray-level.  }
	\label{fig:CO-SLED-FUV-n30}
\end{figure*}
\begin{figure*}[hbt]
	\resizebox{\hsize}{!}{\includegraphics{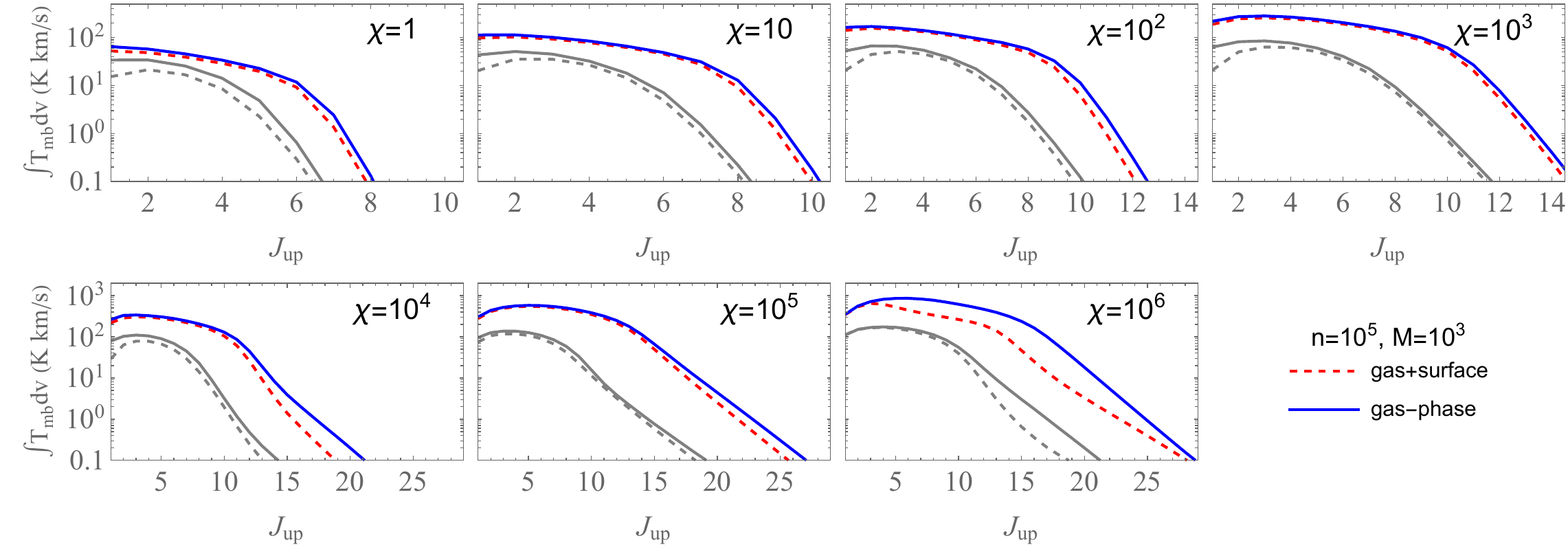}}
	\caption{Same as Fig.~\ref{fig:CO-SLED-FUV-n30} for $n_0=10^5$~cm$^{-3}$. }
	\label{fig:CO-SLED-FUV-n50}
\end{figure*}
\begin{figure*}[hbt]
	\resizebox{\hsize}{!}{\includegraphics{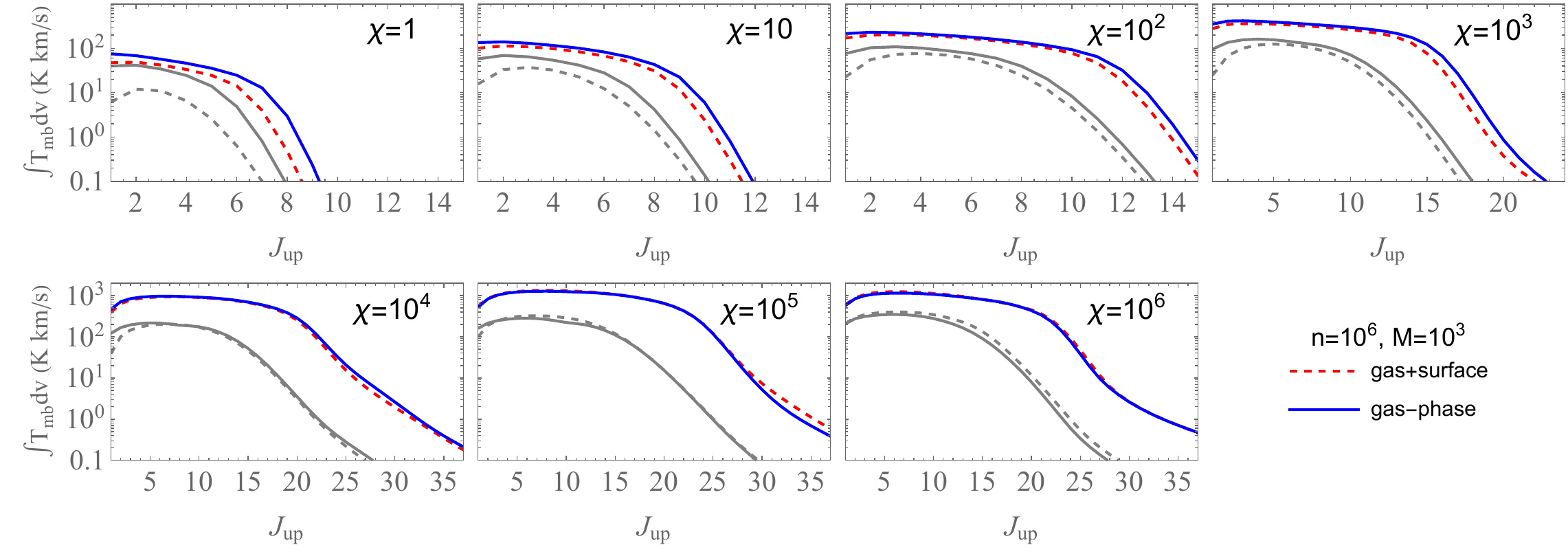}}
	\caption{Same as Fig.~\ref{fig:CO-SLED-FUV-n30} for $n_0=10^6$~cm$^{-3}$. }
	\label{fig:CO-SLED-FUV-n60}
\end{figure*}
We show the spectral line energy distribution (SLED) of \twco{} and \thco{} lines as function of the FUV field strength $\chi$ for $n_0=10^3,10^5,10^6$~cm$^{-3}$ in Figs.~\ref{fig:CO-SLED-FUV-n30}, \ref{fig:CO-SLED-FUV-n50}, \ref{fig:CO-SLED-FUV-n60}, respectively. The effect of the surface chemistry on the line emission is most prominent for high density and higher-J transitions.\footnote{Some low~J transitions show signs of level inversion that are unlikely to appear in practice.} We also show the density profile of selected species affected by the surface chemistry.

\begin{figure*}
	\begin{subfigure}[t]{.49\textwidth}
		\centering
		\includegraphics[width=\linewidth]{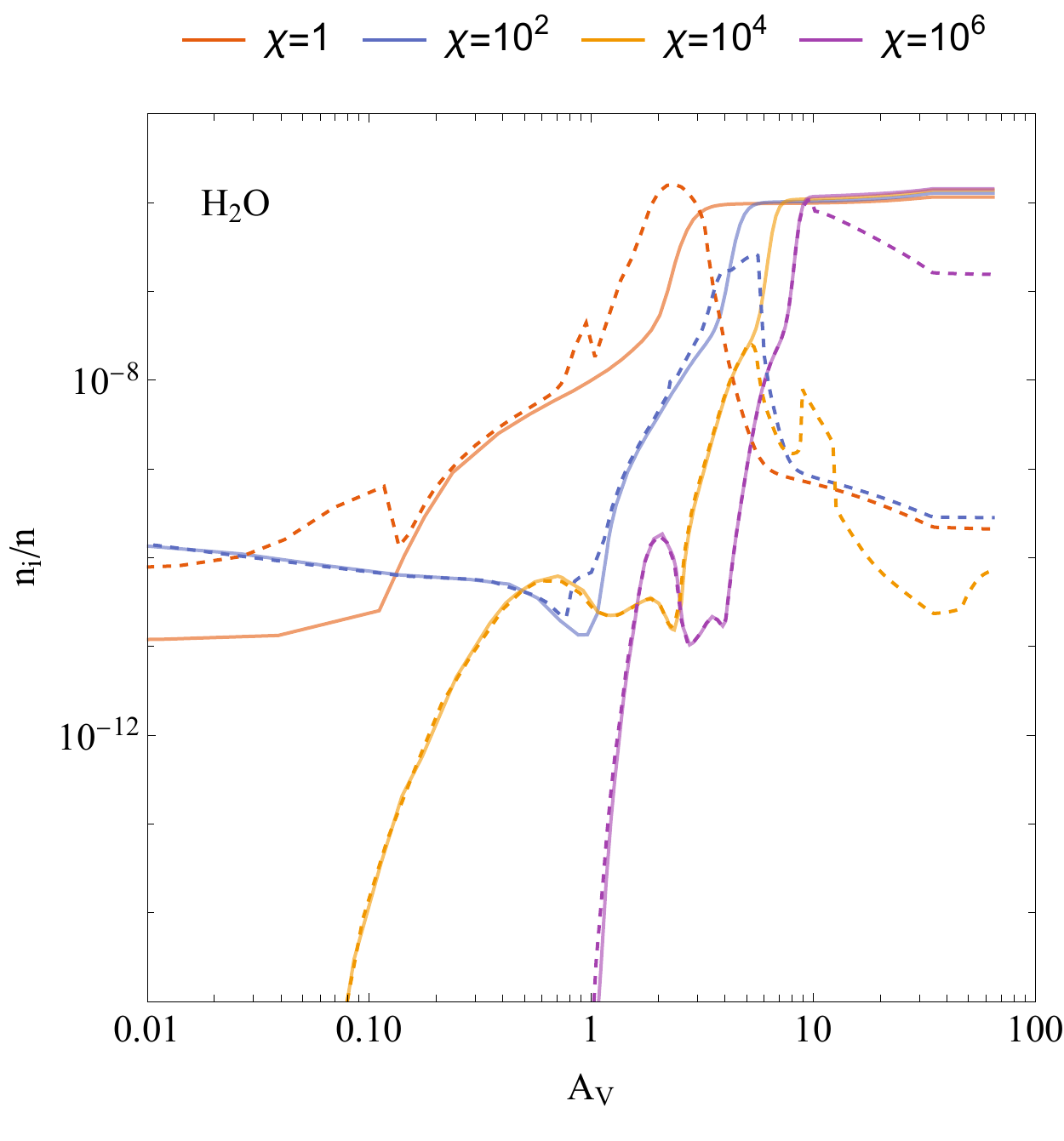}
		\caption{\ce{H2O} density profile.}
		\label{fig:structure_fuv-H2O}
	\end{subfigure}
		\hfill
	\begin{subfigure}[t]{.49\textwidth}
		\centering
		\includegraphics[width=\linewidth]{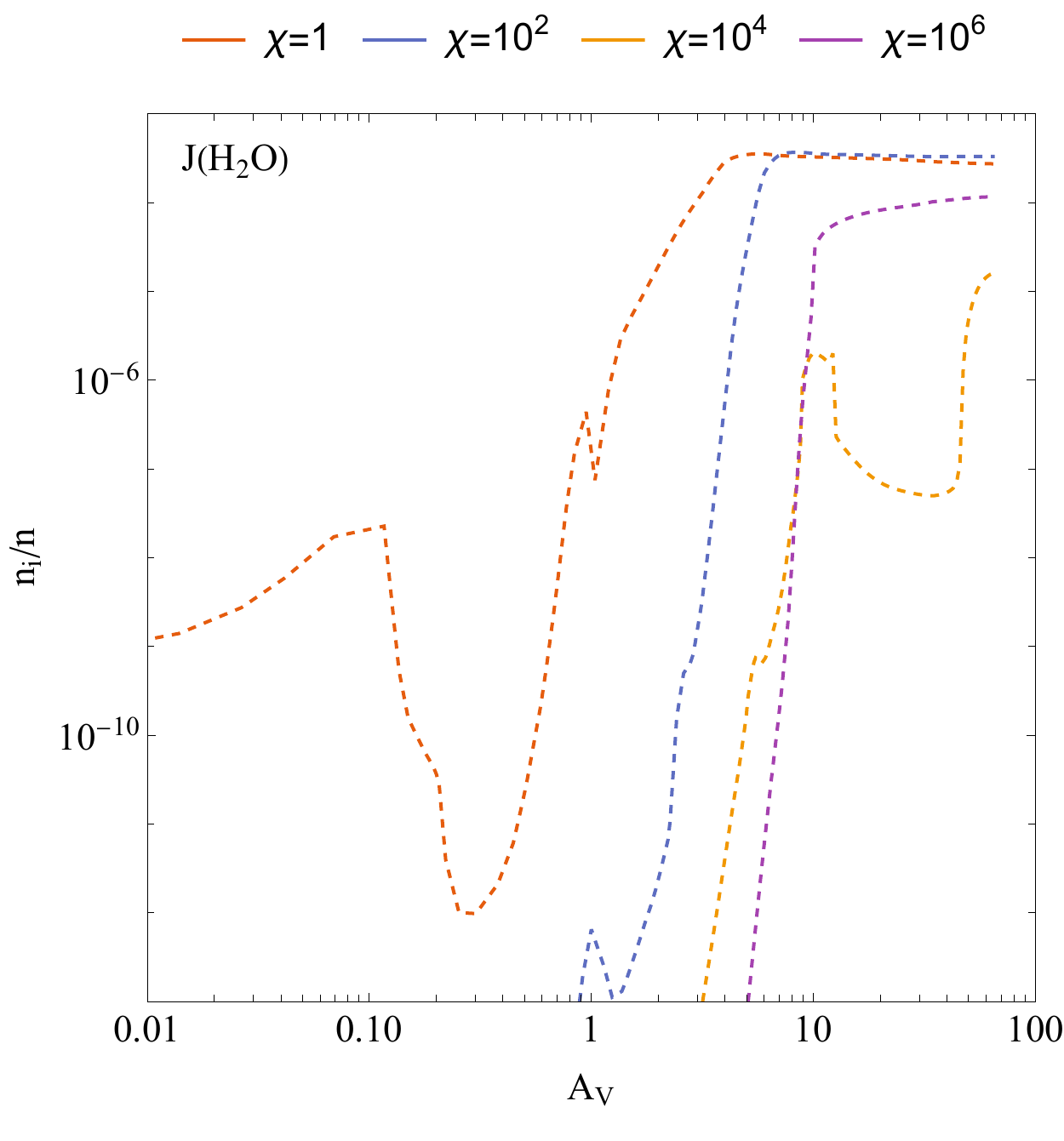}
		\caption{\ce{J(H2O)} density profile.}
		\label{fig:structure_fuv-JH2O}
	\end{subfigure}
	
	\caption{Chemical structure changes with FUV strength.  $n=10^4~\mathrm{cm}^{-3}$. Solid lines indicate pure gas-phase chemistry, dashed lines correspond to the gas+surface chemistry using the CR$_2$ model.}\label{fig:structure_fuv_1} 
\end{figure*}	

\begin{figure*}
	\begin{subfigure}[t]{.49\textwidth}
		\centering
		\includegraphics[width=\linewidth]{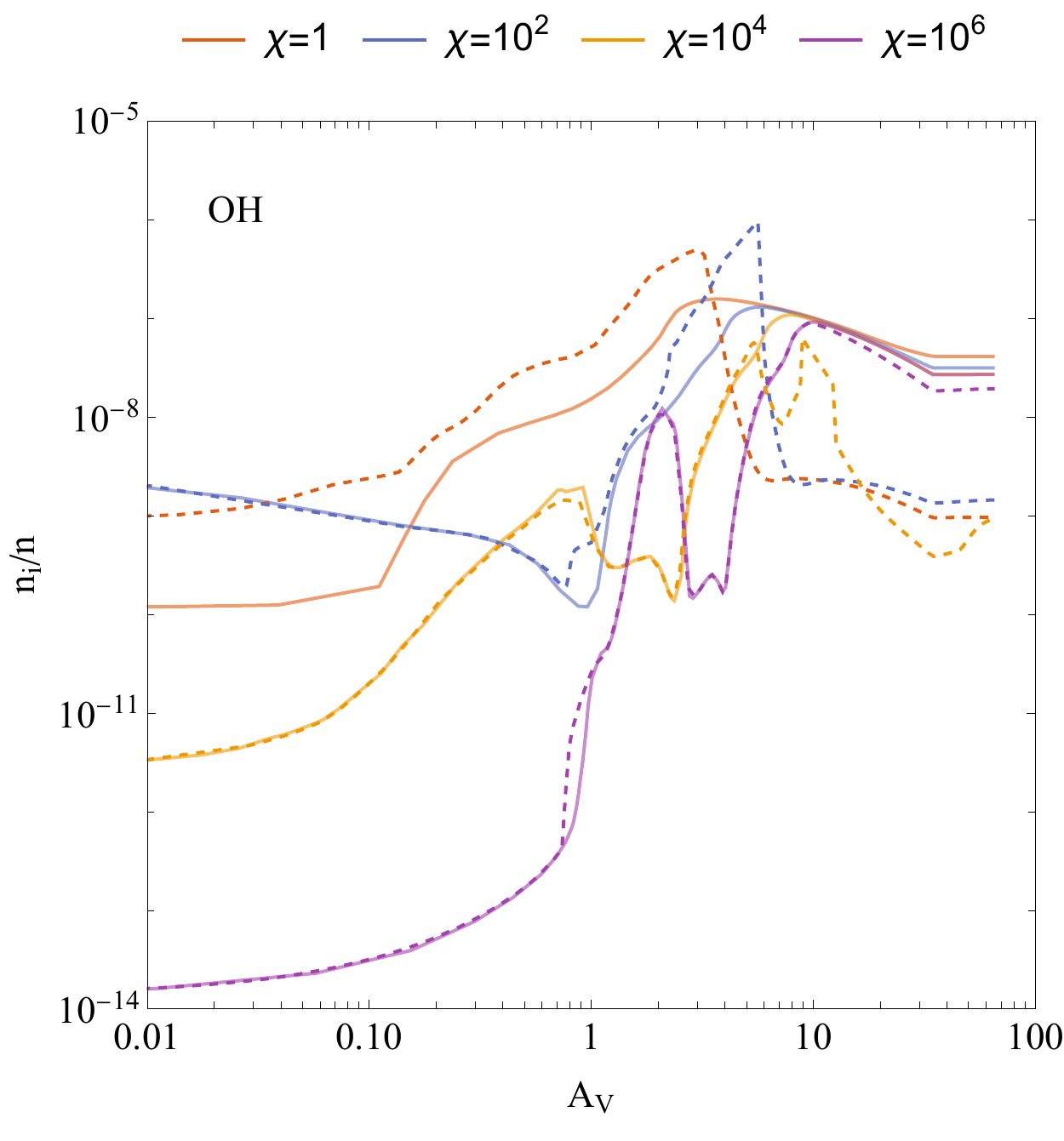}
		\caption{\ce{OH} density profile.}
		\label{fig:structure_fuv-OH}
	\end{subfigure}
		\hfill
	\begin{subfigure}[t]{.49\textwidth}
		\centering
		\includegraphics[width=\linewidth]{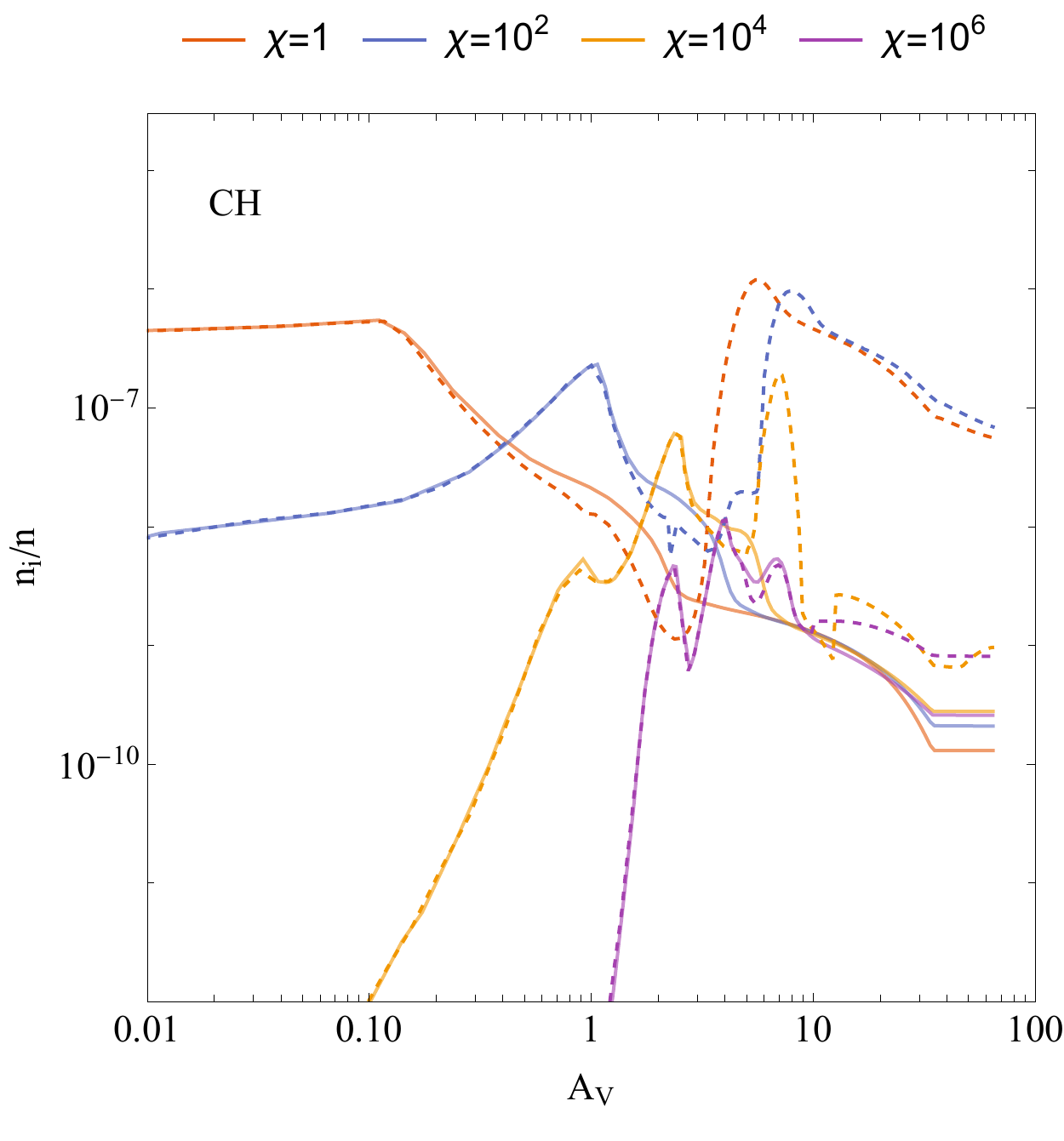}
		\caption{\ce{CH} density profile.}
		\label{fig:structure_fuv-CH}
	\end{subfigure}
	
	\caption{Chemical structure changes with FUV strength.  $n=10^4~\mathrm{cm}^{-3}$. Solid lines indicate pure gas-phase chemistry, dashed lines correspond to the gas+surface chemistry using the CR$_2$ model.}\label{fig:structure_fuv_2} 
\end{figure*}

\section{Numerical details}\label{sectnumerical-details}
Sect.~\ref{sect:spatial-loop} described the spatial loops over the adaptive cloud depth grid.  Input to the Bulirsch-Stoer stepper are the dependent variable vector $y_{1...\mathcal{N}}(z)=N_{1...\mathcal{N}}(z)$ and its derivative $dy_{1...\mathcal{N}}(z)/dz=n_{1...\mathcal{N}}(z)$ at the starting value of the independent variable $z=0$. Also input are the step size to be attempted, \texttt{htry} (we start with $10^6$~cm), the required accuracy, \texttt{eps}, and the vector \texttt{yscal(1:$\mathcal{N}$)} against which the error is scaled to give the desired accuracy $\Delta_0=\texttt{eps}\times\texttt{yscal}_i$. On output, $N$ and $z$ are replaced by their new values. \texttt{hdid} is the step size that was actually accomplished, and \texttt{hnext} is the estimated next step size. By default \Kt{} uses $\texttt{eps}=0.01$ and $\texttt{yscal}_i=|N_i|+|\texttt{htry}\times n_i| + 10^{-21} \mathrm{cm}^{-3}\times\texttt{htry}$ \citep[see][for details]{numericalrecipes}. \footnote{During the first global iteration and for every first spatial step in all subsequent iterations we use a relaxed value of $\texttt{eps}=0.05$.} It is also possible to alter the value of \texttt{htry} depending on local conditions to influence the performance of the iteration in terms of accuracy and total computation time. 

A disadvantage of the adaptive stepping is that testing for global convergence is not straight forward. In contrast to models with fixed spatial grids one cannot easily compare densities and population numbers between each iteration  and check whether a given numerical tolerance is met because the spatial grid changes between iterations. One obvious solution is of course to define a spatial reference grid and interpolate the resulting structure of each iteration to this grid. However, this introduces the additional error of the interpolation to the numerical uncertainty of the solution. We  nevertheless offer this option in a coming version of the code. 

\subsection{Chemical solver details}
\paragraph{Steady-state chemistry}\label{sect:steady-state}
\Kt{} offers a variety of solution approaches to solve Eq.~\ref{eq:newton-raphson} including LU decomposition (Lapack routines \texttt{DGESV} and \texttt{DGESVX}, \citet{lapack}), the minimum norm-solution to a real linear least squares problem: $\min ||\vec{\mathcal{F}} - \mathbb{Q}\cdot\vec{n}||$ using the singular value decomposition (SVD, Lapack routine \texttt{DGELSD}) as well as the linear least squares solver \texttt{DGELS}. Note, that \texttt{DGESV} is the workhorse among these routines and the others provide a fallback option in case the standard approach fails. \footnote{\texttt{DGESV} fails primarily when the matrix $\mathbb{Q}$  is rank deficient, which is usually an effect of a stiff system of ODEs together with problematic numerics from other parts of the computation.} We summarize the chemical solution in \textbf{Algorithm: Chemistry}~\ref{alg:chemistry_solution}.

\setcounter{algorithm}{0}
\begin{algorithm}
	\floatname{algorithm}{Algorithm: Chemistry}
	 \caption{Solution} \label{alg:chemistry_solution}
	\begin{algorithmic}
		\State $T_g \leftarrow T_\mathrm{g,new}$
		\State $n(i)\leftarrow  n(i)^\mathrm{old}$
		\While{$\mathtt{iteration counter}<600$}\Comment{Newton-Raphson iteration}
		\State compute $\vec{ \mathcal{F}}$
		\State compute $\mathbb{Q}$
		\State apply preconditioning
		\State Solve $\mathbb{Q}\cdot \delta\vec{n}=\vec{\mathcal{F}}$ for $\delta\vec{n}$ \Comment{use e.g. \texttt{DGESV} to invert  $\mathbb{Q}$}
		\State $\mathtt{converged}\leftarrow\mathtt{TRUE}$
		\Foreach{species $i$ }
		\If{$n(i) > 10^{-33}$~cm$^{-3}$}
		\State $\eta_i\leftarrow  \delta n_i/n(i)$
		\If{$\eta_i > 0.05$ \texttt{AND} $n(i) > 10^{-15}$~cm$^{-3}$ }
		\State $\mathtt{converged}\leftarrow$ \texttt{FALSE}
		\EndIf
		\State $n(i)^\mathrm{new}\leftarrow n(i)f_{step}(\eta_i)$ \Comment{see Eq.~\ref{eq:newton_stepper}}
		\Else
		\State  $n(i)^\mathrm{new}\leftarrow 10^{-33}$~cm$^{-3}$
		\EndIf	
		\State OPTIONAL: sanity checks on $n(i)$		
		\EndForeach
		\If{$\mathtt{converged}$ }
		\Return $\vec{n}^\mathrm{new}$ \Comment{Chemistry converged}
		\EndIf
		\EndWhile
		\State Apply fallback strategy, e.g. call \texttt{DLSODES}\Comment{not converged}\\
        \State $\vec{n}^\mathrm{new}\leftarrow \;\vec{n}^\mathrm{old}$	\Comment{not converged}	
	\end{algorithmic}
\end{algorithm}

\paragraph{Time-dependent chemistry} \label{sect:time-dependent}
\Kt{} is by default computing the chemical steady-state solution as explained above. In addition we also implemented the possibility to approximate the local equilibrium solution by a time-dependent solution of the chemistry using a long time $t_\mathrm{equil}$. We use the  DLSODES (double precision) solver from the ODEPACK package \citep{odepack} to solve the system of ODEs assuming the previous chemical solution as initial condition.\footnote{We also implemented the option to use other numerical solvers, such as the DLSODA \citep{odepack} and the DVODPK from the Netlib library dvodpk (\url{http://www.netlib.org/ode/}) and MA28, the advance solver of sparse systems of linear equations  from  the  HSL  library (\url{http://www.hsl.rl.ac.uk/}).}  when the steady-state solution can not be found by the code (see  \textbf{Algorithm: Chemistry}~\ref{alg:chemistry_solution}). This approach makes the code  more stable but reduces the overall performance in terms of computation time. Presently we do not store the individual time steps and only use the solution at the equilibrium time. We are currently implementing a fully time-dependent solution in \Kt.

\subsubsection{Benchmark of the chemical solver}
In this section we  compare how the different approaches to solve the chemical network perform in terms of total CPU time and overall solution accuracy. We solve the 8 benchmark problems from \citet{comparison07} and compare the individual results against each other and against the PDR benchmark results. We use the set of model parameters described in \citet{comparison07} and the same chemical database with the addition of state-to-state formation rates of \ce{CH+} and \ce{SH+} \citep{agundez2010,nagy2013}. Changes to the past \Kt{} setup are the updated UV radiative transfer and dust temperature computation \citep{roellig2013dust}, \ce{H2} self-shielding according to \cite{draine1996}, \ce{CO} self-shielding  according to \citet{visser2009}, updated heating and cooling computations and improved numerics. Surface chemistry is ignored and \ce{H2} formation is simplified as described in \citet{comparison07}.

Table~\ref{tab:solver_performance} shows the total CPU time used by the different model setups for the 8 benchmark problems.\footnote{Computed by the Intel FORTRAN routine \texttt{cpu\_time}. The time returned is summed over all active threads. The result is the sum (in units of seconds) of the current process's user time and the user and system time of all its child processes, if any.} The numbers in parenthesis give the results for runs where the time-dependent solver \texttt{DLSODES} was used as backup of the stationary solver in case the local chemical solution did not converge. 

For the models with fixed temperatures, the choice of the steady-state solver does not significantly affect the total CPU time; all models finish with in about half a minute. The fast model convergence results from removing the requirement of a self-consistent temperature solution for the problem. Fig.~\ref{fig:numscheme} shows that this simplifies the "Local iteration" to solving the chemistry and the energy level populations. As a result the global iteration reaches convergence after about 4 iterations. Computing the time-dependent approximation to the steady-state solution converges also after 4-6 iterations. 
Models F1, F3, and F4 required about $10^4$ total chemical iterations while F2 required $>10^5$ iterations.

The variable temperature models show a larger variation in CPU times. In general, the time to find a stable solution is 10-1000 times longer compared to the "F" models. For almost all problems, allowing the code to use a time-dependent solution as fallback in case of convergence difficulties decrease the total CPU time by a significant factor.  This was the case for the V1 problem using \texttt{DSGESV}. The choice of linear system solver moderately affects the total CPU times.  Using \texttt{DGESVX} yields CPU times up to a factor 2 longer compared to \texttt{DGESV}, except for V1. CPU time consumption of \texttt{DSGESV} is also higher than \texttt{DGESV}. Across all solver, the problem V1 required the most CPU time. The time to immediately compute a time-dependent solution (last column in Table~\ref{tab:solver_performance}) instead of the steady-state solution 
is growing from V1 to V4. For the V4 models  \texttt{DLSODES} is about 20 times slower than the other solvers. We can summarize that the approach of using the \texttt{DLSODES} as fallback solver in case the linear solvers fail is successful and leads to a  stabilization of the system, in particular if one considers the additional non-linearity introduced by the choice of the chemical network, as described in Sect.~\ref{sect:numerical_aspects}. It is also surprising that it is difficult to predict computation times depending on the parameter regime especially across different numerical solvers.

\begin{table}[htb]
	\centering
	\caption{Total PDR benchmark CPU time (in units of s) with different chemical solvers. \label{tab:solver_performance}
	}
	\begin{tabular}{lrrrr}
		\hline \hline
		\vrule width 0pt height 2.2ex
		Model & \texttt{DGESV} & \texttt{DGESVX} & \texttt{DSGESV} & \texttt{DLSODES}\\
		\hline%
		\vrule width 0pt height 2.6ex %
		&\multicolumn{4}{c}{\textbf{Fixed} temperature models}\\
		\cline{2-5}
		\vrule width 0pt height 2.2ex  
		F1  & 21 & 22 & 21 & 129 \\
		($n=10^3,\chi=10$)& (24) & (24) & (23) & \\
		F2 & 34  & 36  & 33  & 131  \\	
		($n=10^3,\chi=10^5$) & (32) & (35) & (30) & \\
		F3 & 9  & 8 & 14  & 90 \\	
		($n=10^{5.5},\chi=10$)& (10) & (10) & (10) & \\
		F4 & 10  & 9 & 9  & 210 \\	
		($n=10^{5.5},\chi=10^5$)& (11) & (11) & (11) & \\
		\hline%
		\vrule width 0pt height 2.6ex %
		&\multicolumn{4}{c}{\textbf{Variable} temperature models}\\
		\cline{2-5}
		\vrule width 0pt height 2.2ex  
		V1 & 17449 & 17090 & $\sim$ & 3285 \\
			($n=10^3,\chi=10$)& (466) & (621) & (492) & \\ 
		V2 & 1897 & 4816 & 3371  & 3743 \\
		($n=10^3,\chi=10^5$)& (514) & (645) & (599) & \\     
		V3 & 1524 & 2488 & 4414 & 9803 \\
		($n=10^{5.5},\chi=10$)& (1686) & (3026) & (1411) & \\     
		V4 & 1664 & 1912 & 1812 & 23592 \\
		($n=10^{5.5},\chi=10^5$)& (179) & (1303) & (1489) & \\              
		\hline
	\end{tabular}
	\tablefoot{The numbers in parenthesis give the results for runs where a time-dependent solver was used as backup approach after the stationary solver failed. The last column corresponds to runs where \texttt{DLSODES} was immediately used instead of a steady-state solver. ($\sim$) indicates jobs stopped after 60 iterations. All computations were done on a 8-core Intel(R) Xeon(R) CPU E5-2630 v3 \@ 2.40GHz with 64 GB memory. No parallelization was applied. Densities $n$ are given in units of $\mathrm{cm}^{-3}$.
	}
\end{table}

\subsection{Numerical aspects of solving the chemical problem \label{sect:numerical_aspects}}
Few publications on numerical details of solving the chemical network in the astrochemical context are available in the scientific literature. \citet{nejad2005} gives an extensive overview over available numerical solvers and methods to improve stability and speed of the solution. An important quantifier is the computative complexity of the solving algorithm, i.e. the number of computational operations necessary to solve the problem. In terms of solving Eq.~(\ref{eq:newton-raphson}), we need to invert $\mathbb{Q}$ in order to determine the new density vector $\vec{n}^{new}$. Inverting a $\mathcal{N}\times \mathcal{N}$ matrix using classical Gaussian elimination can be done in $\mathcal{O}(\mathcal{N}^3)$ steps. Note, that this is also the cost of classically multiplying two $\mathcal{N}\times \mathcal{N}$ matrices. More advanced algorithms can perform matrix multiplications in fewer steps. For example, \citet{strassen1969} presented an algorithm that performs matrix multiplication with complexity $\mathcal{O}(\mathcal{N}^{\log_2 7})$. For large problems this is a non-negligible improvement. 

For the LU decomposition applied in the routines \texttt{DGESV} and \texttt{DGESVX} we find an approximate complexity of $\mathcal{O}(\frac{2}{3}\mathcal{N}^3)$. Please note, that the number of chemical reactions does not influence the time to solve the problem, but it does influence the time to set up the system Eq.~(\ref{eq:chemical_rate_equation}). Since the rate coefficients involve expensive operations such as  exponential functions and fractional powers, increasing the number of reactions might lead to increased CPU running times of the code.  This can be partially compensated by pre-computing and storing computationally expansive expression that don't change during the chemical iteration.

In the following we will discuss various strategies to reformulate the chemical problem in a mathematically equivalent but numerically more favorable way.  

\subsubsection{Preconditioning of the Jacobi matrix}\label{sect:precondLR}
Matrix preconditioning in the context of solving stiff systems of ODEs is a wide and technical topic. \citet{nejad2005} gives an extended discussion on preconditioning. Here, we would like to present a simple yet sometimes helpful preconditioning strategy that can be used in \Kt{} if numerical convergence is problematic, e.g. due to round-off errors. With preconditioning, we mean a transformation of the system from Eq.~\ref{eq:newton-raphson} to the following form \citep{viallet2016}:
\begin{equation}\label{eq:preconditioning}
	-(\mathbb{L}^{-1} \mathbb{Q} \mathbb{R}) (\mathbb{R}^{-1} \delta\vec{n})=(\mathbb{L}^{-1} \mathcal{F})
\end{equation}
where $\delta\vec{n}=(\vec{n}^\mathrm{new}-\vec{n}^\mathrm{old})$ and $\mathbb{L}$ and $\mathbb{R}$ are the left and right preconditioning matrices. Various choices for $\mathbb{L}$ and $\mathbb{R}$  are possible and \Kt{} employs the following definition: $\mathbb{L} = \mathbb{R} = diag(\vec{n}^\mathrm{old})$ where $diag(\vec{n})$ specifies a diagonal matrix with  vector $\vec{n}$ as diagonal elements and all other elements set to zero. In index notation we can write:
\begin{equation}
	\tilde{\mathcal{F}}_j=\mathcal{F}_j/L_j=\mathcal{F}_j/n_j
\end{equation}
with the row index $j$ of the right hand side vector $\mathcal{F}$ and $L_j$ the non-zero element in row $j$ of $\mathbb{L}$. The tilde denotes the preconditioned version of the original vector/matrix. The Jacobi matrix is computed with:
\begin{equation}
	\tilde{Q}_{i,j}=Q_{i,j} R_j/L_i=Q_{i,j} n_j/n_i
\end{equation}
where $R_j$ the non-zero element in row $j$ of $\mathbb{R}$. The new system Eq.~(\ref{eq:preconditioning}) solves for $\mathbb{R}^{-1} \vec{\delta n}$ and we have to multiply the solution with $\mathbb{R}$, i.e. with $\vec{n}^\mathrm{old}$ to get the final solution.

Another strategy  discussed in \citet{nejad2005} is to perform a row-reorder in the system Eq.~(\ref{eq:newton-raphson}). The idea is to make use of the sparsity of the problem,  which is common in astrochemical computations because most chemical species react with few reaction partners and because many reaction rates are effectively zero in many circumstances. It is numerically favorable if large sub-matrices of the Jacobi matrix  are null matrices.  Modern solvers of systems ODEs and linear problems are often optimized to invert sparse matrices efficiently.

If we reorder the rows for example by the descending count of zero elements per row in the Jacobi matrix we get a new matrix with large sub-matrices with zero elements which might improve solution speed. \Kt{} allows to perform a row reordering such that rows of $\mathbb{Q}$ with the most zero elements are at the top and the densest rows at the bottom. The inverse scheme would also be an option and we advice any interested modeler to experiment with either scheme.  

Because of the high magnitude difference in chemical densities of the involved species we find a large variance in the magnitudes of the Jacobi matrix entries. This might reduce efficiency of the solution and re-scaling the system is an option to somewhat assist the algorithms. We offer the option to re-scale every row in Eq.~(\ref{eq:linear_system}) such that the maximum entry in the respective Jacobi matrix row is unity.

In cases where \Kt{} tries to approximate a steady-state solution by solving the time-dependent system for a sufficiently long time $t_\mathrm{equil}$, we apply the additional heuristic scaling factor $1/(10^{-2} t_\mathrm{equil} )$ to the row scaling. We mention this here as experimental approach and invite other groups to perform additional tests.

\paragraph{Results}
A performance analysis of the introduced numerical tweaks in full detail is beyond the scope of this paper. Their performance and their applicability  depends on detailed model parameters. We performed test computations against a reference model of $n_0=10^4~\mathrm{cm}^{-3}$ with pure gas-phase chemistry solving for the steady-state without any preconditioning. The following scenarios have been tested:
\begin{enumerate}
	\item preconditioned Jacobian as described in Sect.~\ref{sect:precondLR}
	\item Jacobian and r.h.s vector re-scaled to unity first AND precondition the re-scaled Jacobian
	\item the (already) preconditioned Jacobian and r.h.s vector re-scaled to unity 
	\item resorted Jacobian with dense rows first 
	\item re-scaled AND resorted Jacobian with dense rows first
	\item re-scaled AND resorted Jacobian with sparse rows first 
	\item adaptive choice of elemental conservation rows in  $\mathbb{Q}$, replacing the most abundant species
	\item adaptive choice of elemental conservation rows in  $\mathbb{Q}$, replacing the least abundant species
\end{enumerate}
 In Table~\ref{tab:precond1} we summarize the number of global iterations for each of the described variations plus the total computation time. The first row shows the reference computation applying no preconditioning. Preconditioning the Jacobian with $\mathbb{L}$ and $\mathbb{R}$ (strategy 1) shows the strongest impact in terms of performance while all others lead to comparable computation times or even a much worse performance. This will also depend on the detailed chemical network used and the parameter regime. It demonstrates though that the computational effort to solve the chemical problem depends strongly on the detailed numerical approach even though they are all mathematically equivalent.
\begin{table}[htb]
	\centering
	\caption{Performance comparison of various preconditioning strategies of the chemical problem}\label{tab:precond1}
	\begin{tabular}{ccr}
		\hline \hline
			\vrule width 0pt height 2.2ex
		strategy & iterations &time [min.]\\
		\hline
			\vrule width 0pt height 2.2ex
		reference & 29 & 52.3\\
		1 & 10 & 15.5  \\
		2 & 22 & 54.1 \\
		3 & 32 & 63.5 \\
		4 & 34 & 196.6 \\
		5 & 15 &  74.2 \\
		6 & 45 & 248.0 \\
		7 &  \multicolumn{2}{c}{not converged} \\		
		8 & 29 & 46.6 \\ \hline
	\end{tabular}
\end{table}
From the CPU times shown in Table~\ref{tab:precond1} we conclude that the LR-preconditioning outlined in Sect.~\ref{sect:precondLR} performs best. Currently, the standard setup in \Kt{} is no preconditioning until a more systematic parameter study for typical PDR model parameter ranges has been performed.

\section{Newton-Raphson stepping}\label{sect:app-newton}
\begin{figure}
		\resizebox{\hsize}{!}{\includegraphics{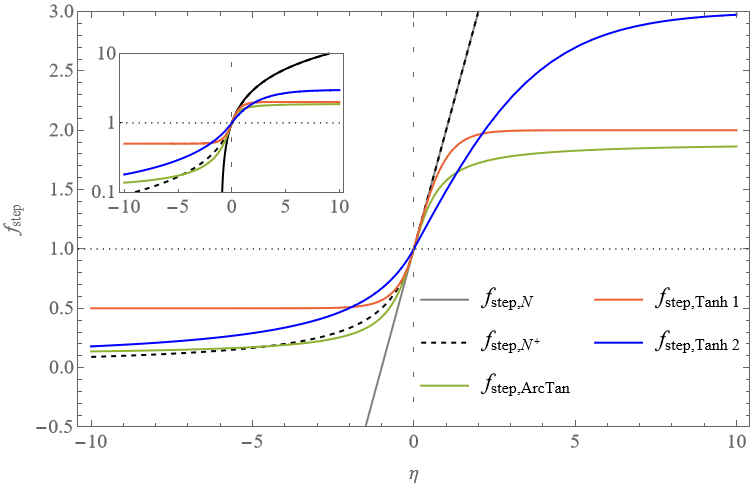}}
		\caption{Comparison of various stepper functions. The inset shows the vertical axis in log-space. $f_\mathrm{step,Tanh 1}$ assumes ($\omega_-=\omega_+=2, \lambda=1$), $f_\mathrm{step,Tanh 2}$ assumes ($\omega_-=20, \omega_+=5, \lambda=0.5$).  }
		\label{fig:stepper_1}
	\end{figure}

Historically, \Kt{} used the following ArcTan stepping:
\begin{equation}\label{eq:damping_arctan}
	f_\mathrm{step,A}(\eta)=1 + \lambda \frac{2}{\pi} \arctan\left(\frac{\pi}{2}\eta \right)
\end{equation}
with $\lambda=0.9$. The scaling by $(2/\pi)$ limits the second term to the range $\pm 1$ and the pre-factor of $\lambda=0.9$ limits $f_\mathrm{step,A}$ to the range $0.1,...,1.9$ and dampens the slope around the origin to 0.9. The stepper in Eq.~\ref{eq:damping_arctan} has several advantages over the linear stepper in Eq.~\ref{eq:newton_stepper}: it prohibits steps into the negative domain and ensures positive densities; it is almost Newtonian for small relative steps $\eta$ and therefore converges fast once it is sufficiently close to the root; it prohibits too wide and too small steps because $1-\lambda \le f_\mathrm{step,A}(\eta)\le 1+\lambda\; \forall \eta$ (see Fig.\ref{fig:stepper_1}).

An alternative approach is a modification of Eq.~\ref{eq:newton_stepper} to only prevent Newton steps through the negative domain:
\begin{equation}\label{eq:positive_newton_stepper}
	f_\mathrm{step,N+}(\eta)= 
	\begin{cases}
		(1-\eta)^{-1}  & \text{if $\eta<0$} \\
		1+\eta & \text{if $\eta\ge 0$} 
	\end{cases} 
\end{equation}
By construction $f_\mathrm{step,N+}(\eta)=f_\mathrm{step,N+}^{-1}(-\eta)$ guarantees a positive stepping factor.  Eq.~\ref{eq:positive_newton_stepper} converges to zero for large negative steps $\eta \ll -1$.
It is also symmetric in the log-domain, e.g. for $\eta=1$ we find $n(i)^\mathrm{new}=2\times n(i)^\mathrm{old}$, and conversely $\eta=-1$ returns $n(i)^\mathrm{new}=n(i)^\mathrm{old}/2$. 
However, this stepper diverges for large positive steps, $\eta \gg 1$ and the symmetry between negative and positive steps may lead to oscillations around the actual solution.

Inheriting from both approaches we construct a more general stepping function  $f_\mathrm{step,Tanh}$ (see Eq.~\ref{eq:damping_symlogtanh}) whose symmetry and limits can be controlled with few numerical parameters.
The min/max values of $f_\mathrm{step,Tanh}$ are given by 
\begin{equation}\label{eq:limits_symlogtanh}
	\lim f_\mathrm{step,Tanh}(\eta)= 
	\begin{cases}
		\left(1+\lambda(\omega_--1)\right)^{-1}	& \text{if $\eta<0$} \\
		1+\lambda(\omega_+-1)   & \text{if $\eta\ge 0$} 
	\end{cases}
\end{equation}

Consequently, $\omega_+$ controls the maximum factor to increase $n(i)^\mathrm{old}$ and $\omega_-$ controls the minimum factor for decrease. Choosing slightly different values for $\omega_+$ and $\omega_-$ avoids oscillations in symmetric problems.

The slope $\partial f_\mathrm{step,Tanh}/\partial \eta$ for small $\eta$, i.e. close to the solution,  is given by $\lambda$. A value close to unity guarantees the quadratic convergence behavior of the Newton-Raphson method for small values of $\eta$. However, far from the minimum of a function $f$, a full Newton-Raphson step not necessarily decreases the function. We only know that the stepping direction initially decreases $f$. Reducing the step width but a damping parameter $\lambda < 1$ then limits non-convergent steps\footnote{We limit ourselves to non-adaptive dampening strategies here. However, adaptive dampening strategies such as line searches and backtracking may further improve global convergence performance.}. 

Figure~\ref{fig:stepper_1} compares the discussed $f_\mathrm{step}$ variants. The log-plot inset shows the symmetry of $f_\mathrm{step,Tanh}$ that can be controlled by with $\omega_+$ and $\omega_-$. We see how the Tanh remains closer to the linear Newton behavior for larger $\eta$ compared to the ArcTan. Note, that the ArcTan stepper $f_\mathrm{step,A}$ can be reasonably well approximated with $\omega_-=11$, $\omega_+=2$, and $\lambda=0.9$.

\subsection{Choosing efficient stepping strategies}\label{sect:efficient_stepping}
Ideally, one would like to find a single optimal stepping strategy to reliably solve the Newton-Raphson scheme. In the numerical PDR context this turns out to be a demanding task. We tested a large variety of different steppers and compound stepping strategies. We computed a small grid of models with 107 species including surface species (see Table~\ref{tab:surface_chemistry_test}). The model density (constant), mass and FUV field were varied on the following regular grid: $n=(10^2,10^4,10^6)$~$\mathrm{cm^{-3}}$, $M=(50,100)$~$\mathrm{M_\odot}$, and $\chi=(1,10^3,10^6)$~$\chi_\mathrm{Draine}$. We set $\zeta_\mathrm{CR}=5\times 10^{-17}$~$\mathrm{s^{-1}}$ and assume $R_V=3.1$. The maximum number of Newton-Raphson steps is set to 600 per solution attempt. Global convergence was determined as described in Sect.\ref{sect:global_convergence}.
	
At first, we test the performance of different choices of ($\omega_-,\omega_+,\lambda$) of the Tanh stepper and compare it against the old stepper $f_\mathrm{step,A}$. 
Table~\ref{tab:bench-steps} shows the total count of Newton-Raphson steps per computed model. Each row corresponds to a particular stepper choice. $f_\mathrm{step,A}$ is the standard ArcTan stepping. The last seven rows correspond to a compound strategy that is discussed below. Models that failed to converge and aborted are marked with a dash. Reasons for abortion are \texttt{NaN}  values in the chemical solution or an array overflow in case of requesting more than 1900 spatial grid points because of too steep chemical gradients.

For some settings we also allow for a more aggressive Newton-Raphson approach where we keep the chemical vector after the final iteration even if the convergence criterion for the local chemistry was not met (indicated by the subscript $()_\mathrm{new}$ in Table~\ref{tab:bench-steps}). This is a  fragile approach because it may fail if the Newton-Raphson stepping ends  far from the true solution, but it may speed up convergence if the steps approached the true solution. The standard approach is to discard the (not converged) chemical vector and use the previous solution. This ensures that a 'true' solution of the chemistry is used.

There is no superior stepper and their individual performance depends on many details of the model numerics and the model parameters. We found that asymmetric choices $\omega_+<\omega_-$  perform better than symmetric setups. Note, that $\omega_+>\omega_-$ is generally not recommended, e.g. \Kt{} frequently failed to finish computation because of density overflow in the case of $\omega_+=5, \omega_-=2$. The  push towards larger densities leads to numerically unstable behavior. The general trend shown in Table~\ref{tab:bench-steps} is the number of steps growing with model density and with model FUV strength. There is no clear correlation with the model mass. 

As $\lambda$ determines the slope of the stepping function for  small arguments one would assume that solutions are found more quickly for $\lambda$ closer to unity. However, the values from Table~\ref{tab:bench-steps} do not generally confirm this assumptions, we find many counterexamples.  At first glance the total number of steps shows no clear correlation with $\lambda$ but we note that missing models distort the row total, e.g. $(10,5,0.5)$ and $(10,5,0.6)$ did not converge for two high density models each. These models typically took significant number of steps to finish and are not included in the row total. When approximately accounting for the missing models there is a slight performance improvement with increasing $\lambda$, excluding the case of $\lambda=1$. Generally, symmetric choices of $\omega_-=\omega_+$ tend to perform worse than runs with $\omega_->\omega_+$.  

A general conclusion from Table~\ref{tab:bench-steps} is that for low to intermediate densities the details of the Newton-Raphson stepping are hardly affecting the number of steps with the exception of a few cases that failed to converge. The high density model performance shows a much higher variance with stepping, with some models performing up to 10 times worse than others and some even more extreme cases.

Computations where we did not discard the density vector after the last Newton-Raphson step even if convergence was not yet reached (subscript $()_\mathrm{new}$)
led to more non-converging models overall but successfully reached convergence in some cases where the standard approach failed. In particular the choice of $\omega_-=11$, $\omega_+=2$, and $\lambda=0.9$ seems to perform better in the $()_\mathrm{new}$ case.
	
	\begin{table*}[htb]
		\centering 
		
		\caption{Total Newton-Raphson step count for different stepper choices.  \label{tab:bench-steps}
		}
		\resizebox{\textwidth}{!}{%
			\begin{tabular}{cccccccccccccccccccc}
				\hline \hline
				\vrule width 0pt height 2.2ex
				density $n$
				& 
				$10^2$   &$10^2$   & $10^2$ &
				$10^2$ &$10^2$ & $10^2$ & 
				$10^4$   &$10^4$   & $10^4$ &
				$10^4$ &$10^4$ & $10^4$ & 
				$10^6$   &$10^6$   & $10^6$ &
				$10^6$ &$10^6$ & $10^6$ & 
				total
				\\
				mass $M$ & 
				$50$   &$50$   & $50$ &
				$10^3$ &$10^3$ & $10^3$ & 
				$50$   &$50$   & $50$ &
				$10^3$ &$10^3$ & $10^3$ & 
				$50$   &$50$   & $50$ &
				$10^3$ &$10^3$ & $10^3$ & 
				steps
				\\
				FUV field $\chi$		 & 
				$1$   &$10^3$   & $10^6$ &
				$1$   &$10^3$   & $10^6$ & 
				$1$   &$10^3$   & $10^6$ &
				$1$   &$10^3$   & $10^6$ & 
				$1$   &$10^3$   & $10^6$ &
				$1$   &$10^3$   & $10^6$ & 
				 
				\\
				\hline%
				\vrule width 0pt height 2.2ex  	
				
				$f_\mathrm{\text{step},A}$ & \text{82.4K} & \text{68K} & \text{922K} & \text{95.1K} & \text{171K} & \text{264K} & \text{8.17M} & \text{246K} & \text{304K} & \text{4.81M} & \text{4.71M} & \text{354K} & \text{4.85M} & \text{11.3M} & \text{29.7M} & \text{6.38M} &
				\text{9.31M} & \text{22.7M} & \text{104M} \\
				\text{(11,2,0.9)} & \text{84.1K} & \text{66.8K} & \text{360K} & \text{97K} & \text{175K} & \text{569K} & \text{12.3M} & \text{167K} & \text{263K} & \text{2.72M} & \text{1.17M} & \text{239K} & \text{7.86M} & \text{11.1M} & \text{21.5M} & \text{5.47M} &
				\text{8.52M} & \text{16.1M} & \text{88.7M} \\
				\text{(10,5,0.5)} & \text{130K} & \text{85.3K} & \text{123K} & \text{139K} & \text{320K} & \text{142K} & \text{6.23M} & \text{444K} & \text{459K} & \text{6.59M} & \text{697K} & \text{1.03M} & \text{44.6M} & \text{11.3M} & - & - & \text{9.55M} & \text{12.8M} &
				\text{94.6M} \\
				\text{(10,5,0.6)} & \text{116K} & \text{93.9K} & \text{98.3K} & \text{120K} & \text{268K} & \text{648K} & \text{8.16M} & \text{724K} & \text{389K} & \text{3M} & \text{3.59M} & \text{345K} & - & \text{8.44M} & \text{10.2M} & - & \text{8.91M} & \text{13M} &
				\text{58.1M} \\
				\text{(10,5,0.7)} & \text{99.3K} & \text{113K} & \text{127K} & \text{157K} & \text{219K} & \text{133K} & \text{6.66M} & \text{783K} & \text{551K} & \text{2.38M} & \text{1.84M} & \text{525K} & \text{6.81M} & \text{7.37M} & \text{9.94M} & \text{12.3M} &
				\text{5.91M} & \text{14.1M} & \text{70M} \\
				\text{(10,5,0.8)} & \text{96.2K} & \text{382K} & \text{117K} & \text{101K} & \text{956K} & \text{549K} & \text{1.17M} & \text{610K} & \text{352K} & \text{6.42M} & \text{1.08M} & \text{245K} & \text{10.6M} & \text{7.5M} & \text{8.39M} & \text{20.1M} &
				\text{7.26M} & \text{11.8M} & \text{77.7M} \\
				\text{(10,5,0.9)} & \text{83.6K} & \text{169K} & \text{505K} & \text{96.6K} & - & \text{377K} & \text{1.45M} & \text{698K} & \text{364K} & \text{2.37M} & \text{2.67M} & \text{367K} & \text{11.5M} & \text{9.17M} & \text{9.8M} & \text{10.6M} & \text{12.4M} &
				\text{8.46M} & \text{71.1M} \\
				\text{(10,5,1.0)} & \text{249K} & \text{89.1K} & \text{622K} & \text{106K} & \text{202K} & - & \text{1.29M} & \text{272K} & \text{244K} & \text{1.54M} & \text{1.16M} & \text{250K} & \text{21.2M} & \text{12M} & \text{15.5M} & \text{8.14M} & \text{7.49M} &
				\text{21.5M} & \text{91.9M} \\
				\hline%
				\vrule width 0pt height 2.2ex 
				$\text{(11,2,0.9})_{\text{new}}$ & \text{84.1K} & \text{66.8K} & \text{348K} & \text{97K} & \text{175K} & \text{272K} & \text{4.21M} & \text{167K} & \text{261K} & \text{9.22M} & \text{692K} & \text{474K} & \text{15.1M} & \text{13.3M} & \text{22.4M} &
				\text{14.7M} & \text{9.13M} & \text{15.8M} & \text{107M} \\
				$\text{(10,5,0.5})_{\text{new}}$ & \text{130K} & \text{75.9K} & \text{108K} & \text{139K} & \text{320K} & \text{115K} & \text{6.23M} & \text{444K} & \text{358K} & \text{6.59M} & \text{588K} & \text{1.03M} & \text{44.6M} & \text{30M} & - & - & \text{9.55M} &
				\text{19.2M} & \text{120M} \\
				$\text{(10,5,0.6})_{\text{new}}$ & \text{116K} & \text{93.9K} & \text{89.7K} & \text{120K} & \text{253K} & \text{350K} & \text{8.16M} & \text{724K} & \text{389K} & \text{3M} & \text{1.4M} & \text{561K} & - & \text{53.6M} & \text{10.8M} & - & \text{7.9M} &
				\text{13.6M} & \text{101M} \\
				$\text{(10,5,0.7})_{\text{new}}$ & \text{99.3K} & \text{115K} & \text{119K} & \text{157K} & \text{246K} & - & \text{4.56M} & \text{783K} & - & \text{8.06M} & \text{7.57M} & \text{457K} & \text{40.8M} & - & \text{10.7M} & \text{12.3M} & \text{16.2M} &
				\text{12.2M} & \text{114M} \\
				$\text{(10,5,0.8})_{\text{new}}$ & \text{96.2K} & \text{320K} & \text{76.4K} & \text{101K} & \text{955K} & \text{549K} & \text{1.17M} & \text{610K} & \text{684K} & \text{6.42M} & - & \text{363K} & - & \text{31.5M} & \text{8.39M} & \text{20.1M} & \text{6.78M} &
				- & \text{78.1M} \\
				$\text{(10,5,0.9})_{\text{new}}$ & \text{83.6K} & \text{173K} & \text{469K} & \text{96.6K} & - & \text{303K} & \text{1.23M} & \text{698K} & \text{419K} & \text{1.21M} & \text{7.13M} & \text{261K} & - & \text{17M} & \text{10.1M} & \text{18.8M} & \text{12.4M} &
				\text{8.1M} & \text{78.4M} \\
				$\text{(10,5,1.0})_{\text{new}}$ & \text{249K} & \text{76.8K} & \text{470K} & \text{106K} & \text{182K} & \text{655K} & \text{986K} & \text{272K} & \text{280K} & \text{1.36M} & - & \text{178K} & - & \text{12M} & \text{15.5M} & - & \text{7.49M} & \text{21.5M} &
				\text{61.4M} \\
				\hline%
				\vrule width 0pt height 2.2ex 
				\text{(20,5,0.5)} & \text{131K} & \text{69.7K} & \text{71.5K} & \text{130K} & \text{237K} & \text{87.3K} & \text{3.49M} & \text{520K} & \text{350K} & \text{15.4M} & \text{754K} & \text{493K} & \text{8.12M} & \text{10.2M} & \text{6.84M} & \text{4.8M} &
				\text{10.2M} & \text{9.98M} & \text{71.9M} \\
				\text{(20,5,1.0)} & \text{143K} & \text{64.1K} & \text{502K} & \text{91.3K} & \text{172K} & \text{317K} & \text{2.92M} & \text{317K} & \text{284K} & \text{8.89M} & \text{2.77M} & \text{302K} & \text{6.96M} & \text{8.02M} & \text{13.8M} & \text{92M} &
				\text{5.79M} & \text{12M} & \text{155M} \\
				\text{(2,5,0.5)} & \text{586K} & \text{286K} & \text{143K} & \text{594K} & \text{813K} & \text{525K} & - & \text{1.24M} & \text{553K} & - & \text{67.9M} & \text{791K} & - & \text{12.6M} & \text{23.8M} & - & \text{11.5M} & \text{18.9M} & \text{140M} \\
				\text{(2,5,1.0)} & \text{1.02M} & \text{800K} & \text{650K} & \text{1.76M} & \text{2.25M} & - & \text{8.93M} & \text{1.5M} & \text{341K} & \text{11.6M} & \text{10.8M} & \text{273K} & - & \text{12.5M} & \text{17.1M} & \text{13.3M} & \text{87.1M} & - &
				\text{170M} \\
				\text{(3,3,0.3)} & \text{460K} & \text{421K} & \text{318K} & \text{2.05M} & \text{1.13M} & \text{585K} & \text{83.3M} & \text{10.4M} & \text{846K} & \text{74.3M} & - & \text{582K} & \text{141M} & \text{19.9M} & \text{10.9M} & \text{181M} & \text{23.6M} &
				\text{14.9M} & \text{565M} \\
				\text{(3,3,1.0)} & \text{289K} & \text{158K} & \text{362K} & \text{227K} & \text{755K} & \text{343K} & \text{26.4M} & \text{1.46M} & \text{276K} & - & - & \text{294K} & \text{109M} & \text{11.3M} & \text{34.8M} & \text{313M} & \text{8.62M} & \text{31.9M} &
				\text{538M} \\
				\text{(5,2,0.5)} & \text{135K} & \text{69.5K} & \text{107K} & \text{140K} & \text{234K} & \text{61.2K} & \text{14M} & \text{515K} & \text{333K} & \text{20.9M} & \text{1.6M} & \text{815K} & - & \text{9.55M} & \text{39.7M} & \text{12.8M} & \text{8.86M} &
				\text{44.1M} & \text{154M} \\
				\text{(5,2,1.0)} & \text{86.1K} & \text{61.1K} & \text{743K} & \text{91.2K} & \text{174K} & \text{501K} & \text{3.39M} & \text{273K} & \text{510K} & \text{4.4M} & \text{4.84M} & \text{327K} & \text{106M} & - & \text{17.3M} & \text{11.3M} & \text{6.52M} &
				\text{16.8M} & \text{173M} \\
				\text{(5,5,0.5)} & \text{1.14M} & \text{707K} & \text{215K} & \text{786K} & \text{1.32M} & \text{552K} & \text{39.6M} & \text{765K} & \text{763K} & - & \text{5.57M} & \text{1.08M} & \text{266M} & \text{51.5M} & \text{11.6M} & \text{242M} & \text{11.8M} &
				\text{15.2M} & \text{650M} \\
				\text{(5,5,1.0)} & \text{720K} & \text{314K} & \text{228K} & \text{282K} & \text{1.15M} & \text{203K} & - & \text{729K} & \text{294K} & \text{4.34M} & - & \text{320K} & \text{281M} & \text{12.7M} & - & \text{276M} & \text{13M} & - & \text{591M} \\
				\hline%
				\vrule width 0pt height 2.2ex 
				\text{adaptive 1} & \text{83.7K} & \text{68.2K} & \text{1.12M} & \text{98.4K} & \text{171K} & \text{377K} & \text{6.11M} & \text{187K} & \text{268K} & \text{4.65M} & \text{2.45M} & \text{211K} & \text{3.5M} & \text{6.65M} & \text{6.17M} & \text{3.39M} &
				\text{6.05M} & \text{7.32M} & \text{48.9M} \\
				 \hline
			\end{tabular}
		}
		\tablefoot{The Tanh stepper $f_\mathrm{step,Tanh}$ is tested with different values ($\omega_-,\omega_+,\lambda$) as indicated in the first column. The subscript $()_\mathrm{new}$ indicates no reset to previous density solution if convergence was not met. See text for details. Dashes mark models that did not converge at all. The last column gives the row total excluding not converged models. (Large numbers are abbreviated with K and M indicating $10^3$ and $10^6$, respectively).}
	\end{table*}

For a given set of PDR parameters it is possible to find an optimized set of stepping parameters ($\omega_-,\omega_+,\lambda$) that will minimize the computing time.  Changing the PDR parameters  may significantly change the topography of the Newtonian vector field $\delta \vec{n}$ and consequently may require a modified stepping strategy to succeed.	
To mitigate these complications we tested a compound strategy, named  \textbf{adaptive~1} in Table~\ref{tab:bench-steps}. In this approach we change the stepping parameters along the Newton search to find a chemical solution. Our general approach is to use a relatively robust  stepping strategy for the first 100 Newton-Raphson steps. Typically, $99.99\%$ of all solution attempts require less than 30-40 steps. If the standard stepper is not successful after 100 steps, it will most likely not converge at all, e.g. because of oscillating or global non-convergent behavior. 
To dampen oscillations we then progressively reduce step widths. First we reduce $\omega_-$ for the next 100 steps, followed by stronger dampening by reducing $\lambda$. If the more careful stepping is unable to escape a local minimum we allow for a more aggressive stepping, e.g. by allowing for larger values of $\omega_{+,-}$ or even $\omega_-<\omega_+$.  Algorithm  \textbf{adaptive~\ref{alg:adaptive_1}} shows the details of the  adaptive strategy.  

\setcounter{algorithm}{0}
\begin{algorithm}
    \caption{}
	\label{alg:adaptive_1}
	\begin{algorithmic}
		\State {start from $\vec{n}_0$ }
		\If  {\# steps $ \leq 100$}
		\State use 	$ f_\mathrm{step,A}$
		\ElsIf {$101 \leq $ \# steps $\leq 200$}
		\State  $\vec{n}_{101} \gets \vec{n}_o$ \Comment{start again from $\vec{n}_0$ }
		\State use 	$ f_\mathrm{step,Tanh}$ with $(\omega_-=5, \omega_+=2,\lambda=1.0$)
		\ElsIf {$201 \leq $ \# steps $\leq 400$}
		\State use 	$ f_\mathrm{step,Tanh}$ with $(\omega_-=5, \omega_+=2,\lambda=0.5$)
		\Else  
		\State  $\vec{n}_{401} \gets \vec{n}_o$ \Comment{start again from $\vec{n}_0$ }
		\State use 	$ f_\mathrm{step,Tanh}$ with $(\omega_-=5, \omega_+=5,\lambda=1.0$)
		\EndIf
	\end{algorithmic}
\end{algorithm}
	
 In its final stepping attempt  \textbf{adaptive~1} applies  a more aggressive stepper setup. In addition, it restarts the root search from the initial density after 100 and again after 400 steps.
	
The large number of Newton steps computed in some models highlights the importance of the stepper choice.  Comparing the adaptive and the single stepper strategies shows better performance for the adaptive strategy across all PDR parameters. \textbf{adaptive~1} performs best with 1/2  total Newton steps for the whole grid compared to the reference stepper $f_{step,A}$. From Table~\ref{tab:bench-steps} we conclude that  \textbf{adaptive~1} is the most successful algorithm.

\section{Chemical network}\label{sect:chem_appendix}

\subsection{Gas-phase chemistry}
The gas-phase chemistry in \Kt{} is based upon the UDfA2012 database with updates listed in Table~\ref{tab:chemical_updates}. It contains the following reaction types:
For two-body reactions the Arrhenius form for the rate coefficient is
\begin{equation}
	k=\alpha \,\mathrm{cm^3s^{-1}}\,\left(T/300K\right)^\beta\exp\left(-\gamma /T\right) \; 
\end{equation}
with the parameters $\alpha, \beta, \gamma$ tabulated in the database.
UDfA12 distinguished between the following two-body reactions: neutral-neutral, ion-neutral, charge exchange, ion-ion neutralization, dissociative recombination, radiative recombination, associative electron detachment and radiative association. In addition it contains two direct photo-processes: photo-dissociation and photo-ionization both described by:
\begin{equation}\label{eq:photo_reaction}
	k=f_{ss}\mathcal{P}_0\exp\left(-\gamma A_V\right)
\end{equation}
with the  photo-rate in the unshielded interstellar ultra-violet radiation field $\mathcal{P}_0$, the unit-less (self-)shielding factor $f_{ss}$, and the extinction by interstellar dust at visible wavelengths $A_V$.

\begin{table}[ht]
\caption{\label{tab:chemical_updates} Changes and addition to the UDfA12 reaction network}
\resizebox{\columnwidth}{!}{%
\begin{tabular}{lrrrl}
		\hline\hline
		\vrule width 0pt height 2.2ex
	reaction&\textalpha&\textbeta&\textgamma&Ref\\
&$(\;\times10^{-10})$&&\\
	\hline
	\vrule width 0pt height 2.6ex
\ce{CS2 + $h\nu$ -> CS2+ + e- } &$1.74$&0.00&3.6&(1),(2)\\ 
\ce{N2O + $h\nu$ -> N2O+ + e- } &$1.70$&0.00&4.1&(3) \\	
\ce{H3+ + F     -> H2F+ + H}    &$4.80$&0.00&0.0& (4) \\
\ce{CH3O + O    -> OH + H2CO}   &0.06&0.00&0.0& (5) \\
\ce{CH3O + O    -> O2 + CH3}    &0.190&0.00&0.0& (5) \\
\ce{CH3OH + O   -> OH + CH3O}   &0.166&0.00&2.4& (6) \\
\ce{CH3   + OH  -> H + CH3O}    &$2.0\times 10^4$ & 0.00 & 6990.0& (7) \\   
\ce{CH3OH + OH  -> H2O + CH3O}  &0.006 &-1.02&0.0& (5) \\
\ce{CH3O  + H2O -> OH + CH3OH}  &$3.88\times 10^4$ &3.80&5790.0& (6) \\
\ce{CH3O  + CO  -> CO2 + CH3 }  &0.261 &0.00&5940.0& (6) \\
\ce{CH3O  + HCO -> CO + CH3OH}  &1.5 &0.00&0.0& (6) \\
\ce{CH3O  + H   -> H2 + H2CO }  &0.3 &0.00&0.0& (5) \\
\ce{CH3O  + H   -> OH + CH3 }   &0.03 &0.00&0.0& (5) \\
\ce{CH3O  + CH2 -> CH3 + H2CO}  &0.003 &0.00&0.0& (6) \\
\ce{CH3O  + CH3 -> CH4 + H2CO}  &0.4 &0.00&0.0& (6) \\
\ce{CH3O  + C2H3 -> C2H4 + H2CO}&0.4&0.00&0.0& (6) \\
\ce{CH3OH + H   -> H2 + CH3O}   &0.664 &0.00&3070.0& (6) \\
\ce{CH3OH + CH2 -> CH3 + CH3O}  &$1.14\times 10^{-5}$&3.10&3490.0& (6) \\
\ce{CH3OH + CH3 -> CH4 + CH3O}  &$1.18\times 10^{-4}$&3.45&4020.0& (6) \\
\ce{CH3O  + NO  -> HNO + H2CO}  &0.023 & -0.70&0.0& (8) \\
\ce{CH3O  + C   -> CO + CH3 }   & 3.0 &0.00&0.0& (5) \\

\hline
\end{tabular}	
}
\tablebib{
	(1) {\citet{grosch2015}};
	(2) {\citet{keller2013}}; 
	(3) {\citet{cook1968}}; 
	(4) {\citet{neufeld2005}}; 
	(5) {\citet{ruaud2015}};
    (6) {\citet{hebrard2009}};
    (7) {\citet{baulch2005}};
    (8) {\citet{atkinson2006}}
    }
\end{table}

Direct cosmic-ray ionization is described by 
\begin{equation}
	k=\zeta_\mathrm{H_2} 
\end{equation}
and cosmic-ray induced photo-processes with a rate
\begin{equation}
	k=\zeta_\mathrm{H_2}\left(T/300K\right)^\beta\gamma/(1-\omega) 
\end{equation}
where $\zeta_\mathrm{H_2}$ is the direct CR ionization rate per \ce{H2}, $\gamma$ describes the probability per CR ionization that a photo-process takes place and $\omega$ is the dust grain albedo in the FUV \citep{udfa12}. We assume $\omega=0.5$.

For reactions with isotopologues we take the same rate coefficients as for the standard isotopologue with the addition of the fractionation reactions listed in Table~\ref{tab:chemical_updates2}. Reaction rates for the reactions \ce{c-C3H2 + HE+},  \ce{c-C3H + C+}, \ce{c-C3H + H+}, and \ce{c-C3H + He+} have been replace with rate coefficients from KIDA \citep{kida}. New reactions involving the linear isomers \ce{l-C2H3+}, \ce{l-C3H2+}, \ce{l-C3H2}, \ce{l-C3H} have also been taken from KIDA.

\begin{table}[htb]
	\setlength{\tabcolsep}{4pt}
\centering

\caption{\label{tab:chemical_updates2} Updated fractionation reactions}
\resizebox{\columnwidth}{!}{%
\begin{tabular*}{\columnwidth}{lccc}
		\hline\hline
    \vrule width 0pt height 2.2ex
	reaction&\textalpha&\textbeta&\textgamma\\
	&$(\;\times10^{-10})$&&\\
	\hline
    \vrule width 0pt height 2.6ex 
	\ce{^{13}C+ + CO ->  C+  +  ^{13}CO }& $4.42$& -0.29& 0.0	 \\
	\ce{C+ + ^{13}CO    ->   ^{13}C+ +  CO} & $4.42$& -0.29&34.5	 \\
	\ce{^{13}C+ +  C^{18}O ->  ^{13}C+ + C^{18}O} & $ 4.42$& -0.29& 0.0	 \\
	\ce{C+ + ^{13}C^{18}O   -> ^{13}C+ +  C^{18}O } &  $ 4.42$& -0.29&35.4	 \\
	\ce{HCO+ +  ^{13}CO    ->   H^{13}CO+ + CO} & $2.83$& -0.26& 0.0	 \\
	\ce{H^{13}CO+ + CO  ->  HCO+ + ^{13}CO}&$2.83$& -0.26&17.8	 \\
	\ce{HC^{18}O+ + ^{13}C^{18}O    -> H^{13}C^{18}O+  + C^{18}O  } &  $2.83$& -0.26& 0.0	 \\
	\ce{H^{13}C^{18}O+ +  C^{18}O ->  HC^{18}O+ + ^{13}C^{18}O } & $2.83$& -0.26&17.8	 \\
	\ce{HCO+ +  C^{18}O ->  HC^{18}O+ + CO    } & $ 2.81$& -0.29& 0.0	 \\
	\ce{HC^{18}O+ + CO  ->  HCO+  + C^{18}O } & $  2.81$& -0.29& 6.4	 \\
	\ce{H^{13}CO+ + ^{13}C^{18}O   ->  H^{13}C^{18}O+ +  ^{13}CO  } &  $2.81$& -0.29& 0.0  \\
	\ce{H^{13}C^{18}O+  + ^{13}CO ->  H^{13}CO+ + ^{13}C^{18}O}&   $2.81$& -0.29& 6.4	 \\
	\ce{HCO+  + ^{13}C^{18}O  -> H^{13}C^{18}O+ +  CO   } &   $3.14$& -0.27& 0.0	 \\
	\ce{ H^{13}C^{18}O+ +  CO ->   HCO+ +  ^{13}C^{18}O} &  $3.14$& -0.27&24.2	 \\
	\ce{H^{13}CO+ + C^{18}O -> HC^{18}O+ + ^{13}CO } &   $2.87$& -0.22& 0.0  \\
	\ce{HC^{18}O+ +  ^{13}CO ->  H^{13}CO+ + C^{18}O } &   $2.87$& -0.22&11.4	 \\	
	\hline
\end{tabular*}
}
\tablebib{
\citet{mladenovic2014}}
\end{table}

\subsection{Surface chemistry}\label{sect:surface_appendix}

We follow the rate equation approach as described in \citet{hasegawa1992, hasegawa1993} and account for competing surface processes \citep{garrod2011} and chemically inactive bulk ice. The species on the surface need to be mobile in order to scan the surface for a suitable reaction partner. The thermal hopping time scale to move between adjacent binding sites on the surface is given by \citet{hasegawa1992} as
\begin{equation}\label{eq:hopping_time}
	t_\mathrm{hop,i}=\nu_{0,i}^{-1}\exp\left(\ E_{B,i}/T_d \right) 
\end{equation} 
with the characteristic vibration frequency for the adsorbed species
\begin{equation}\label{eq:characteristic_frequency}
	\nu_{0,i}=(2n_\mathrm{site}k_B E_{D,i}/\pi^2 m_{i})^{1/2} 
\end{equation}
where $n_\mathrm{site}$ is the surface density of binding sites and $m$ the  mass of the particle and $k_B$ is the Boltzmann constant. We assume $n_\mathrm{site}=1.5\times10^{15} \mathrm{cm^{-2}}$ \citep{tielens1987}. $E_{D,i}$ is the desorption energy for physical desorption, $E_{B,i}$ is the energy barrier between binding sites. Note, that in this section all energies $E_D,\,E_B,\,E_a$ are given in units of K. Following \citet{hasegawa1992} we assume $E_{B,i}\approx0.3 E_{D,i}$.  For \ce{H} and \ce{H2} tunneling might also be a more rapid migration process compared to thermal hopping with a tunneling time:
\begin{equation}\label{eq:tunneling_time}
	t_\mathrm{tunnel,i}=\nu_{0,i}^{-1}\exp\left[(2a/\hbar)(2 m_i k_B E_{B,i})^{1/2}\right]\;\; \mathrm{s}
\end{equation} 
where  $a=2 \text{\AA}$ is the energy barrier width assuming a rectangular barrier \citep{garrod2011}. Note, that Eq.~\ref{eq:tunneling_time} is not valid for particles heavier than \ce{H2}.  The time for an adsorbed particle to sweep over  a number of binding sites is $t_\mathrm{diff,i}$ and the diffusion rate $R_\mathrm{diff,i}$ is defined as the inverse of the diffusion time $t_\mathrm{diff,i}=N_S\times max (t_\mathrm{hop,i}, t_\mathrm{tunnel,i})\;[\mathrm{s}]$. $N_S$ is the total number of surface sites per grain. The surface reaction rate coefficient  can then be written as:
\begin{equation}\label{eq:surface_rate}
	K_{ij}=f_\mathrm{eff,ij}\left( R_\mathrm{diff,i}+R_\mathrm{diff,j}\right)/(n_\mathrm{grain}N_s) \;\; \mathrm{cm^3\,s^{-1}}
\end{equation}
where $n_\mathrm{grain}$ is the number density of dust grains. The denominator gives the total number of surface binding sites. The probability that a reaction occurs is given by $f_\mathrm{eff,ij}$ and is assumed to be unity for an exothermic reaction without activation energy.\footnote{Notation: We denote surface reaction rate coefficients with a capital $K_{ij}$ and gas phase reaction rate coefficients with a lower-case $k_{ij}$.} For an exothermic reaction between surface species $i$ and $j$ with activation energy $E_{a,ij}$ the probability of reaction during a single collision between the reactants can be expressed as a Boltzmann factor $\exp\left[-E_{a,ij}/ T_d \right]$ or using the quantum mechanical tunneling probability. Later probability is only relevant for light species and we apply it only if one of the reactants is \ce{J(H)}  or \ce{J(H2)}.
\begin{equation}
	\label{eq:tunneling}
	\kappa_{ij}=\max \left( \exp\left[-E_{a,ij}/ T_d \right], \exp\left[-2 (a/\hbar)(2\mu_{ij} k_B E_{a,ij})^{1/2}\right]\right)
\end{equation}
where $\mu_{ij}=m_i m_j/(m_i+m_j)$ is the reduced mass. During the collision between $i$ and $j$ overcoming the activation energy barrier is in competition with the migration of either reactant to a neighbor site and with desorption of either reactant. We use the reaction probability $f_\mathrm{eff,ij}$ as described by \citet{chang2007,garrod2011} to account for the competing processes in case an activation energy barrier exists
\begin{equation}
	\label{eq:efficiency}
	f_\mathrm{eff,ij}=\frac{\nu_{0}}{\nu_0 \kappa_{ij} + K_\mathrm{diff,i} +K_\mathrm{diff,j} + K_\mathrm{des,i} +K_\mathrm{des,j}}
\end{equation}
with $\nu_0=\max(\nu_{0,i},\nu_{0,j})$ and the species dependent diffusion rate coefficient $K_\mathrm{diff,\_}=\nu_{0,\_} t_\mathrm{diff,\_}^{-1}$ ($\_$) is a placeholder for species $i$ and $j$).  $K_\mathrm{des,\_}$ is the sum of all desorption probabilities. 

\paragraph{Accretion}
The adsorption or accretion rate coefficient depends on the thermal velocity of the incoming particle $v_{th,i}$, the total effective cross section area of all target grains $\sigma_\mathrm{dust} (\mathrm{cm^{2}})$, i.e. the cross section per grain times the dust number density $n_d$, and the sticking coefficient $S(T,T_d)$ as:
\begin{equation}\label{eq:accretion}
	k_\mathrm{acc,i}=\sigma_\mathrm{dust} v_\mathrm{th,i} S(T,T_d)  
\end{equation}
with the mean thermal velocity
\begin{equation}\label{eq:thermal_velocity}
	v_\mathrm{th,i}=\sqrt{\frac{8 k_B T}{\pi m_i}}\;\; 
\end{equation}
For the sticking of \ce{H} we use the coefficient provided by \citet[][their Eq.~(3.7)]{hollenbach79}, in case of  \ce{H2} we 
assume $S(T,T_d)=0.5$ \citep{acharyya2014} and for any other species we assume $S(T,T_d)=1/3$ \citep{willacy1993}.
\paragraph{Thermal desorption}
Species on the grain surface are removed depending on the grain surface and the their binding energy $E_{D,i}$, which is itself a function of the grain material and ice composition. The rate coefficient is given by
\begin{equation}
	K_\mathrm{evap,i}=\nu_{0,i}\exp\left(-E_{D,i}/T_d \right) 
\end{equation}
where $\nu_{0,i}$ is from Eq.~(\ref{eq:characteristic_frequency}).

\paragraph{\ce{H2} formation induced desorption}
The \ce{H2}  binding energy of 4.48~eV is released during the formation process. It is commonly assumed that the energy is distributed evenly between internal energy and kinetic energy of the desorbed molecule and lattice energy of the surface. This local "hotspot"-heating of the surface may be sufficient to desorb additional species. Following \citet{willacy1994} we apply this only for species with a binding energy of less than 1210~K \citep{roberts2007}. The rate coefficient is:
\begin{equation}
	K_\mathrm{\ce{H2}-des}=\epsilon R_{\ce{H2}}/n_s(tot) 
\end{equation}
where $R_{\ce{H2}} (\mathrm{cm^{-3}s^{-1}})$ is the \ce{H2} formation rate on grains \citep{roellig2013dust} and $\epsilon$ measures the number of atoms and molecules that are desorbed by this process. The value of $\epsilon$ is uncertain and we assume a low efficiency of $\epsilon =0.01$ . Division by the total density of all surface species $n_s(tot)=\sum_in_s(i)$  distributes the binding energy across all species on the grain.
Depending on the value of $\epsilon$ this process can dominate other desorption processes and we found it to significantly contribute to the overall desorption. For a more detailed analysis see \citet{roberts2007} and more recently \citet{prochaska2021}.
\paragraph{Photo-desorption}
\begin{table}
	\caption{\label{tab:yield} Photo-desorption yields}
	\centering
	\resizebox{\columnwidth}{!}{%
	\begin{tabular}{lll}
		\hline\hline
		\vrule width 0pt height 2.2ex
		species & $Y_i$&Ref\\
		\hline
		\vrule width 0pt height 2.2ex
		\ce{O},\ce{^{18}O} & $10^{-4}$ &(1)\\
		\ce{H2O},\ce{H2{}^{18}O}&$10^{-3}(1.3+0.032\, T_d)\left(1-e^{-x/l(T)}\right)$ &(2)\\	
		\makecell[cl]{ \ce{H2O -> OH + H} \\ \ce{H2{}^{18}O -> ^{18}OH + H}}
		&$2\times 10^{-3}$ &(3)\\
		\ce{CO},\ce{^{13}CO},\ce{C^{18}O},\ce{^{13}C^{18}O} & $10^{-2}$ &(4)\\	\hline
	\end{tabular}
	}
	\tablefoot{
		$Y_i=10^{-3}$ for all other species. $x$ is the ice thickness in ML, $l(T)~0.6+0.024 T$.}
		\tablebib{
		(1) {\citet{hollenbach2009}};
		(2) {\citet{oeberg2009}};
		(3) {\citet{andersson2008}};
		(4) {\citet{fayolle2011}}
	}
\end{table}
The absorption of UV photons at the dust surface can sufficiently increase the internal energy of the surface species to induce desorption. Following \citet{cuppen2017} we write the rate coefficient for photo-desorption as:
\begin{equation}\label{photo_desorption}
	K_\mathrm{ph-des,i}=\frac{1}{n_s(tot)} \sigma_\mathrm{dust} Y_i\,  \chi\, F_\mathrm{Draine} \left\langle f_\mathrm{ss} \exp\left(-\gamma_i A_V \right)\right\rangle\;\;\mathrm{s^{-1}}
\end{equation}
where $Y_i$ is the photo-desorption yield per UV photon. Experimental data on $Y_i$ is only available for very few species and we assume a general value of $Y_i=10^{-3}\, \mathrm{(molecules/photon)}$ except for the species listed in Table~\ref{tab:yield}. For \ce{H2O} we use the expressions given by \citet{oeberg2009} accounting for ice thickness and gas and dust temperature (their Eq.~10).
$\chi F_{Draine}$ is a measure of the local UV energy density with $F_\mathrm{Draine}=1/h \int_{91.2 nm}^{205nm}\lambda u_\lambda d\lambda=1.921\times 10^8 \; \mathrm{photons\,cm^{-2}\, s^{-1}}$. 
  $f_\mathrm{ss}$ is the self-shielding factor which is unity for all species except \ce{CO} where the shielding of the ice species is also provided by the column in the gas.  $\left\langle \exp\left(-\gamma_i A_V \right)\right\rangle$ accounts for the dust attenuation of FUV radiation from the cloud surface  to the local position. The $\langle \rangle$ indicates an average over the full solid angle.

\paragraph{Total desorption}
The total desorption rate coefficient is the sum of all individual desorption rate coefficients:
\begin{multline}\label{eq:total_desorption}
	K_\mathrm{des,i} = f_\mathrm{des} \left( K_\mathrm{evap,i} + K_\mathrm{CR-des,i} + K_{\ce{H2}-des,i} +  \right) + \\  K_\mathrm{ph-des,i} + f_\mathrm{surf} K_\mathrm{chem-des,i} 
\end{multline}
where $f_\mathrm{des}$ describes the fraction of all surface species that are considered candidates for desorption:
\begin{equation}\label{eq:desorbable_fraction}
	f_{des}=\begin{cases} 
		1 &,\, n_s(tot) \le \sum n_\mathrm{site} \\
		2\sum n_\mathrm{site}/n_s(tot)&,\, n_s(tot) > \sum n_{site} 
	\end{cases} 
\end{equation} 
where  $\sum n_\mathrm{site}$ gives the total number of all grain surface binding sites per volume summed over all grains \citep[see also][]{aikawa1996,woitke2009}. In Eq.~(\ref{eq:desorbable_fraction}) we make the assumption, that surface species from the top two layers of the ice mantle can desorb to the gas phase \citep{aikawa1996}. For the chemical desorption we assume $f_\mathrm{surf}=f_\mathrm{des}$, where $f_\mathrm{surf}$ is the fraction of particles that can undergo surface reactions. Note, that the effects of ice opacity to FUV radiation and the fact that photo-desorption only occurs from the top few monolayers of the ice is implicitly included in the photo-desorption yields \citep{cuppen2017}.  

\tablefirsthead{		
	\hline\hline
	\vrule width 0pt height 2.2ex
	reaction&$\Delta H_R (\mathrm{eV})$&BR&$E_a (K)$\\
	\hline}
\tablehead{		
	\hline\hline
	\vrule width 0pt height 2.2ex
	reaction&$\Delta H_R (\mathrm{eV})$&BR&$E_a (K)$\\
	\hline}
\tabletail{\\ \hline}
\tablelasttail{\hline }
\topcaption{\label{tab:chemical_desorption1} Chemical desorption reactions.}
\begin{supertabular}{lccc}
	\hline 
	\ce{\text{J}(O) + \text{J}(H)  -> OH }&4.44&0.39&0\\
	\ce{\text{J}(O) + \text{J}(H) -> \text{J}(OH)}&4.44&0.61&0\\
	\hline
	\ce{\text{J}(OH) + \text{J}(H) -> H2O}&5.17&0.22&0\\
	\ce{\text{J}(OH) + \text{J}(H) -> \text{J}(H2O)}&5.17&0.78&0\\
	\hline   	   
	\ce{\text{J}(CO) + \text{J}(H) -> HCO}&0.66&0.01&2000\\
	\ce{\text{J}(CO) + \text{J}(H) -> \text{J}(HCO)}&0.66&0.99&2000\\
	\ce{\text{J}(CO) + \text{J}(H) -> HOC}&0.66&0.01&2000\\
	\ce{\text{J}(CO) + \text{J}(H) -> \text{J}(HOC)}&0.66&0.99&2000\\        
	\hline   
	\ce{\text{J}(O2) + \text{J}(H) -> O2H}&2.24 &0.02 &0\\  
	\ce{\text{J}(O2) + \text{J}(H) -> \text{J}(O2H)}&2.24 &0.98 &0\\
	\hline   
	\ce{\text{J}(O2H) + \text{J}(H) -> OH + OH}&1.47 &0 &0\\  
	\ce{\text{J}(O2H) + \text{J}(H) -> \text{J}(OH) + OH}&1.47 &0.01 &0\\
	\ce{\text{J}(O2H) + \text{J}(H) -> \text{J}(OH) + \text{J}(OH)}&1.47 &0.99 &0\\
	\hline 
	\ce{\text{J}(O2H) + \text{J}(H) -> H2O2}&3.69 &0.01 &0\\  
	\ce{\text{J}(O2H) + \text{J}(H) -> \text{J}(H2O2)} &3.69 &0.99 &0\\      
	\hline
	\ce{\text{J}(HCO) + \text{J}(H) -> H2CO}&3.91 &0.04 &0\\  
	\ce{\text{J}(HCO) + \text{J}(H) -> \text{J}(H2CO)}&3.91 &0.96 &0\\
	\hline
	\ce{\text{J}(HCO) + \text{J}(H) -> CO + H2}&3.85 &0.43 &130\\
	\ce{\text{J}(HCO) + \text{J}(H) -> \text{J}(CO) + H2 }&3.85 &0.35 &130\\
	\ce{\text{J}(HCO) + \text{J}(H) -> CO + \text{J}(H2)}&3.85 &0.12 &130\\
	\ce{\text{J}(HCO) + \text{J}(H) -> \text{J}(CO) + \text{J}(H2) }&3.85 &0.10 &130\\
	\hline
	\ce{\text{J}(HOC) + \text{J}(H) -> CHOH }& 0.66 & 0.04 & 2000\\
	\ce{\text{J}(HOC) + \text{J}(H) -> \text{J}(CHOH) }& 0.66 & 0.96 & 2000 \\        
	\hline  
	\ce{\text{J}(O3) + \text{J}(H) -> O2 + OH} &3.33&0.02&480\\
	\ce{\text{J}(O3) + \text{J}(H) -> \text{J}(O2) + OH}&3.33&0.02&480\\
	\ce{\text{J}(O3) + \text{J}(H) -> O2 + \text{J}(OH)}&3.33&0.39&480\\
	\ce{\text{J}(O3) + \text{J}(H) -> \text{J}(O2) + \text{J}(OH)}&3.33&0.57&480\\
	\hline
	\ce{\text{J}(H2CO) + \text{J}(H) -> HCO + H2 }&0.61&0&2200\\
	\ce{\text{J}(H2CO) + \text{J}(H) -> \text{J}(HCO) + H2}&0.61&0.17&2200\\
	\ce{\text{J}(H2CO) + \text{J}(H) -> HCO + \text{J}(H2)}&0.61&0&2200\\
	\ce{\text{J}(H2CO) + \text{J}(H) -> \text{J}(HCO) + \text{J}(H2)}&0.61&0.83&2200\\
	\hline
	\ce{\text{J}(H2CO) + \text{J}(H) -> CH3O }&0.88&0&2000\\
	\ce{\text{J}(H2CO) + \text{J}(H) -> \text{J}(CH3O)}&0.88&1.00&2000\\
	\hline
	\ce{\text{J}(CH3O) + \text{J}(H) -> CH3OH }&0.88&0.01&0\\
	\ce{\text{J}(CH3O) + \text{J}(H) -> \text{J}(CH3OH)}&0.88&0.99&0\\
	\hline
	\ce{\text{J}(CH3O) + \text{J}(H) -> H2CO + H2} &0.88&0.01&150\\
	\ce{\text{J}(CH3O) + \text{J}(H) -> \text{J}(H2CO) + H2}&0.88&0.71&150\\
	\ce{\text{J}(CH3O) + \text{J}(H) -> H2CO + \text{J}(H2)}&0.88&0.01&150\\
	\ce{\text{J}(CH3O) + \text{J}(H) -> \text{J}(H2CO) + \text{J}(H2)}&0.88&0.27&150\\
	\hline
	\ce{\text{J}(CH3OH) + \text{J}(H) -> CH3O + H2 }&-0.04&0&3200\\
	\ce{\text{J}(CH3OH) + \text{J}(H) -> \text{J}(CH3O) + H2} &-0.04&0&3200\\
	\ce{\text{J}(CH3OH) + \text{J}(H) -> CH3O + \text{J}(H2)} &-0.04&0&3200\\
	\ce{\text{J}(CH3OH) + \text{J}(H) -> \text{J}(CH3O) + \text{J}(H2)} &-0.04&1.00&3200\\
	\hline
	\ce{\text{J}(H2O2) + \text{J}(H) -> H2O + OH} &2.95&0&1000\\
	\ce{\text{J}(H2O2) + \text{J}(H) -> \text{J}(H2O) + OH}&2.95&0.04&1000\\
	\ce{\text{J}(H2O2) + \text{J}(H) -> H2O + \text{J}(OH)}&2.95&0.01&1000\\
	\ce{\text{J}(H2O2) + \text{J}(H) -> \text{J}(H2O) + \text{J}(OH)}&2.95&0.95&1000\\
	\hline
	\ce{\text{J}(O) + \text{J}(O) -> O2 }&5.16&0.70&0\\
	\ce{\text{J}(O) + \text{J}(O) -> \text{J}(O2)} &5.16&0.30&0\\
	\hline
	\ce{\text{J}(O2) + \text{J}(O) -> O3 }&1.10&0&0\\
	\ce{\text{J}(O2) + \text{J}(O) -> \text{J}(O3)} &1.10&1.00&0\\
	\hline
	\ce{\text{J}(CO) + \text{J}(O) -> CO2} &5.51&0.18&650\\
	\ce{\text{J}(CO) + \text{J}(O) -> \text{J}(CO2) }&5.51&0.82&650\\
	\hline
	\ce{\text{J}(O) + \text{J}(HCO) -> CO2 + H} &4.85&0.08&0\\
	\ce{\text{J}(O) + \text{J}(HCO) -> \text{J}(CO2) + H }&4.85&0.61&0\\
	\ce{\text{J}(O) + \text{J}(HCO) -> CO2 + \text{J}(H)} &4.85&0.04&0\\
	\ce{\text{J}(O) + \text{J}(HCO) -> \text{J}(CO2) + \text{J}(H) }&4.85&0.27&0\\
	\hline
	\ce{\text{J}(H2CO) + \text{J}(O) -> CO2 + H2} &5.45&0.10&350\\
	\ce{\text{J}(H2CO) + \text{J}(O) -> \text{J}(CO2) + H2} &5.45&0.67&350\\	
	\ce{\text{J}(H2CO) + \text{J}(O) -> CO2 + \text{J}(H2) }&5.45&0.03&350\\
	\ce{\text{J}(H2CO) + \text{J}(O) -> \text{J}(CO2) + \text{J}(H2) }&5.45&0.20&350\\
	\hline
	\ce{\text{J}(OH) + \text{J}(OH) -> H2O2} &5.51&0&0\\
	\ce{\text{J}(OH) + \text{J}(OH) -> \text{J}(H2O2) }&5.51&1.00&0\\
	\hline
	\ce{\text{J}(N) + \text{J}(N) -> N2 }&9.79&0.860&0\\
	\ce{\text{J}(N) + \text{J}(N) -> \text{J}(N2) }&9.79&0.14&0\\
	\hline
	\ce{\text{J}(O2H) + \text{J}(O) -> O2 + OH} &2.20&0.01&0\\
	\ce{\text{J}(O2H) + \text{J}(O) -> \text{J}(O2) + OH} &2.20&0.03&0\\
	\ce{\text{J}(O2H) + \text{J}(O) -> O2 + \text{J}(OH)} &2.20&0.25&0\\
	\ce{\text{J}(O2H) + \text{J}(O) -> \text{J}(O2) + \text{J}(OH)} &2.20&0.65&0\\
	\hline
	\ce{\text{J}(OH) + \text{J}(CO) -> CO2 + H }&1.07&0&400\\
	\ce{\text{J}(OH) + \text{J}(CO) -> \text{J}(CO2) + H }&1.07&0.19&400\\
	\ce{\text{J}(OH) + \text{J}(CO) -> CO2 + \text{J}(H)} &1.07&0&400\\
	\ce{\text{J}(OH) + \text{J}(CO) -> \text{J}(CO2) + \text{J}(H) }&1.07&0.73&400\\
	\hline
	\ce{\text{J}(OH) + \text{J}(HCO) -> CO2 + H2 }&4.93&0.08&0\\
	\ce{\text{J}(OH) + \text{J}(HCO) -> \text{J}(CO2) + H2} &4.93&0.67&0\\
	\ce{\text{J}(OH) + \text{J}(HCO) -> CO2 +  \text{J}(H2)} &4.93&0.03&0\\
	\ce{\text{J}(OH) + \text{J}(HCO) -> \text{J}(CO2) +  \text{J}(H2)} &4.93&0.22&0\\
	\hline
	\ce{\text{J}(OH) + \text{J}(O) -> O2 + H }&0.72&0.01&0\\
	\ce{\text{J}(OH) + \text{J}(O) -> \text{J}(O2) + H }&0.72&0.13&0\\
	\ce{\text{J}(OH) + \text{J}(O) -> O2 + \text{J}(H) }&0.72&0.05&0\\
	\ce{\text{J}(OH) + \text{J}(O) -> \text{J}(O2) + \text{J}(H) }&0.79&0.82&0\\
	\hline
	\ce{\text{J}(O3) + \text{J}(O) -> O2 + O2} &4.06&0.16&2500\\
	\ce{\text{J}(O3) + \text{J}(O) -> \text{J}(O2) + O2 }&4.06&0.48&2500\\
	\ce{\text{J}(O3) + \text{J}(O) -> \text{J}(O2) + \text{J}(O2) }&4.06&0.36 &2500\\
	\hline
	\ce{\text{J}(CO2) + \text{J}(H) -> CO + OH }&-1.00&0&10000\\
	\ce{\text{J}(CO2) + \text{J}(H) -> \text{J}(CO) + OH}&-1.00&0&10000\\
	\ce{\text{J}(CO2) + \text{J}(H) -> CO + \text{J}(OH)}&-1.00&0&10000\\
	\ce{\text{J}(CO2) + \text{J}(H) -> \text{J}(CO) + \text{J}(OH)}&-1.00&1.00&10000\\
	\hline
	\ce{\text{J}(H2O) + \text{J}(H) -> OH + H2 }&-0.65&0&9600\\
	\ce{\text{J}(H2O) + \text{J}(H) -> \text{J}(OH) + H2}&-0.65&1.00&9600\\
	\hline
\end{supertabular}

Table~\ref{tab:chemical_desorption1} list all chemical desorption reactions with their respective branching ratios. $\Delta H_R$ is the exothermicity of the reaction, BR is the branching ratio and $E_a$ denotes the activation barrier of the reactions. The BR are computed per reaction (left-hand side). In case of multiple product channels for reaction \ce{\text{J}(A) + \text{J}(B)} they have to be re-scaled accordingly.  We assumed the following desorption energies $E_{D,i}$  for the reaction products in the Table: \ce{H}(600K), \ce{H2}(430K), \ce{CO}(1300K), \ce{N2}(1100K), \ce{O2}(1200K), \ce{O3}(1800K), \ce{OH}(4600K), \ce{CO2}(2600K), \ce{H2O}(5600K), \ce{HCO}(2400K), \ce{O2H}(3650K), \ce{CH3O}(4400K), \ce{H2CO}(4500K), \ce{H2O2}(6000K), \ce{CH3OH}(5000K). 
See Table~\ref{tab:binding_energies} and \citet{kida} (\url{https://kida.astrochem-tools.org/} for references.)

\section{Heating and cooling processes} \label{sect:heatingcooling_appendix}

Table~\ref{tab:heating_cooling} lists all cooling and heating processes implemented in \Kt{} with their references. For those that are not copied from the literature we specify their details here. We chose the notation such that heating and cooling processes are denoted with $\Gamma_\mathrm{ID}$ and $\Lambda_\mathrm{ID}$, respectively with $\mathrm{ID}$ taken from Table~\ref{tab:heating_cooling}. Processes which can act as heating or cooling processes are named according to their dominating behavior.

\paragraph{\ce{H2} heating and cooling}
Table~\ref{tab:heating_cooling} lists four heating and cooling processes related to \ce{H2} which are all based on the transfer of internal molecular energy to kinetic  gas energy by means of inelastic collision.

UV radiation pumps \ce{H2} molecules to excited electronic states (Lyman and Werner bands) which decay quickly back to ground electronic states (except for $\eta\approx$10\% of the excited states which decay into the continuum as pointed out by Solomon (1965) \citep[cited in][]{field1966}). The decay leads to a population of superthermal states and collisional de-excitation leads to heating of the gas with the rate:
\begin{equation}
	\label{eq:collisional_deexcitation}
	\Gamma_{1}=\sum_i n_{\ce{H2},i} \sum_j C(i\rightarrow j)(E_i-E_j)\;\; \mathrm{erg\,s^{-1}\, cm^{-3}}
\end{equation}   
where the sums over $i$ and $j$ are over all considered states. $n_{\ce{H2},i}$ is the level population of the $i$-th level, $C(i\rightarrow j)$ is the total collision rate from level $i$ to $j$ and $E_i-E_j$ is the energy difference between both levels \citep[see also][]{sternberg1989}. Depending on the level population this process can also cool the gas. Various approximations to this computation have been published, e.g. by \citet{tielens1985} and \cite{roellig06}.

Photo-dissociation of \ce{H2} molecules produces two energetic atoms which transfer their energy to the gas. The heating rate depends on the FUV pumping rate $P$, the dissociation probability $\eta$  and the average kinetic energy $\langle E_j* \rangle\approx 0.4 \mathrm{eV}$ of the atoms produced by the spontaneous decay out of the excited states of the Lyman and Werner electronic bands \citep{stephens1973}:
\begin{equation}
	\label{eq:photodissociation_heating}
	\Gamma_{2}=\sum_i  n_{\ce{H2},i} \sum_{j^*} P(i\rightarrow j^*)\eta(j^*)\langle E_j* \rangle \;\; \mathrm{erg\,s^{-1}\, cm^{-3}}
\end{equation} 
where $i$ indicates all considered electronic ground state levels of \ce{H2} and $j^*$ indicates all  electronically excited levels.  

The formation of \ce{H2} molecules releases the binding energy of $\Delta H_{\ce{H2},b}=7.2\times 10^{-12}\mathrm{erg}$.  Under the assumption of energy equipartition between grain lattice energy, inner energy and kinetic energy, 1/3 of this energy is converted to kinetic energy. With the \ce{H2}-formation rate $R_f(T)$ we find for the heating rate:
\begin{equation}
	\label{eq:formation_heating}
	\Gamma_{3}=R_f(T) n\,n(H) \frac{1}{3}\Delta H_{\ce{H2},b}\;\; \mathrm{erg\,s^{-1}\, cm^{-3}}
\end{equation}
Details on the \ce{H2} formation treatment in \Kt{} are given in \citet{roellig2013dust}.

In hot gas, \ce{H}--\ce{H2} and \ce{H2}--\ce{H2} collisions are energetic enough to dissociate molecular hydrogen. This removes the amount of $E_{\ce{H2}b}$ from the kinetic gas energy \citep{lepp83}. The cooling rate can be written as:
\begin{equation}
	\label{eq:kinetic_dissociation}
	\Lambda_{21}=E_{\ce{H2},b} n(\ce{H2})\left(k_{\ce{H + H2}}+k_{\ce{H2 + H2}}\right) \;\; \mathrm{erg\,s^{-1}\, cm^{-3}}
\end{equation}
where the rate coefficients for the collisional dissociation reactions \ce{H + H2 -> H + H + H}  and \ce{H2 + H2 -> H2 + H + H} are taken from UDfA12:
\begin{align}
	\label{eq:kinetic_dissociation_rates}
	k_{\ce{H + H2}}  &=4.67\times 10^{-7}(T/300K)\exp\left(-55000K/T\right) \; \mathrm{ cm^{3}s^{-1}}\\
	k_{\ce{H2 + H2}} &=1.00\times 10^{-8}(T/300K)\exp\left(-84100K/T\right) \; \mathrm{ cm^{3}s^{-1}}
\end{align}

\paragraph{Cosmic ray heating}
The interaction of energetic cosmic ray protons with \ce{H} and \ce{H2} produces hot electrons that transfer their energy to the ISM in the following collisions. A detailed treatment as been presented by \citet{glassgold2012}. Generally, the heating rate of cosmic rays can be written as:
\begin{equation}
	\Gamma_7=\zeta_{CR} Q\, n \;\; \mathrm{erg\,s^{-1}\, cm^{-3}}
\end{equation}
where $\zeta_{CR}$ is the CR ionization rate per H nucleus in $\mathrm{s^{-1}}$ and $Q$ is the average energy deposited as heat per ionization. \citet{glassgold2012} provide values of $Q$ for different interstellar environments as function of total density (their Table~6). $Q$ can then be approximated by the following expression:
\begin{equation}
	Q(x)=1.60218\times 10^{-12} \frac{4.37}{0.238\, +0.4624 e^{-0.42 x}} \;\; \mathrm{erg}
\end{equation}
with $x=\log_{10}(n/cm^{-3})$. The pre-factor converts from eV to erg.

\paragraph{Grain photo-electric heating}
The heating by hot electrons released during photo-electric (PE) absorption of FUV photons  by dust particles is the most important PDR heating process under most circumstances. The photo-electrons can also recombine with the grains and cool the gas with a rate $\Lambda_8$ , therefore $\Gamma_\mathrm{PE}=\Gamma_8-\Lambda_8$.

\Kt{} offers a choice of how to treat PE heating  $\Gamma_\mathrm{PE}$. The user can decide which approximation of the PE heating (and electron recombination) to use. We implemented the description given by \citet[][(their Eqs.~(41),(43),(44))]{bt94} and by \citet[][(their Eqs.~(44),(45) and Table~2 and 3)]{wd01pe}. They mainly differ in the applied dust size distributions and the estimates for photoelectric thresholds, yields, and electron capture rates. See \citet{roellig2013dust} for more details.

\paragraph{Gas-grain collisions}
Collision between gas particles and dust grains can transfer energy between the two. If the gas is hotter than the grains the collisions cool the gas. If the gas is cooler the collisions act as gas heating.  \citet{burke1983} give the rate as 
\begin{equation}\label{eq:gas-grain}
	\Lambda_{18}= n_g n  \sigma_g \left(\frac{8kT}{\pi m_H}\right)^{1/2}\tilde{\alpha}_T 2 k_B ( T_g - T_d)\;\; \mathrm{erg\,s^{-1}\, cm^{-3}}
\end{equation}
We apply an approximate accommodation factor $\tilde{\alpha}_T\approx0.3$ when using the PE heating description by \citet{bt94}. In case of the PE heating described by \citet{WD01PEH} we employ a modified accommodation factor given by \citet{groenewegen1994}:
\begin{equation}\label{eq:gas-grain2}
	\tilde{\alpha}_T=0.35 \exp\left(-\sqrt{\frac{T_d-T_k}{500 K}}\right)+0.1
\end{equation}

\paragraph{Carbon ionization}
The ionization of atomic carbon releases hot electrons with an energy of about $1~\mathrm{eV} = 1.602\times10^{-12} \mathrm{erg}$. We approximate the heating rate by
\begin{equation}\label{eq:carbon_ionization}
	\Lambda_{22}= 2.2\times 10^{-22}\mathrm{erg\,s^{-1}} n(C) \chi_\mathrm{FUV} 
\end{equation}

\begin{table*}[htb]
	\caption{Fine-structure and rotational line cooling data}
	\label{tab:cooling_data}
	\centering
	\begin{tabular}{lccrllccl}
		\hline \hline
		\vrule width 0pt height 2.2ex
		species & $i$ & $j$ & $\lambda_{ij}$ & $A_{ij}$ & collision &$\alpha$ & $\beta$ & Ref \\
		&     &     & $(\mathrm{\mu m})$& $(\mathrm{s^{-1}})$ & & & & \\
		\hline
		\vrule width 0pt height 2.2ex
		\ce{C+} & $^2P_{3/2}$ & $^2P_{1/2}$ & 157.7 & $2.29(-6)$& \ce{C+ + H} & $3.71(-10)$&0.15& (1)  \\
		&   &  &  &  & \ce{C+ + o/p-H2} & $4.55(-10)$&$1.60(-10)$&  (2), (3)  \\ 
		\hline
		\vrule width 0pt height 2.2ex
		\ce{C} & $^3P_{1}$ & $^3P_0$ & 609.1 & $7.93(-8)$& \ce{C+ + H} & $1.01(-10)$&0.117& (4)   \\	
		&   &  &  &  & \ce{C + p-H2} & $8.00(-11)$&$0$&(5)  \\
		&   &  &  &  & \ce{C + o-H2} & $7.50(-11)$&$0$& (5)\\
		& $^3P_{2}$ & $^3P_0$ & 230.3 & $1.71(-14)$& \ce{C+ + H} & $4.49(-11)$&0.194&(4) \\	
		&   &  &  &  & \ce{C + p-H2} & $9.00(-11)$&$0$& (5)\\
		&   &  &  &  & \ce{C + o-H2} & $3.54(-11)$&$0.167$&(5)\\	 
		& $^3P_{2}$ & $^3P_1$ & 370.4 & $2.65(-7)$& \ce{C+ + H} & $1.06(-10)$&0.234&(4)\\	
		&   &  &  &  & \ce{C + p-H2} & $2.00(-11)$&$0$&(5)\\
		&   &  &  &  & \ce{C + o-H2} & $5.25(-11)$&$0.244$&(5)\\	  
		\hline
		\vrule width 0pt height 2.2ex
		\ce{O} & $^3P_{1}$ & $^3P_2$ & 63.2 & $8.91(5)$& \ce{C+ + H} & $1.87(-11)$&0.539& (6), (7)\\	
		&   &  &  &  & \ce{O + p-H2} & $3.46(-11)$&$0.316$&(5)\\
		&   &  &  &  & \ce{O + o-H2} & $2.70(-11)$&$0.362$&(5)\\
		& $^3P_{0}$ & $^3P_2$ & 44.1 & $1.34(-10)$& \ce{C+ + H} & $2.10(-12)$&0.841& (6) , (7)\\	
		&   &  &  &  & \ce{O + p-H2} & $7.07(-11)$&$0.268$& (5)\\
		&   &  &  &  & \ce{O + o-H2} & $5.49(-11)$&$0.317$& (5)\\	 
		& $^3P_{0}$ & $^3P_1$ & 145.5 & $1.75(-5)$& \ce{C+ + H} & $2.52(-10)$&0.169& (6) , (7)\\	
		&   &  &  &  & \ce{O + p-H2} & $1.44(-11)$&$1.109$& (5)\\
		&   &  &  &  & \ce{O + o-H2} & $4.64(-11)$&$0.976$& (5)\\	
		\hline
		\vrule width 0pt height 2.2ex            
		\ce{Si+} & $^2P_{3/2}$ & $^2P_{1/2}$ & 34.8 & $2.20(-4)$& \ce{Si+ + H} & $6.50(-10)$&0&(8)\\
		\hline
		\vrule width 0pt height 2.2ex            
		\ce{CO}, \ce{^{13}CO} & $J+1$ & $J$ & \multicolumn{2}{c}{tabulated}& \ce{CO + o/p-H2} & \multicolumn{2}{c}{tabulated}&(9), (10)\\
		\ce{OH} & $J,\Omega,p$ & $J',\Omega',p'$ & \multicolumn{2}{c}{tabulated}& \ce{OH + H2} & \multicolumn{2}{c}{tabulated}& (11), (12) \\
		\hline
	\end{tabular}
	\tablefoot{
		$a(-b)$ corresponds to $a\times 10^{-b}$. Collision rates are parameterized using the form $C_{ij}=\alpha T^\beta$ with the exception of the collisions \ce{C+ + o/p-H2} which use the formula $C_{ij}=\alpha + \beta \exp(-100/T)$. Collision rates for \ce{CO + H2} and \ce{OH + H2} are interpolated based on tabulated data. $J$ is the rotational angular momentum quantum number, $\Omega$ is the projection of $J$ on the inter-nuclear axis, $p$ is the parity.    
		\textbf{References:} (1) \citet{barinovs2005}, (2) \citet{wiesenfeld2014}, (3) \citet{lique2013}, (4) \citet{launay1977}, (5) \citet{schroder1991}, (6) \citet{vieira2017}, (7) \citet{abrahamsson2007}, (8) \citet{TH85}, 
		(9) \citet{cdms}, (10) \citet{yang2010}, (11) \citet{JPL}, (12) \citet{dewangan1987} 
	}
\end{table*} 

\paragraph{Line cooling}
Details on the computation of atomic and molecular line cooling can also be found elsewhere \citep[e.g.][]{tielens1985,sternberg1989,roellig06,woitke2009,papadopoulos2011}. Following \citet{sternberg1988} the general expression for the cooling rate due to a transition from level $i$ to level $j$ of species $x$ is:
\begin{equation}\label{eq:line_cooling}
	\Lambda^x_{ij}=n_i A_{ij} h\nu_{ij}\beta(\tau_{ij})-B_{ij} h\nu_{ij}E_{ij}(n_j-n_i)\beta(\tau_{ij}) \;\;\mathrm{ erg\,  cm^{-3}\,s^{-1}}
\end{equation}
where $A_{ij}$ and $B_{ij}$ are the Einstein coefficients, $n_i$ is the level population of level $i$, $E_{ij}$ is the external radiation at the transition frequency $\nu_{ij}$ and $\beta(\tau_{ij})$ is the escape probability at the optical depth $\tau$.  For all cooling lines we compute $\tau$ from the perpendicular column density of the respective species. We use the expression for a spherical cloud given by \citet{stutzki1985}:
\begin{equation}\label{eq:optical_depth}
	\beta(\tau)=\frac{1}{\sqrt{\pi}}\int_{-\infty}^{+\infty}e^{-z^2}e^{-\tau {e^{-z^2}}}dz
\end{equation}
More details are also given in \citet{stoerzer1996}. The energy level population is computed by solving the system of rate equations balancing all populating and de-populating processes. Using the escape probability approximation these equilibrium equations take a simple form. For the level $i$ we can write:
\begin{equation}\label{eq:statistical_equlibrium}
	\frac{dn_i}{dt}=\sum_{k\neq i}n_k C_{ki} - n_i\left(\sum_{k \neq i} C_{ik}-\sum_{j<i}\beta(\tau_{ij}) A_{ij} \right) = 0
\end{equation}
where $\sum_{j>i}$ sums over all levels with $E_j<E_i$. The first term describes all populating collisions, the second term all depopulating collisions and emission events. For two and three-level systems the solution to Eq.~(\ref{eq:statistical_equlibrium}) can be given analytically. Larger $n$-level systems are solved  by LU decomposition. Presently, \Kt{} does not consider pumping by external radiation and we set $E_{ij}=0$ in Eq.~(\ref{eq:line_cooling}).

We also added two high-temperature cooling processes: O {\scriptsize I} 6300 \AA\ \, emission of the meta-stable $^1D$ level of atomic oxygen, and H {\scriptsize I} Lyman-\textalpha\, emission, which become relevant for $T>5000$~K.
We use the following expressions \citep{sternberg1988}:
\begin{align}\label{eq:metastable}
	\Lambda_{16}&=7.3\times 10^{-19} n(\ce{e^-}) n(\ce{H}) e^{-118400K/T}\;\;\mathrm{erg\, cm^{-3}\,s^{-1}}\\
	\Lambda_{20}&=1.8\times 10^{-24} n(\ce{O}) \left[n(\ce{H})+n(\ce{H2})\right]e^{ -22800K/T}\;\;\mathrm{erg\, cm^{-3}\,s^{-1}}
\end{align}

\section{The clumpy \Kt{} setup}\label{sect:clumpy_appendix}
\begin{table*}[htb]
	\centering
	\caption{Properties of a discrete example ensemble
		\label{tab:ensemble_properties}}
	\[
	\begin{array}{rrrrccccccrc}
		\hline \hline
		\vrule width 0pt height 2.6ex 
		M_i  &  N_i & &n_{0,i} & R_i & N_i\times\mathrm{area} & & V_i &  N_i\times V_i   & & N_i\times M_i & \\
		(M_\odot)&   & &(\mathrm{cm^{-3}})& (\mathrm{pc})& (\mathrm{pc^2}) & &  (\mathrm{cm^3}) &  (\mathrm{cm^3})  & & (M_\odot)& \\ 
		\hline
		\vrule width 0pt height 2.6ex 		
		0.001 & 24250.4 &( 0.842) & 3.7\times 10^5 & 0.0030 & 0.67 & (0.084) & 3.2\times 10^{48} & 7.75\times 10^{52} & (<10^{-3}) & 24.3 &
		(0.024 )\\
		0.01 & 3843.44 &( 0.133 )& 1.9\times 10^5 & 0.0081 & 0.78 & (0.099) & 6.44\times 10^{49} & 2.48\times 10^{53} & (0.002) & 38.4 &
		(0.038) \\
		0.1 & 609.144 & (0.021) & 9.2\times 10^4 & 0.0219 & 0.92 &( 0.116) & 1.30\times 10^{51} & 7.91\times 10^{53} & (0.006) & 60.9 & (0.061)
		\\
		1. & 96.5428 &( 0.003) & 4.6\times 10^4 & 0.0597 & 1.08 & (0.136) & 2.62\times 10^{52} & 2.53\times 10^{54} & (0.021) & 96.5 & (0.097 )\\
		10. & 15.301 & (<10^{-3}) & 2.3\times 10^4 & 0.162 & 1.27 &( 0.159) & 5.27\times 10^{53} & 8.07\times 10^{54} & (0.067) & 153.0 & (0.153) \\
		100. & 2.42504 & (<10^{-4}) & 1.1\times 10^4 & 0.442 & 1.49 & (0.187) & 1.06\times 10^{55} & 2.58\times 10^{55} & (0.215) & 242.5 & (0.243) \\
		1000. & 0.384344 & (<10^{-5}) & 5.6\times 10^3 & 1.2 & 1.75 &( 0.22) & 2.14\times 10^{56} & 8.23\times 10^{55} & (0.687) & 384.3 & (0.384) \\ \hline
		\vrule width 0pt height 2.6ex 
		\mathrm{total} & 28817.7 &  &  &   & 7.96 &   &  & 1.2\times 10^{56} &   & 1000.0 &   \\
		\hline
	\end{array}\]
	 \tablefoot{
	 We use the following ensemble parameters: $M_{ens}=1000~M_\odot$, $n_{ens}=10^4\mathrm{cm^{-3}}$, $\alpha=1.8$, and $\gamma=2.3$. Numbers in parenthesis indicate the fraction in the column total}
\end{table*}
\citet{stutzki1998} showed that the observed clump-mass spectrum of the ISM can be described as power-law spectrum:
\begin{equation}
	\label{eq:clump-mass-spectrum}
	\frac{dN_{cl}}{dM_{cl}}=A M_{cl}^{-\alpha}
\end{equation}
giving the number of clumps $dN_{cl}$ in the mass bin $dM_{cl}$. In addition
the masses of the clumps are related to their radii $R_{cl}$ by the
mass-size relation
\begin{equation}
	\label{eq:mass-size-relation}
	dM_{cl}=C R_{cl}^{\gamma}
\end{equation}
The power-law indices $\alpha$ and $\gamma$
can be derived from observations. \citet{kramer1998} present clump mass spectra of seven different molecular clouds and find that $\alpha=1.6...1.8$ for all clouds from their sample. As a consequence low-mass clumps are much more numerous compared to massive ones. \citet{schneider2004} find similar results with a slightly larger variation of $\alpha=1.6...2.1$ and with a tentative trend of higher values of $\alpha$ in regions with higher gas density. Other studies \citep{elmegreen1996} find $\langle\gamma\rangle=2.2-2.5$ over a large sample of clouds.
\citet{heithausen1998} studied the Polaris Flare to derive the power law spectrum over a mass range of at least 5 orders of magnitude, down to masses less than $10^{-3}~M_\odot$. They find $\alpha=1.84$ and 
$\gamma = 2.31$ and the default setup in \Kt{} assumes these values. Assuming that the clump masses in an ensemble are all within the range $m_l\le M_{cl}\le m_u$ we can write the total number of clumps in an ensemble $N_{ens}$ as
\begin{equation}
	\label{eq:ensemble_number}
	N_{ens}=\frac{A}{\alpha-1}\left(m_l^{1-\alpha}-m_u^{1-\alpha}\right) \;\;\; \mathbf{for}\,\alpha\ne 1\;,
\end{equation}
and the total ensemble mass
\begin{equation}
	\label{eq:ensemble_mass}
	M_{ens}=\frac{A}{2-\alpha}\left(m_l^{2-\alpha}-m_u^{2-\alpha}\right) \;\;\; \mathbf{for}\,\alpha\ne 2\;.
\end{equation}
Taking $M_{ens}$ as model parameter of the clumpy Eq.~(\ref{eq:ensemble_mass}) gives the constant $A$. The constant $C$ can be written as:
\begin{equation}
	C=\left(\frac{4\pi}{3}\frac{2-\alpha}{1+3/\gamma-\alpha}\frac{m_u^{(1+3/\gamma-\alpha)}-m_l^{(1+3/\gamma-\alpha)}}{m_u^{2-\alpha}-m_l^{2-\alpha}}\rho_{ens}   \right)^{\gamma/3}
\end{equation}
where $\rho_{ens}$ is the averaged ensemble density.

Note, that $A_d$ scales linearly with $M_{ens}$ and therefore all ensemble quantities that result from the linear superposition of the individual clumps also scale with the ensemble mass. Eq.~(\ref{eq:mass-size-relation}) leads to smaller clumps being denser while Eq.~(\ref{eq:clump-mass-spectrum}) returns many more low-mass clumps compared to high-mass clumps. To illustrate the consequences for the ensemble we summarize some ensemble properties in Table~\ref{tab:ensemble_properties}.

Table~\ref{tab:ensemble_properties} shows that the clump number distribution is heavily dominated by the smallest clumps ($~84\%$) and the mass-size relation states that smaller, i.e. less massive clumps, have a higher density. The most mass however is concentrated in the higher mass clumps, e.g. 62\% of the total ensemble mass is in the 3 most massive clumps of the ensemble. The volume distribution is even more skewed to the massive clumps. Over 90\% of the total ensemble volume is filled be the three most massive clumps and 40\% of the projected area.

Within a single ensemble we assume here that the clumpy model ensemble is spatially unresolved and ignore mutual shielding and absorption within the ensemble. The beam $(\Omega_B)$ averaged emission of the clumpy model is then just the weighted superposition of the individual clumps. The weighting factor $\omega_{cl}=\frac{dN_{cl}}{dM_{cl}}\Omega_{cl}$ is the total solid angle of all clumps in the mass bin $dM_{cl}$ with $\Omega_{cl}\approx\pi R_\mathrm{cl}^2/D^2$ being the solid angle of a single clump at a distance $D$, assuming $D>>R$.

\begin{equation}
	\label{eq:ensemble_intensiy}
	I_{ens}=\frac{1}{\Omega_{B}}\int_{m_l}^{m_u} I_{cl}\omega_{cl} dM_{cl} \;\;\mathrm{erg s^{-1} cm^{-2} Hz^{-1} sr^{-1}}
\end{equation}
The integral is taken over all masses in the range $[m_l,m_u]$. Full radiative transfer is implemented in {\Kt}-3D \citep{andree2017}.

\section{WL-PDR - A sandbox PDR model code written in Wolfram Mathematica}\label{sect:wl-pdr}

It is useful to reduce the numerical complexity of a model in order to better study numerical or physical aspects of the model. For some numerical tests in this paper we use the simple 1-D, semi-infinite plane-parallel PDR model code WL-PDR. 
The code is ported to Mathematica from PyPDR \citep{pypdr}. PyPDR is used as an educational tool to learn about the basic physics and some numerical internals of PDR codes. We ported PyPDR to Mathematica by Wolfram Research \citep{Mathematica} to seamlessly integrate the toy PDR model into our research infrastructure and to make use of the advanced analysis, numerics and visualization capabilities of Mathematica.  

The features of WL-PDR are :
\begin{itemize}
	\item Plane-parallel, semi-infinite model geometry
	\item One-side, uni-directional, perpendicular FUV illumination
	\item Non-LTE excitation of \oi, \cii, \ci, CO, $^{13}$CO, CH, CH$^+$, and HCO$^+$ using an escape probability approach. The molecular data is stored in the format of the LAMDA database \citep{lamda}.
	\item Time dependent chemistry: The chemical abundances are solved by evolving the chemical rate equations up to a certain chemical age.
	\begin{itemize}
		\item AccuracyGoal$\rightarrow |\log_{10}(n\times 10^{-10})|$, i.e. \texttt{atol=1e-10*n}
		\item PrecisionGoal$\rightarrow 10$, i.e. \texttt{rtol=1e-10}
		\item MaxSteps$\rightarrow 10^9$
	\end{itemize}
	\item The thermal balance is solved self-consistently with the chemistry iteratively.
	\item Self-shielding factors for CO are from \citet{visser2009}.
	\item Heating and cooling rates are implemented for:
	\begin{itemize}
		\item H$_2$ processes (pumping \citep{roellig06}, line cooling \citep{roellig06}, formation- and dissociation processes \citep{sternberg1989,jonkheid2004}
		\item Gas-grain heating/cooling \citep{tielens2005}
		\item Photoelectric heating and recombination cooling \citep{bt94}
		\item Ly-\textalpha cooling \citep{sternberg1989}
		\item Optical Oxygen-6300 \AA cooling \citep{sternberg1989}
		\item Heating by C-ionization and cosmic rays \citep{jonkheid2004}
		\item Line cooling by \oi, \cii, {\ci} (fine structure), CO, $^{13}$CO, CH, CH$^+$, and HCO$^+$ (rotational) assuming a thermalized H$_2$ ortho/para ratio
	\end{itemize}
	The model output is:
	\begin{itemize}
		\item $A_V$ grid
		\item Abundances (cm$^{-3}$)
		\item Column densities to the edge of the PDR (cm$^{-2}$)
		\item Gas and dust temperature (K)
		\item Heating and cooling rates (erg s$^{-1}$ cm$^{-3}$)
		\item Total heating and cooling rates and sum of them (erg s$^{-1}$ cm$^{-3}$)
		\item Surface brightness of CO, $^{13}$CO, CH, CH$^+$, HCO$^+$,  \oi, \cii, and {\ci} (erg s$^{-1}$ cm$^{-2}$ sr$^{-1}$)
	\end{itemize}
	
\end{itemize}
The code can be downloaded at \url{https://github.com/markusroellig/WL-PDR}.
\end{appendix}
\end{document}